\documentclass[1p, preprint, onecolumn]{elsarticle}
\usepackage[T1]{fontenc}
\usepackage{lineno}
\usepackage{xcolor}
\usepackage{amsthm,amsmath,amssymb,amsfonts}
\usepackage{tabularx}
\usepackage{multirow}
\usepackage{algorithm,algorithmic}
\usepackage{color}
\usepackage[export]{adjustbox}
\usepackage[breaklinks=true]{hyperref}
\usepackage{xurl}
\usepackage{xcolor}
\usepackage{bbm}
\usepackage{booktabs}
\usepackage{subcaption}
\usepackage{graphicx}
\usepackage{pdflscape}

\biboptions{sort&compress}

\usepackage{comment}

\DeclareRobustCommand{\VAN}[3]{#2}
\let\VANthebibliography\thebibliography
\def\thebibliography{\DeclareRobustCommand{\VAN}[3]{##3}\VANthebibliography}

\journal{International Journal of Fatigue}

\begin{document}
\begin{frontmatter}
		
\title{Discovery of Fatigue Strength Models via Feature Engineering and automated eXplainable Machine Learning applied to the welded Transverse Stiffener}

\author[TUD]{Michael A. Kraus}
\author[RWTH]{Helen Bartsch}
	             
\affiliation[TUD]{organization={Professur Baustatik, ISMD, TU Darmstadt},
	            addressline={Franziska-Braun-Str. 3}, 
	            city={Darmstadt},
                postcode={64287},
                country={Germany}}

\affiliation[RWTH]{organization={Institut für Stahlbau, RWTH Aachen University},
	             addressline={Mies-van-der-Rohe-Str. 1}, 
	             city={Aachen},
	             postcode={52074},
	             country={Germany}
	             }

\begin{abstract}
This research introduces a unified approach that combines Automated Machine Learning (AutoML) with Explainable Artificial Intelligence (XAI) techniques for the prediction of fatigue strength in welded transverse stiffener details. The methodology integrates expert-based feature engineering with algorithmic feature creation to enhance the fidelity and transparency of the model.
An initial data curation phase was conducted to clean and harmonize an extensive fatigue test database in accordance with EN 1993-1-9. Subsequently, multiple regression models—including gradient boosting machines, random forests, and neural networks—were trained via AutoML across three model configurations, each varying in their level of domain-informed feature parametrization. This setup enabled a comparative assessment of feature selection strategies rooted in engineering knowledge versus algorithmic heuristics.

From the ML perspective, ensemble methods (e.g., CatBoost, LightGBM) showed the most robust performance, while among the formulated and tested hypothesis models, \(\mathcal{M}_2\) delivered the best trade-off achieving a test RMSE $\approx 30.6$~MPa and \(R^2_{\mathrm{Test}}\approx0.780\) across the full \(\Delta\sigma_{c,50\%}\) range, and RMSE $\approx 13.4$~MPa with \(R^2_{\mathrm{Test}}\approx0.527\) within the engineering-relevant 0 - 150~MPa domain. Models with denser feature sets (\(\mathcal{M}_3\)) showed marginal training gains but poorer generalization, while models with base features (\(\mathcal{M}_1\)) performed on par, confirming the stability of simpler configurations. 
XAI analyses (SHAP and feature importance) identified stress ratio \(R\), stress range \(\Delta\sigma_i\), yield strength \(R_{eH}\), and post-weld treatment—especially TIG dressing vs. as-welded—as dominant predictors, supporting established fatigue mechanics. Secondary influence of geometric descriptors (plate width, throat thickness, stiffener height) emphasizes local design effects on fatigue life.

The proposed framework demonstrates that AutoML combined with XAI can yield accurate, interpretable, and robust fatigue strength models for welded steel structures. The approach bridges data-driven modeling with domain-centric validation, offering new ways for AI-assisted design and assessment in structural engineering. Future work will extend the framework to probabilistic fatigue life prediction and integration into digital twin applications.
\end{abstract}

\begin{keyword}
Fatigue Strength Prediction \sep Machine and Deep Learning \sep explainable AI \sep Fatigue Data Base \sep Fatigue Details \sep Transverse Stiffener \sep Automated Machine Learning

\end{keyword}

\end{frontmatter}


\section{Introduction}
The fatigue design of steel components presents a significant challenge in various sectors of infrastructure, particularly affecting the lifespan of critical structures such as bridges and wind turbine towers. The fatigue resistance of welded details is typically assessed through fatigue analysis according to EN 1993-1-9\cite{prEN2023}, where the fatigue strength values are taken from a detail table. This detail table provides reference values for the fatigue strength $\Delta \sigma_{c}$ of various construction details, typically in a generalized form. However, it does not account for important distinctions such as local geometry, fabrication characteristics (e.g. weld shape or welding process), material properties, environmental conditions, and others. As a result, overly conservative limits for fatigue resistance are often applied. Therefore, the generalized fatigue strength values provided in the detail table often lead to overly conservative design, as they do not account for important influencing factors. A more precise fatigue evaluation that considers these influencing factors leads to a more economical design with longer-lasting structures, which supports sustainable resource use and reduces CO$_2$ emissions in construction. However, this requires a thorough understanding of the influencing factors. To address this challenge, we propose an automated machine learning framework in combination with domain expertise in the remainder of this paper.

\section{State-of-the-Art and Literature Review}
\label{sec:SotA_LitRev}
\subsection{To Fatigue Strength of the Transverse Stiffener}
\label{sec:SotA_fatigue}
\textbf{Fatigue testing }of important welded details, such as the transverse stiffener, together with predictive model development was subject to intensive research over the past decades, covering classical mechanics-based or phenomenological methods and increasingly advanced data-driven machine learning and deep learning approaches. This section aims to provide a comprehensive overview of the state-of-the-art in this field with emphasis on recent work not older than 10 years and a focus on phenomenological-statistical modelling rather than mechanics-based simulations, excluding details with a topical focus on classical or ML- or DL-based modelling of fatigue crack growth, remaining fatigue life, ultra low cycle fatigue life, and continuous damage mechanics.

Classical fatigue testing methods have long been the foundation of structural integrity assessments for welded components, including transverse stiffeners. These methods typically involve systematic cyclic loading tests to evaluate the fatigue life and performance of structural elements. 

While very elementary, fundamental experiments on transverse stiffener were carried out in the 60s and 70s of the last century (e.g. \cite{gurney1962},  \cite{fisher1974}), current investigations concentrate on more specialised topics. 
Many studies include the influence of important post-treatment methods, such as HFMI, where favourable compressive stresses are introduced by hammering \cite{duerr2017,alden2020}. Post-treatment methods are also often investigated on high-strength steels, as they are particularly efficient here \cite{gkatzogiannis2021}.  Other specific studies focus, for instance, on the impact of low temperatures on fatigue \cite{braun2020}, the effects of hot-dip galvanizing \cite{ferraz2024} or weld imperfections \cite{Bartsch2023b}. 

In addition to experimental approaches, \textbf{numerical simulations} have become increasingly important in fatigue analysis. Nascimento et al. performed extensive nonlinear numerical simulations to investigate the axial forces in transverse stiffeners, revealing discrepancies between design assumptions and experimental observations \cite{nascimento2023}. Studies on the positive effects of post-treatment methods have also already been analyzed numerically \cite{loschner2024}. It is important to note that while many influences can be investigated numerically, test data still provide reliable results, especially when it comes to specific production and material-related factors. This is why current research is focusing on data-driven methods. 

\subsection{To Data-Driven Investigations and Modelling of Fatigue}
\label{sec:SotA_DDI_fatigue}
Since fatigue is a complex problem with many influencing factors, data-driven investigations together with \textbf{Machine Learning (ML) and Deep Learning (DL)} modelling methods are predestined to be applied. These advanced computational techniques introduce novel paradigms for predictive modelling and model development revealing complex patterns and relationships within (large) datasets that are often challenging to observe through traditional statistical or analytical methods.

As far as the \textbf{application in the area of fatigue} is concerned, data-driven surrogate modelling of fatigue is investigated since three decades. The applicability of artificial neural networks (ANNs) for the prediction of the fatigue resistance and the fatigue crack growth has been investigated by several different studies. \cite{Han1995} introduced the neural network method to evaluate the fatigue life of welds with welding defects. The results showed that the method is applicable for evaluate the fatigue live of welds, if the ANN is trained by sufficiently good defect images. In \cite{Pleune2000} an ANN was trained by a data base of 1036 fatigue tests, and then used to predict fatigue life for specified sets of loading and environmental conditions. From the results it was concluded that the ANN can find the fatigue life for any set of conditions due to finding trends and patterns in the data. Furthermore, it is stated that ANN can interpolate effects by learning trends and patterns when data is not available. The strain-life fatigue properties were predicted using ANN in \cite{Genel2004}, demonstrating that a well-trained network is a reliable tool for the prediction of fatigue properties. It is stated that the best prediction quality is observed for a network with five inputs and one output. Integration of ML with structural health monitoring systems has been explored by Ghahremani et al., towards fatigue testing of retrofitted web stiffeners on steel highway bridges \cite{ghahremani2013}. \cite{zhu2018} used artificial neural networks to predict fatigue crack propagation in steel plates. In \cite{wang2019} support vector machines have been applied to estimate fatigue damage in offshore steel structures. Furthermore, \cite{zhao2020} used extreme gradient boosting (XGBoost) for fatigue crack growth prediction in steel structures. Yang \cite{yang2020} implemented a hybrid ML approach combining genetic algorithms and neural networks for fatigue life assessment of steel components. In \cite{lee2020} a random forest algorithms was implemented to predict fatigue crack growth in steel bridges. Moreover, \cite{zhang2021} developed a deep learning model using convolutional neural networks for fatigue life prediction of welded steel joints.  \cite{li2021} developed a transfer learning-based method for fatigue strength prediction of steel welded joints. \cite{braun2022comparison} demonstrates that gradient boosted trees combined with SHAP analysis outperform traditional stress concentration factor (SCF)-based methods in predicting both fatigue failure location and number of cycles to failure for small-scale butt-welded joints, while also quantifying the mutual influence of weld geometry and other contributing factors. \cite{wu2022} developed a graph neural network approach for fatigue life prediction of steel joints. Also, \cite{liu2023a} applied transformer networks for time-series fatigue data analysis in steel structures. A multi dimensional multi scale composite neural network with multi depth is introduced in \cite{Pan2024} to predict multiaxial fatigue life properties. \cite{Zhang2024} proposed a Symbolic Regression Neural Network framework for the prediction of multiaxial fatigue properties, which combines traditional MSE with insights derived from linear regression. The underlying physical relationships are integrated through symbolic regression into the loss function, whereby the predictive accuracy is significantly improved compared to conventional models and standard neural networks.

As far as the \textbf{innovative development of the methodology} is concerned, several works propose probabilistic methods for uncertainty quantification, as well as generative design and design optimization. \cite{chen2019} applied Bayesian neural networks for probabilistic fatigue life prediction of steel bridges. Moreover, \cite{tan2020} implemented a deep belief network for fatigue reliability assessment of steel bridges. Additionally, \cite{wang2022} applied federated learning for collaborative fatigue analysis across multiple steel structure datasets. \cite{zhang2023} developed an attention-based neural network for multiaxial fatigue life prediction in steel components. Generative AI techniques have emerged as powerful tools for design space exploration and automated concept generation \cite{balmer2024design,bucher2023performance}. Silva-Campillo et al. proposed design criteria for scantling of connections under fatigue loading, utilizing machine learning algorithms to optimize structural performance \cite{silva2021}. Liu et al. employed generative adversarial networks (GANs) to evaluate the fatigue performance of stiffener-to-deck plate welds in orthotropic steel decks, showcasing the potential of machine learning to enhance traditional fatigue assessment methods \cite{liu2023}. This approach allows for the identification of complex patterns in fatigue data that may not be readily apparent through classical analysis. \cite{park2021} used reinforcement learning for optimizing inspection schedules for fatigue-prone steel components.

Recent studies have integrated physics-informed neural networks (PINNs) into structural analysis, particularly for fatigue property prediction in steel structures \cite{kraus2020physik}. \cite{kim2021} developed a PINN for fatigue damage estimation in steel beams, while Zhou et al. \cite{Zhou2023} introduced a probabilistic PINN framework incorporating the physics of fatigue mechanisms. Jing et al. \cite{Jing2024} proposed hierarchical neural network models optimized for limited experimental data, achieving high accuracy and robust extrapolation within a twofold error band. Dong et al. \cite{Dong2025} combined the M-integral fatigue model with neural network fitting, deriving a physically consistent loss function that enhances predictive performance and generalization. Feng et al. \cite{Feng2024} integrated defect-sensitive constraints into a PINN to predict high-cycle fatigue life of SLM 316L steel, demonstrating superior accuracy when partial differential equations of fatigue mechanics were embedded. Li et al. \cite{Li2024} applied a modified PINN to orthotropic steel deck joints, incorporating implicit activation functions and explicit physical constraints into the loss function, improving accuracy compared to conventional PINNs. Bartosak et al. \cite{Bartosak2025} investigated laser powder bed fusion 316L steel, implementing the Smith-Watson-Topper damage model within a PINN to capture defect-driven fatigue behavior. Additionally, Halamka et al. \cite{Halamka2023} predicted fatigue life under multiaxial loading using a hybrid PINN with a logarithmic activation function, enforcing the power-law relationship between damage and fatigue life. He et al. \cite{He2023} compared three multiaxial fatigue life models—shear strain critical plane (FS), critical plane (SW), and maximum normal strain (SWT)—for PINN-based fatigue prediction. While FS and SW provided accurate predictions across various loading paths, SWT exhibited non-conservative results for torsional and out-of-phase conditions.

Decision-tree-based ML models, such as Random Forest (RF), Gradient Boosting Decision Trees (GBRT), and XGBoost, have demonstrated strong predictive capabilities in fatigue analysis. However, their lack of interpretability has driven the integration of Explainable Artificial Intelligence (XAI) techniques, particularly SHAP (Shapley Additive Explanations), to provide insights into model decision-making. Recent research applies these XAI-driven tree models to structural health monitoring (SHM) datasets for fatigue prediction. Ma et al. \cite{ma2023} employed XGBoost with SHAP to analyze fatigue crack propagation in orthotropic steel decks, revealing critical features such as bridge age and loading conditions. Sun et al. \cite{sun2023} used GBRT for fatigue damage prediction in suspension bridges, demonstrating superior accuracy ($R^2 \approx 0.9$) compared to artificial neural networks (ANN) and support vector machines (SVM). Similarly, Li and Song \cite{li2023} trained an XGBoost model on National Bridge Inventory (NBI) data, utilizing SHAP to rank influential factors affecting bridge deck deterioration while addressing data imbalance with adaptive synthetic sampling (ADASYN). In industrial fatigue applications, Serradilla et al. \cite{serradilla2023} leveraged RF with explainability techniques like LIME and ELI5 to estimate the remaining useful life (RUL) of fatigued components, aligning expert knowledge with ML-driven insights. Beyond infrastructure, XAI-based fatigue analysis extends to wind turbine damage diagnostics, where Movsessian et al. \cite{movsessian2023} integrated SHAP with ML models to enhance interpretability. Peralta et al. \cite{peralta2021,peralta2022} applied SHAP-based anomaly detection to pedestrian bridge monitoring, though fatigue-specific insights remain underdeveloped. Comparative studies by Ma et al. \cite{ma2023} and Sun et al. \cite{sun2023} benchmarked XGBoost, RF, and GBRT against ANN and SVM, consistently highlighting the superior trade-off between interpretability and predictive accuracy in structural fatigue assessments. SHAP-based feature importance analyses further validated the effectiveness of decision-tree ensembles for real-world applications. Despite these advances, research gaps persist in data scarcity, temporal modeling, and uncertainty quantification. Ma et al. \cite{ma2023} noted the limited availability of fatigue crack data in bridge inspections, impacting model generalization, while Li and Song \cite{li2023} mitigated data imbalance through synthetic oversampling. However, fatigue is inherently time-dependent, and traditional tree-based models lack sequential learning capabilities necessary for lifetime estimation. 

\subsection{To the presented study}
\label{sec:SotA_thisstudy}

Based on these previous studies, in this paper we propose for the first time an end-to-end automated machine learning (AutoML) framework that leverages domain knowledge for data imputation and empirical relationships to formulate model hypotheses for accurate fatigue life prediction, exemplary for the transverse stiffener detail according to EN 1993-1-9 \cite{prEN2023}. First, a dataset comprising fatigue-related features and strength values of the transverse stiffener detail based on publications belonging to the background of EN 1993-1-9 as well as newer international research studies, cf.~Sec.~ \ref{sec:Methods_DataCollection} has been edited. Second, domain expertise guided the design of data approaches for handling and imputing missing values. Third, fatigue strength models were developed by benchmarking multiple ML and DL algorithms across three model hypotheses within an AutoML framework. The AutoML benchmarking process also incorporated automated feature engineering and creation methods, alongside an evaluation of feature importance using Shapley Additive Explanations (SHAP) and permutation feature importance. These results were subsequently analysed through domain knowledge to assess their validity and interpretability, providing novel insights into predictive modelling of fatigue strength.

\section{Methods}
\begin{figure}[h]
  \centering
  \includegraphics[width=0.45\textwidth]{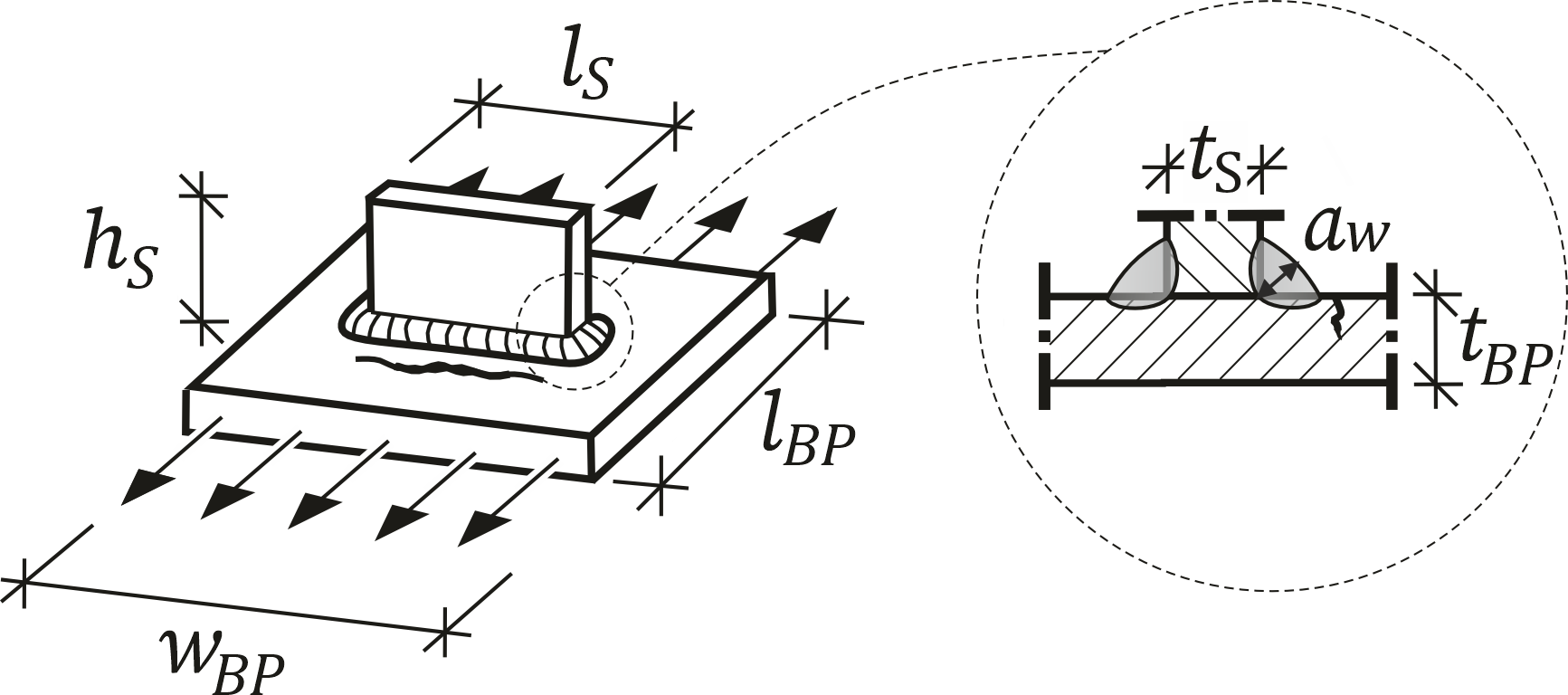}
  \caption{Designations of geometry parameters of transverse stiffener}
  \label{fig:TraverseStiffener}
\end{figure}

\begin{figure}[h]
  \centering
  \includegraphics[width=0.300\textwidth]{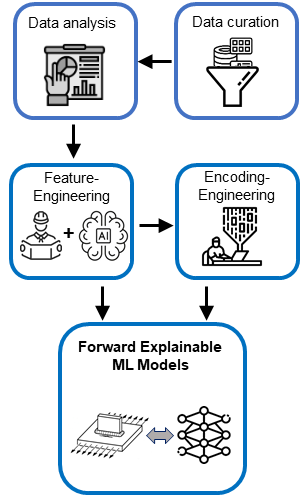}
  \caption{Research procedure.}
  \label{fig:procedure}
\end{figure}

In order to systematically develop holistic yet robust fatigue strength prediction models for the fatigue detail under consideration, cf.~Fig.~\ref{fig:TraverseStiffener}, a multi-stage research procedure combining data-driven as well as domain-expertise-based feature engineering with XAI modelling has been employed as shown in Fig.~\ref{fig:procedure}.

The starting point for this research is a raw set of fatigue testing data from the a European Database, cf.~Sec.~\ref{sec:Methods_DataCollection} for the welded transverse stiffener. Initially, an exploratory data analysis (EDA) approach was used to evaluate data quality, detect outliers, and identify potential correlations among relevant fatigue-related variables, cf.~Sec.~\ref{sec:Methods_FeatureEng}. From the raw data we curated the dataset by eliminating erroneous entries, filling missing values using data imputation strategies motivated by domain-knowledge and aligning data into a standardized format. Subsequent feature engineering combined algorithmic and statistical methods with expert-informed background, thereby capturing essential factors known to influence fatigue phenomena. The feature encoding procedures were tailored to each variable’s characteristics, ensuring consistent representation and mitigating biases from incomplete records, cf.~Sec.~\ref{sec:Methods_FeatureEng}. Once the dataset was refined, we employed AutoML pipelines, featuring diverse explainable baseline, tree and forest as well as neural network algorithms, to efficiently traverse the model space and identify optimal configurations and hyperparameters and gain insight into driving features. Beyond predictive performance, interpretability was ensured through model-agnostic XAI methods. Permutation-based feature importance and SHAP (SHapley Additive Explanations) values were computed to elucidate each model’s reasoning, quantify the global and local impact of input variables, and validate alignment with established fatigue mechanisms. This integrated pipeline - spanning from structured data curation to interpretable learning -combines computational efficiency with domain-aligned insight, ultimately yielding robust models that reflect the complex, nonlinear nature of fatigue for this important detail in civil engineering structures.

\subsection{Data Collection}
\label{sec:Methods_DataCollection}

The basis of this study is the European Fatigue Database, which was developed through previous research efforts \cite{Bartsch2018,Feldmann2019,Bartsch2019, Bartsch2020a, Bartsch2020b, Bartsch2021a, Bartsch2021b}. This database encompasses comprehensive test data from fatigue experiments conducted on a wide variety of specimens, each with distinct properties. Detailed documentation is provided for all relevant characteristics of the test specimens and procedures, including geometry, material, loading conditions, weld properties, and environmental factors, alongside the corresponding test results. Consequently, the database serves as a robust foundation for the analyses presented in this study. For the purpose of this analysis, the representative detail of the transverse stiffener is selected.

 The purpose of the database originally was to re-evaluate the detail catalogue in EN 1993-1-9 using standardized methods. The foundation was primarily based on the background of EN 1993-1-9, with additional new publications added. The database includes not only test results but also all potentially relevant properties of the test specimens:
 
\begin{itemize}
    \item[-] general
    \item[-] geometric properties
    \item[-] material properties
    \item[-] welding properties
    \item[-] loading type and condition
    \item[-] environmental factors
    \item[-] fatigue strength test results
\end{itemize}

The data base contains 23,601 test results for all details included. For the study presented, the transverse stiffener has been selected, one of the most commonly investigated fatigue details with 2,470 test results in total. Previous studies have demonstrated the potential of the fatigue test database for re-evaluating fatigue strength values using established statistical \cite{Bartsch2023b}, numerical \cite{Bartsch2021c, Bartsch2022a}, and data-driven \cite{BartschNN2024} approaches. With this paper, we add a data scientific treatment to the dataset at hand outlined in the following sections.

\begin{landscape}

\begin{table}[ht!]
\centering
\caption{Features and targets of the data set: symbols, descriptions, units, data types, value ranges, and ratio of missing data.}
\begin{tabular}{|c|l|l|l|l|c|}
\hline
\textbf{Symbol}         & \textbf{Description}                                   & \textbf{Unit}   & \textbf{Data Type}  & \textbf{Value Range}           & \textbf{Missing Values [\%]} \\ 
\hline
\multicolumn{6}{|c|}{\textbf{Features} $\mathbf{X}$} \\
\hline
\textit{Scale}          & Test Scale                                           & -               & Binary              & \{small, large\}               & 0.00                          \\ 
\textit{Loading}           & Loading Type                                         & -               & Binary              & \{axial, bending\}             & 0.00                          \\ 
\textit{$I_A$}           & Constant or variable Amplitude                                         & -               & Binary              & \{yes, no\}             & 4.30                          \\
\textit{$f_T$}           & Testing Frequency                                         & Hz             & Float               & [1, 32.0]             & 58.9                          \\
$R_{eH}$                   & Yield Strength                                      & MPa             & Float               & [235, 1125]                     & 6.15                      \\ 
$R_m$                   & Tensile Strength                                   & MPa             & Float               & [275, 1420]                     & 5.14                      \\ 
\textit{Pre-Treat}      & Weld Pre-Treatment                             & -               & Categorical         & \{e.g. none, heat, other\}          & 0.00                          \\ 
\textit{Post-Treat}   & Weld Post-Treatment                                & -               & Categorical         & \{e.g. none, heat, grind\}          & 0.00                          \\ 
Weld type                   & Weld Type                                & -               & Categorical         & \{fillet weld, butt weld\}                    & 25.3                     \\ 
$R_{eH,filler}$           & Yield Strength Filler Material                 & MPa             & Float               & [200, 800]                     & 89.8                     \\ 
$R_{m,filler}$            & Tensile Strength Filler Material               & MPa             & Float               & [300, 900]                     & 89.8                     \\ 
$l_{BP}$                   & Base Plate Length                                         & mm              & Float               & [50, 2000]                     & 16.6                     \\ 
$w_{BP}$                   & Base Plate Width                                          & mm              & Float               & [10, 500]                      & 7.49                       \\ 
$t_{BP}$                   & Base Plate Thickness                                      & mm              & Float               & [1, 100]                       & 0.324                      \\ 
$h_S$                   & Stiffener Height                                    & mm              & Float               & [5, 300]                       & 14.4                     \\ 
$l_S$                   & Stiffener Length                                    & mm              & Float               & [10, 1000]                     & 7.45                      \\ 
$t_S$                   & Stiffener Thickness                                 & mm              & Float               & [1, 50]                        & 0.00                          \\ 
$a_w$                   & Weld Thickness                                     & mm              & Float               & [1, 20]                        & 8.98                      \\ 
\textit{Corrosion}      & Corrosive Conditions                                 & -               & Binary         & \{yes, no\}                    & 0.00                          \\ 
$R$                     & Stress Ratio                                              & -               & Float               & [-1.0, 0.80]                         & 9.80                      \\ 
$\Delta\sigma_i$          & Stress Range                                     & MPa             & Float               & [50, 1125]                       & 0.00                          \\ 
\hline
\multicolumn{6}{|c|}{\textbf{Targets} $\mathbf{Y}$} \\
\hline
$N_i$        & Number of Cycles until Failure                     & -               & Float               & [0, 1]                         & 0.00                          \\ 
$\Delta\sigma_{c,50\%}$        & Fatigue Strength at 2 000 000 Cycles                                  & MPa             & Float               & [0, 500]                       & 0.00                          \\ 
\hline
\end{tabular}
\label{tab:feature-description-expanded}
\end{table}

\end{landscape}

\subsection{Data Cleaning, Exploratory Data Analysis, Handling Missing Data, and Feature Engineering}
\label{sec:Methods_FeatureEng}

The raw data was initially screened and specific data types were assigned based on the nature of the specific feature as well as engineering judgement. Features representing categorical data, e.g. "Welding Process Groups", were cast as string types, while numerical columns like "Base Plate Thickness" and "Tensile Strength" were defined as float types. This ensured proper handling during subsequent analysis. The features $\mathbf{X}$ as well as targets $\mathbf{Y}$ included in the data set are summarized in Table \ref{tab:feature-description-expanded} with its corresponding formula letter, description, and unit.


Exploratory Data Analysis (EDA), incorporating descriptive statistics and custom visualizations such as histograms, was employed to systematically assess the structure and quality of the dataset and to detect potential anomalies or inconsistencies \cite{tukey1977exploratory}. These analyses informed the selection of relevant features and target variables through a combination of domain expertise and data-driven inspection, laying the foundation for robust model development. Particular attention was paid to handling missing data, which was addressed through feature-specific strategies based on visual inspection, statistical summaries, and engineering judgment. Where default values were known but not recorded, entries were filled accordingly; in cases of uncertainty, targeted imputation methods were applied (cf.~Sec.~\ref{sec:Methods_DataImputation}) \cite{little2019statistical}. Furthermore, features lacking engineering significance—such as placeholders or highly incomplete variables—were systematically excluded to ensure a streamlined and meaningful feature space for subsequent modeling efforts.

\subsubsection{Feature Engineering and Golden Feature Creation}
\label{sec:Methods_GoldenFeatureEng}

Feature engineering was carried out to refine the dataset and enhance its suitability for predictive modeling, with transformations grounded in domain-specific knowledge and tailored to project requirements \cite{guyon2003introduction, vanderplas2016python}. Each feature underwent detailed analysis to assess its relevance, data quality, and proportion of missing values. One of the key steps involved combining related categorical variables, such as "Weld Post-Treatment 1" and "Weld Post-Treatment 2," into a single feature named "Weld Post-Treatment": in some fatigue tests, more than one post-treatment has been performed, accordingly the data base contained two columns for these characteristics. For the present study however, two single features are not reasonable. This aggregation reduced dimensionality and consolidated information into broader, more interpretable categories for downstream analysis. Similarly, categorical features like "Weld Process Group" were simplified by collapsing subcategories into larger groups based on expert knowledge, such as merging process classifications into a single representative class for efficiency and clarity. Numerical features, including "Yield Strength," "Tensile Strength," and "Testing Frequency," were systematically examined for outliers and missing values. Based on multicollinearity checks, cf. Sec.~\ref{sec:Methods_Multic} we also derived features.

Golden Features are engineered features derived from the original dataset that exhibit high predictive power. These features are constructed by exploring combinations of original features using mathematical operations such as subtraction and division. The \texttt{mljar-supervised} package includes a dedicated step for Golden Feature discovery. The process begins by generating all unique pairs of features from the original dataset. For each feature pair, a new feature is generated using subtraction or division. To assess the utility of each new feature, a Decision Tree with a maximum depth of 3 is trained using only the constructed feature. The evaluation process involves randomly selecting samples from the dataset for training and testing. The predictive performance of each feature is measured using log loss for classification tasks or mean squared error for regression tasks. Features are ranked based on their performance scores, with lower scores indicating better predictive capabilities. The top-ranked features, typically constituting 5\% of the number of original features (but not fewer than 5 or more than 50), are selected as Golden Features. These features are then incorporated into the training dataset, and their construction details are saved for later inspection and model interpretation. This automated process enables the identification of highly informative features, enhancing the model's performance by leveraging novel feature interactions. A post-hoc inspection process of the data-driven golden feature creation and selection process is applied.

\subsubsection{Correlation Analysis and Multicollinearity Assessment}
\label{sec:Methods_Multic}

Correlation analysis was performed to identify highly correlated features, facilitating the removal of multicollinear features that could adversely impact model performance \cite{ guyon2003introduction,vanderplas2016python, james2023introduction}. To assess multicollinearity among continuous features, the Variance Inflation Factor (VIF) was calculated. VIF is a widely used diagnostic tool in regression analysis that quantifies the extent to which the variance of a coefficient is inflated due to linear relationships with other predictors in the model. While a VIF value greater than 1 indicates some degree of multicollinearity, values exceeding 5 or 10 are typically regarded as critical thresholds, suggesting significant multicollinearity. Such high values can impair the stability of regression coefficients and reduce the interpretability of results.

The VIF for a given feature was calculated using the following formula:

\begin{equation}
\text{VIF}_i = \frac{1}{1 - R_i^2}
\end{equation}

where \( R_i^2 \) is the coefficient of determination obtained by regressing feature \( i \) against all other features in the dataset. For each continuous feature, a regression model was constructed with the respective feature treated as the dependent variable and the remaining features as independent variables. The \( R_i^2 \) value from this model was then used to compute the VIF.

The analysis was conducted exclusively on a subset of continuous features identified during the feature engineering process. Features with a VIF value exceeding 5 were flagged for further investigation, and features with excessively high VIF values were considered for removal to improve the robustness of the dataset. The results of the VIF calculations were summarized in a table and exported for evaluation and documentation. This approach ensured that the final dataset was free from problematic multicollinearity, enhancing the stability and interpretability of downstream regression models. For additional information on VIF and its applications, refer to \cite{james2023introduction}.

\subsubsection{Data Imputation Strategies}
\label{sec:Methods_DataImputation}

To address the remaining missing values in the dataset, a comprehensive and tailored imputation strategy was developed and implemented using the \texttt{scikit-learn} library together with domain expertise towards the kind of imputation strategy. The approach combined multiple imputation methods, including median imputation, random imputation, and constant value imputation, depending on the characteristics of the feature. These methods were incorporated into a preprocessing pipeline to ensure a systematic and reproducible data transformation process.

Median imputation was applied to features with relatively symmetric distributions, such as \texttt{Scale} (Test Scale) and \texttt{Loading} (Loading Type), using the \texttt{SimpleImputer} class. For numerical features exhibiting variability and complex distributions, such as \texttt{$R_{eH}$}, a custom random imputation strategy was employed. The random imputation class was built on \texttt{SimpleImputer}, where missing values were filled with randomly sampled values from the existing, non-missing entries of the respective feature. This ensured that the imputed values preserved the underlying variability and did not introduce bias.

Constant value imputation was used for categorical features where domain knowledge suggested fixed replacements. For example, missing entries in \texttt{Post-Treat} were filled with \texttt{"no weld post-treatment"}, while missing values in \texttt{Weld Type} were assigned a default class of \texttt{"Fillet Weld"}. One reason for this assumption is that any post-treatment would typically be documented. Additionally, fillet welds are considered the standard weld seam shape for transverse stiffeners, as these stiffeners are non-load-bearing components, making butt welds too costly and unnecessary in practice. Similar strategies were applied to other categorical features, such as process groups.

A \texttt{ColumnTransformer} was used to assign the appropriate imputation method to each group of features, ensuring scalability and modularity. The imputation process was integrated into a preprocessing pipeline, which also included steps for feature encoding, power transformations, and scaling. This pipeline approach allowed for seamless imputation and transformation of the dataset in a single operation, significantly reducing preprocessing complexity.

The imputation strategy was validated through exploratory data analysis to confirm that the imputed values aligned with the observed distributions of the original data. The result was a cleaned dataset that retained its original structure and variability, ready for downstream modelling.

\subsubsection{Data Transformations}
\label{sec:Methods_DataTrafo}

Data transformation was implemented through a structured pipeline to ensure consistency and scalability. The pipeline integrated multiple preprocessing steps, including feature encoding (as described in the previous section), power transformations, and scaling. Encoding was achieved using a \texttt{FunctionTransformer} for binary and one-hot encoding of categorical variables. The target, the average value of the fatigue strength at 2,000,000 cycles $\Delta\sigma_{c,50\%}$, was first transformed using the decadic logarithm, then a Yeo-Johnson-power transformation was employed to stabilize variance in the target, and numerical features were scaled using standardization.

\subsection{Performance Quantities and Evaluation Metrics for AI models}
\label{sec:Methods_evaluation_metrics}

For training of the ML resp. DL models, only the root mean squared error (RMSE) loss was chosen for its sensitivity to large errors, which is crucial in this regression task. The post-hoc model quality evaluation is conducted using mean absolute error (MAE), coefficient of determination ($R^2$) together with the error standard deviation $\sigma_E$ using:

\begin{equation}
    \text{MAE} = \frac{1}{n} \sum_{i=1}^{n} |y_i - \hat{y}_i|
\end{equation}

\begin{equation}
    \text{MSE} = \frac{1}{n} \sum_{i=1}^{n} (y_i - \hat{y}_i)^2
\end{equation}

\begin{equation}
    \text{RMSE} = \sqrt{\text{MSE}}
\end{equation}

\begin{equation}
    R^2 = 1 - \frac{\sum_{i=1}^{n} (y_i - \hat{y}_i)^2}{\sum_{i=1}^{n} (y_i - \bar{y})^2}
\end{equation}

where $n$ is the size of the training resp. test set, $y$ is the target quantity, $\hat{y}$ the model prediction and $\bar{y}$ the mean of the targets in the dataset. The model quality evaluation is done using the above defined metrics for the whole dataset as well as a filtered dataset in the range of $\Delta\sigma_{c,50\%} \in [0;150]$~MPa. For ex-post analysis visualisation, the model's predictions are compared to actual values using scatter plots, with additional lines indicating ±1.5 and ±2 standard deviations of the error. The standard deviation of the error is calculated for both training and test sets to assess the model's precision.

\subsection{Fatigue Model Hypotheses \texorpdfstring{$\mathcal{M}_1$}{M1}, \texorpdfstring{$\mathcal{M}_2$}{M2}, \texorpdfstring{$\mathcal{M}_3$}{M3}}
\label{sec:Methods_FAT_Models}

In the course of this paper, three fatigue model hypotheses \texorpdfstring{$\mathcal{M}_1$}{M1}, \texorpdfstring{$\mathcal{M}_2$}{M2}, \texorpdfstring{$\mathcal{M}_3$}{M3} are formulated via selection of a subset of features based on engineering domain knowledge towards fatigue and benchmarked with several ML / DL algorithms in an AutoML style, enhanced with automated feature generation as described in the above sections.

All model hypothesis \texorpdfstring{$\mathcal{M}_i$}{Mi} within this paper use different sets of features to regress the transformed (cf. Sec.~ \ref{sec:Methods_DataTrafo}) target $\Delta\sigma_{c,50\%}$. These features include geometric properties, material properties, loading characteristics, weld types, and various post-treatment methods. Three model variants were developed with progressively expanded feature sets:

\subsubsection{Model \texorpdfstring{$\mathcal{M}_1$}{M1}: Base Configuration}
$\mathcal{M}_1$ incorporates 14 core features: geometric parameters ($l_{BP}$, $h_S$, $t_{BP}$, $t_S$, $a_w$), material properties ($R_{eH}$, $R_m$), loading conditions ($R$, $\Delta\sigma$), and categorical descriptors (\textit{Loading}, \textit{Scale}). Post-treatment effects are captured through one-hot encoded variables including Post-Treatment.

\subsubsection{Model \texorpdfstring{$\mathcal{M}_2$}{M2}: Position-Augmented}
Building on $\mathcal{M}_1$, $\mathcal{M}_2$ adds 2 new features: welding position (acc. to EN ISO 6947) 
and welding process (acc. to EN ISO 4063). This expands the model's capacity to account for spatial orientation effects and fabrication methods. The geometric and material parameters remain identical to $\mathcal{M}_1$.

\subsubsection{Model \texorpdfstring{$\mathcal{M}_3$}{M3}: Full Process Model}
$\mathcal{M}_3$ introduces 3 additional process-sensitive features: Testing Frequency $f_T$, filler material properties (e.g. $R_{eH,filler}$), and Pre-Treatment methods. With 19 total features, this configuration provides complete process chain representation - from base materials ($R_{eH}$, $R_m$) through fabrication (welding characteristings) to post-treatment. The model explicitly captures interactions between mechanical loads (e.g. $\Delta\sigma$), geometric constraints (e.g. $l_{BP}$, $h_S$), and manufacturing variables (e.g welding process, welding position).

\subsection{Machine and Deep Learning Algorithms}
\label{sec:Methods_MLDL_Algos}

This study evaluates eight baseline machine learning algorithm families for regression-based prediction of fatigue strength ($\log_{10} ( \Delta\sigma_{c,50\%})$). The benchmark models comprise: (1) linear regression, (2) tree-based ensembles (Random Forest, Extra Trees, LightGBM, XGBoost, CatBoost), and (3) a feedforward neural network. Each algorithm undergoes rigorous hyperparameter optimization, cf. Sec.~\ref{sec:Hyperparametertuning} using 5 random initialisations.

\subsubsection{Baseline Model 1: Baseline Model}
\label{sec:Methods_baseline_model}

The baseline model serves as a point of reference for evaluating the performance of more sophisticated algorithms. This model predicts the target variable by computing the mean of the target values in the training data. Although simplistic, baseline models are crucial in machine learning approaches to establish a minimal benchmark that any predictive algorithm must exceed to be considered effective. They provide insights into the inherent difficulty of the problem and the potential room for improvement through more complex modeling approaches \cite{hyndman2018forecasting, smyth1997stacked}. By implementing a baseline model, we ensure that advanced algorithms meaningfully outperform simple heuristic predictions, which is a critical step in validating the added complexity of machine learning techniques.

\subsubsection{Baseline Model 2: Linear Regression}
\label{sec:Methods_linear_regression}

Linear regression is one of the simplest and most interpretable machine learning models, designed to establish a linear relationship between the features and the target variable. This paper uses ordinary least squares Linear Regression, which is defined by the equation \( y = X\beta + \epsilon \), where \( y \) is the target variable, \( X \) represents the input features (augmented with an intercept and corresponding basis "1"), \( \beta \) the coefficients, and \( \epsilon \) the error term \citep{montgomery2021introduction}. Despite its simplicity, linear regression is sensitive to multicollinearity among features, which can inflate coefficient variances and affect interpretability. Regularization techniques such as Ridge or Lasso regression can mitigate this issue but were not employed in this baseline setup \citep{james2023introduction}, instead we use the VIF method for feature selection a priori, cf. Sec~\ref{sec:Methods_Multic}.

\subsubsection{Baseline Model 3: Random Forest}
\label{sec:Methods_random_forest}

Random forest is an ensemble learning method that constructs multiple decision trees during training and aggregates their predictions to enhance model accuracy and robustness. It uses bootstrapped datasets and random feature selection for tree construction, reducing overfitting and variance compared to individual trees. This algorithm is particularly well-suited for handling datasets with missing values and non-linear relationships \cite{breiman2001random, hastie2009elements}. In this study, random forest is employed to explore the interaction effects between features while maintaining a balance between interpretability and predictive power. Its ability to rank feature importance further aids in feature selection for downstream analysis.

The hyperparameters optimized in this study include:
\begin{itemize}
    \item \texttt{n estimators}, set to 100, determines the number of decision trees in the forest. Higher values improve robustness at the cost of increased computation.
    \item \texttt{max depth}, varied from 4 to 12, controls the depth of each tree, balancing model complexity and risk of overfitting.
    \item \texttt{min samples split}, set to 2, specifies the minimum number of samples required to split an internal node.
    \item \texttt{max features}, varied from 0.5 to 1, limits the number of features considered at each split, introducing randomness and improving generalization.
\end{itemize}
Random forest also provides feature importance scores, aiding in feature selection and model interpretability \citep{hastie2009elements}.

\subsubsection{Baseline Model 4: Extra Trees}
Extra Trees, or Extremely Randomized Trees, is an ensemble learning method that builds multiple decision trees to enhance predictive performance. Unlike traditional random forest algorithms, Extra Trees introduces additional randomness during tree construction by selecting split thresholds at random instead of optimizing splits based on feature importance. This randomized approach reduces the risk of overfitting and often results in faster training times compared to other tree-based methods \cite{geurts2006extremely}. Key hyperparameters were tuned to optimize its performance:
\begin{itemize}
    \item \texttt{n estimators}, set to 100, specifies the number of trees in the ensemble. A higher value improves robustness but increases computational cost.
    \item \texttt{max depth}, varied from 4 to 12, limits the depth of individual trees to control model complexity and prevent overfitting.
    \item \texttt{min samples split}, varied from 10 to 50, determines the minimum number of samples required to split a node.
    \item \texttt{min samples leaf}, set to 1, ensures that leaf nodes have at least one sample, promoting detailed splits in the data.
    \item \texttt{max features}, varied from 0.5 to 1, restricts the number of features considered at each split to introduce additional randomness and reduce correlation between trees.
\end{itemize}

\subsubsection{Baseline Model 5: LightGBM}
\label{sec:Methods_lightgbm}

LightGBM is a gradient boosting framework designed for efficiency and scalability. Unlike traditional gradient boosting, it uses histogram-based learning and leaf-wise tree growth to reduce training time and memory usage. These features make it particularly effective for large datasets with numerous categorical and numerical variables \cite{ke2017lightgbm}. In this study, LightGBM is used to capture complex, non-linear relationships between features and the target variable. Its support for categorical encoding and optimized performance ensures it remains a competitive candidate in regression tasks. The key hyperparameters tuned included:
\begin{itemize}
    \item \texttt{num leaves}, varied from 3 to 31, controls the number of leaves per tree. Larger values increase model complexity and the risk of overfitting.
    \item \texttt{learning\_rate}, varied within $[0.05, 0.075, 0.1, 0.15]$, determines the step size for weight updates.
    \item \texttt{bagging fraction}, varied within $[0.8, 0.9, 1.0]$, limits tree depth to prevent overfitting.
    \item \texttt{min data in leaf}, varied from 5 to 50, specifies the minimum number of data points in a leaf.
\end{itemize}
LightGBM's speed and scalability make it particularly suitable for large datasets with a mix of categorical and numerical features.

\subsubsection{Baseline Model 6: eXtreme Gradient Boosting (XGBoost)}
\label{sec:Methods_baseline_models_xgboost}

The eXtreme Gradient Boosting (XGBoost) algorithm \citep{chen2016xgboost} was employed for its superior performance in handling structured data and robustness to outliers. As a scalable tree-boosting framework, XGBoost combines gradient descent optimization with $L_1$/$L_2$ regularization to minimize overfitting, while its sparsity-aware split finding efficiently handles missing values \citep{chen2016xgboost}. The model's computational efficiency is enhanced through parallelized tree construction and block-based data storage, critical for large-scale datasets \citep{chen2016xgboost}. A systematic hyperparameter tuning was conducted to optimize the following key parameters:

\begin{itemize}
    \item \textbf{tree complexity}: \texttt{max depth}, varied from 1 to 4, and \texttt{min child weight}, varied from 1 to 10, to balance model expressiveness and generalization
    \item \textbf{ensemble size}: \texttt{n estimators}, varied from 10 to 100 with early stopping (50 rounds) to prevent overfitting
    \item \textbf{Learning dynamics}: \texttt{learning rate} $\eta \in [0.01,0.5]$ for gradient step size and \texttt{subsample} $\rho \in [0.3,1.0]$ for stochastic sampling
\end{itemize}

\subsubsection{Baseline Model 7: Gradient Boosting with Categorical Support (CatBoost)}
\label{sec:Methods_baseline_models_CatBoost}

Gradient Boosting with Categorical Support (CatBoost) is a gradient boosting algorithm specifically designed to handle categorical variables efficiently without requiring extensive preprocessing. It employs ordered boosting to mitigate prediction bias and supports GPU acceleration for faster training. CatBoost is particularly effective in datasets with a high proportion of categorical features, as it eliminates the need for traditional one-hot encoding \cite{prokhorenkova2018catboost}. Its application in this study helps address the challenges posed by categorical variables in the dataset, providing a competitive edge in modeling accuracy. The key hyperparameters tuned included:

\begin{itemize}
    \item \texttt{depth}, varied from 2 to 6, controls the tree complexity
    \item \texttt{learning rate}, varied within $[0.05, 0.1, 0.2]$, determines the step size for weight updates.
    \item \texttt{n estimators}, set to 1000, specifies the number of trees in the ensemble.
    \item \texttt{rsm; subsample}, varied within 0.7 to 1, which is the percentage of features to use at each split selection.
    \item \texttt{min data in leaf}, varied from 5 to 50, specifies the minimum number of data points in a leaf.
\end{itemize}

\subsubsection{Baseline Model 8: Feed Forward Neural Networks}
\label{sec:Methods_baseline_models_nn}

The last baseline model consists of a standard feed forward neural network (NN) for regression tasks. The NN is structured with two fully-connected layers with ReLU activation function and dropout layers. Neural networks excel in handling high-dimensional data but require careful hyperparameter tuning and longer training times compared to other algorithms \cite{goodfellow2016deep, bishop2006pattern}, the hyperparameter tuning investigated specifically:

\begin{itemize}
    \item \texttt{dense 1 size}, varied within $[16, 32, 64]$, specifies the number of neurons in the first dense layer.
    \item \texttt{dense 2 size}, varied within $[4, 8, 16, 32]$, defines the neuron count in the second dense layer.
    \item \texttt{dropout}, varied within $[0, 0.1, 0.25]$, represents the fraction of input units to drop during training to prevent overfitting.
    \item \texttt{learning rate}, varied within $[0.01, 0.05, 0.08, 0.1]$, controls the step size at each iteration while moving toward a minimum of the loss function.
    \item \texttt{Momentum}, varied within $[0.85, 0.9, 0.95]$, used in conjunction with the learning rate to accelerate training by dampening oscillations.
    \item \texttt{Decay}, varied within $[0.0001, 0.001, 0.01]$, refers to the rate at which the learning rate decreases over time, helping to fine-tune the convergence process.
\end{itemize}

\subsection{AI Training and Hyperparameter Tuning}
\label{sec:Hyperparametertuning} 

In this study, a systematic AutoML approach was adopted for training and hyperparameter tuning of machine learning models 1 to 8, ensuring consistency and comparability across all algorithms. After imputation, the dataset was split into a fixed training and testing subsets, where 90\% of the data was allocated to training and 10\% to testing. The primary rationale for using a fixed split lies in maintaining a uniform evaluation protocol over all fatigue model hypotheses, enabling fair comparison of model performance without introducing variability due to random data partitioning. Furthermore, the fixed split preserves the integrity of the test set, which remains unseen during training and tuning, providing an unbiased assessment of model generalization. Loss and performance metrics for ML/DL algorithm training are defined in Sec.~\ref{sec:Methods_evaluation_metrics}. Hyperparameter tuning was performed using \texttt{Optuna}, a state-of-the-art hyperparameter optimization library, with a time budget of 3,600 seconds for each model. The AutoML framework followed a 5-fold cross-validation strategy during training, ensuring robust evaluation of model performance. Cross-validation was stratified and shuffled, further enhancing the reliability of the results.

\subsection{Explainability Measures}
\label{sec:Methods_XAI} 

This study employs several explainability methods to ensure comprehensive interpretability of the obtained machine learning models. 

\textbf{Learning Curves:} For each model, learning curves were plotted to display evaluation metric values across training iterations for both training and validation datasets. These curves provide insights into the model's learning dynamics and potential overfitting or underfitting issues. A vertical line indicates the optimal iteration number, which is subsequently used for making predictions.

\textbf{Feature Importance:} Utilizing a permutation-based method, feature importance was computed to identify the most influential predictors in each model. The top features were visualized in importance plots, and comprehensive results were saved in corresponding csv files. This analysis aids in discerning the variables that significantly impact model predictions.

\textbf{Linear Model Coefficients:} For linear models, the coefficients were extracted. This transparency allows for a clear interpretation of how each feature contributes to the model's output.

\textbf{SHAP Explanations:} SHapley Additive exPlanations (SHAP) values were computed to provide a unified measure of feature attribution. The following SHAP-based analyses were conducted:

\begin{itemize}
    \item \textbf{SHAP Importance:} Illustrated the average impact of each feature on model output magnitude.
    \item \textbf{SHAP Dependence Plots:} Depicted the relationship between individual feature values and their corresponding SHAP values, highlighting interaction effects.
    \item \textbf{SHAP Decision Plots:} Presented detailed explanations for the top 10 best and worst predictions, offering a granular view of how feature values influenced specific model decisions.
\end{itemize}

These explainability measures collectively enhance the transparency and interpretability of the models developed in this study, facilitating a deeper understanding of the underlying patterns and contributing factors in the data.

\section{Results}
\label{sec:Results}

\subsection{Data Collection, Preprocessing and Exploratory Data Analysis}
\label{sec:Results_DataCollectionPreprocessing}

\begin{figure}
    \centering
    \includegraphics[width=0.95\linewidth]{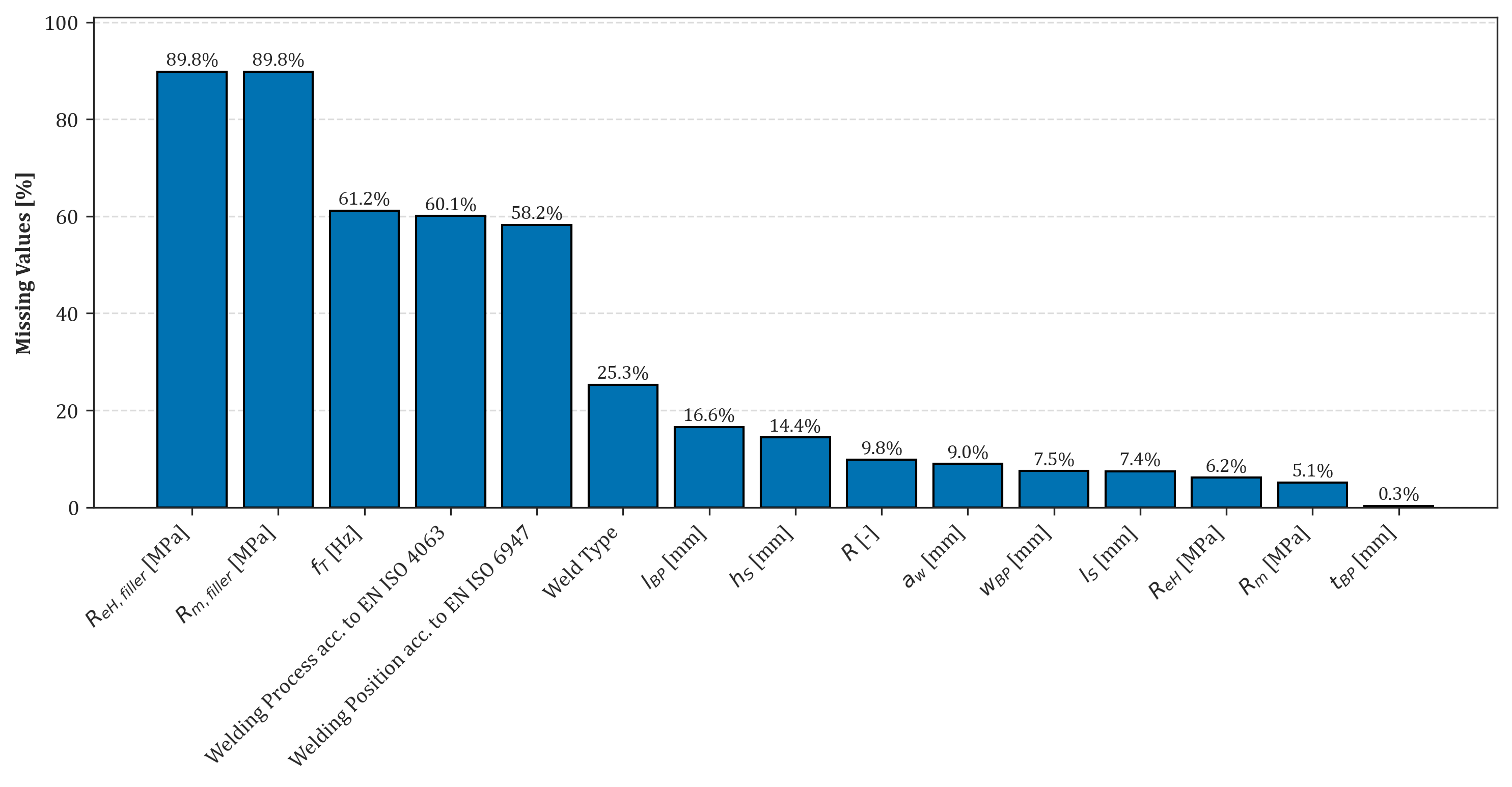}
    \caption{Overview of Missing Data in the Dataset}
    \label{fig:res_data_missing}
\end{figure}

The results of the data cleaning and preprocessing steps reveal several important characteristics of the dataset. The initial data preview revealed a complex dataset comprising 78 columns and 2,861 entries, encompassing a wide range of material properties, test conditions, and geometric parameters relevant to welded structural details. The dataset contained both numerical and categorical data types, with a substantial portion of the columns exhibiting missing values, as evident from non-null counts below the total entry count (cf.~Fig.\ref{fig:res_data_missing}). A total of 52 numerical features—including key variables such as testing frequency, material strengths, and fatigue-related geometric descriptors—formed the basis for subsequent quantitative analysis. To assess data quality and distribution characteristics, histograms were generated for selected features and target variables. These visualizations facilitated the identification of outliers and unusual distribution patterns, while also reporting the proportion of missing data per feature to guide imputation decisions. A comprehensive statistical summary further quantified dataset characteristics by reporting central tendencies, spread, and frequency statistics for both numerical and categorical features. This initial exploratory data analysis established a solid foundation for the cleaning, imputation, and normalization steps required for reliable model development and interpretability in later stages of the study.

\subsubsection{Continuos and Categorical Features}
\label{sec:Results_DataCollectionPreprocessing_Continous}

\begin{figure*}[htbp]
    \centering
    \includegraphics[width=0.45\textwidth]{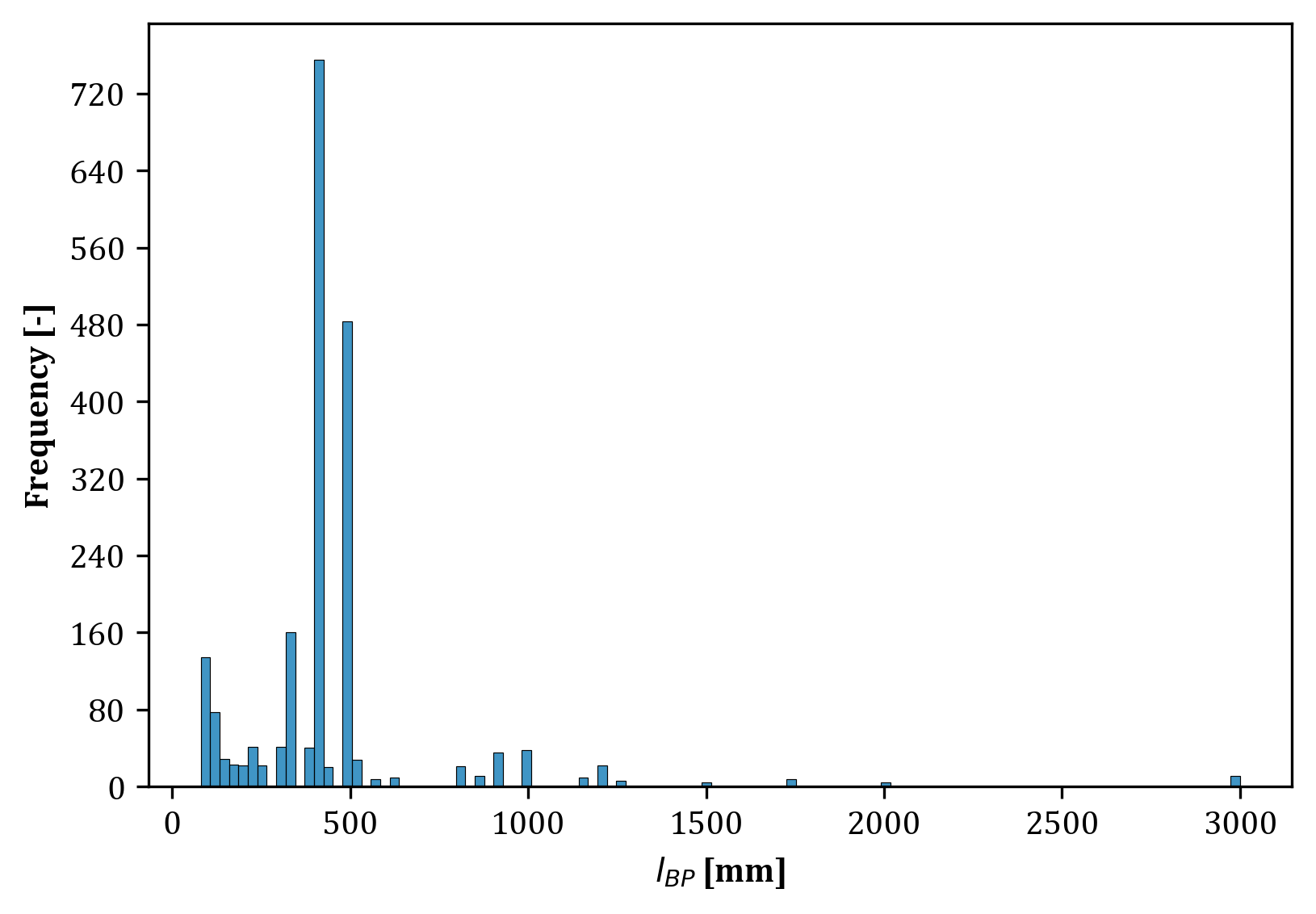}
    \includegraphics[width=0.45\textwidth]{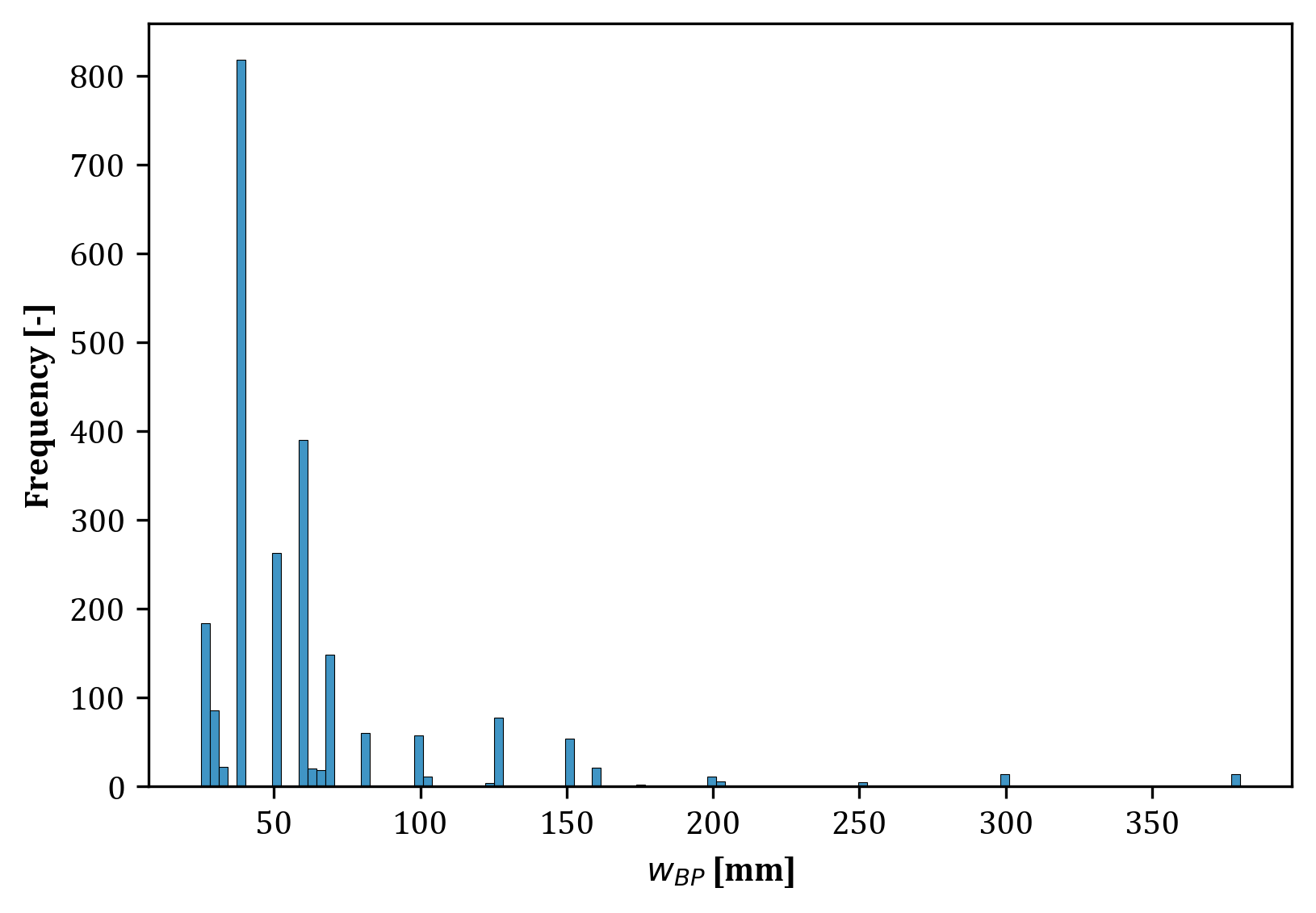}
    \includegraphics[width=0.45\textwidth]{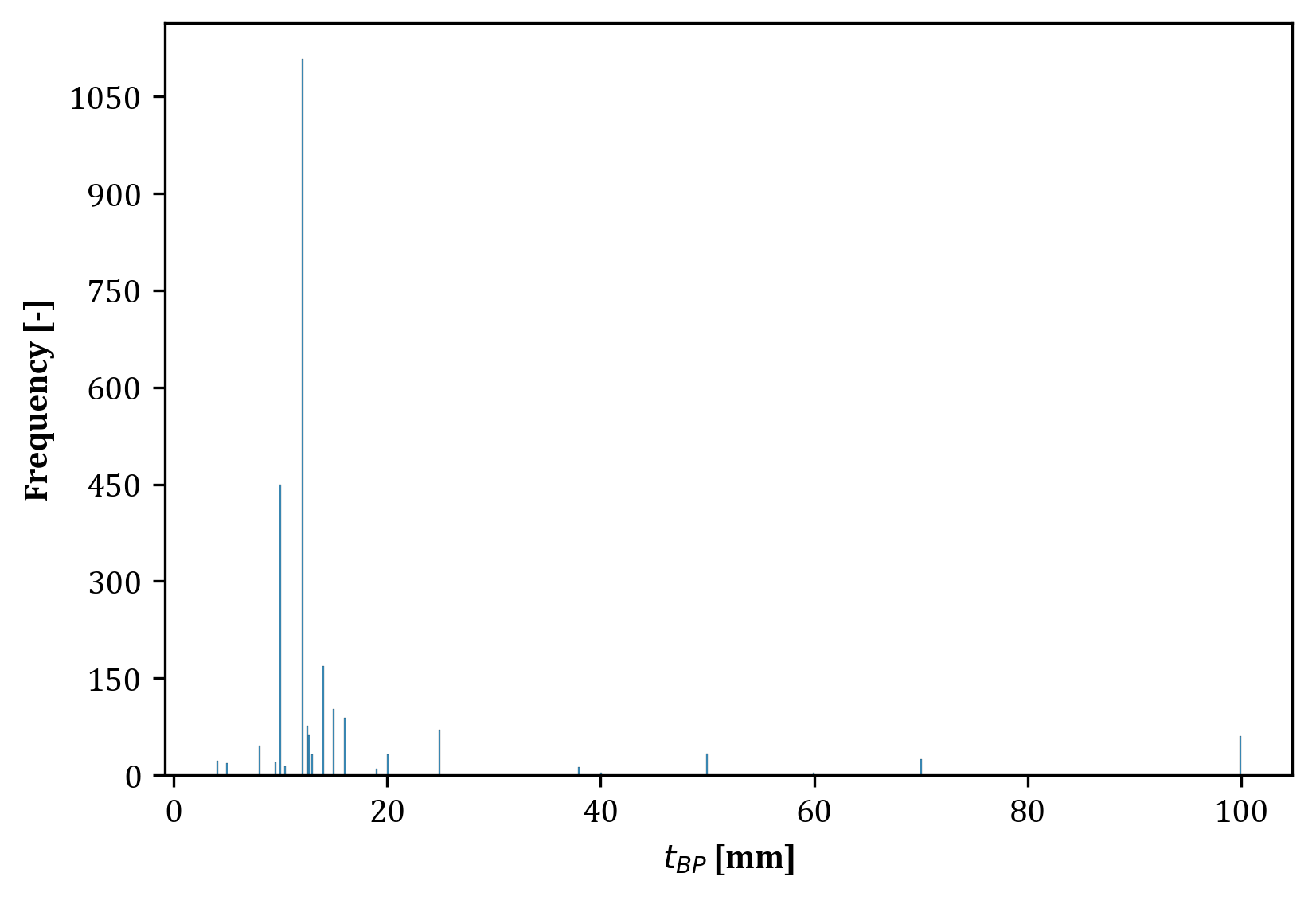}
    \includegraphics[width=0.45\textwidth]{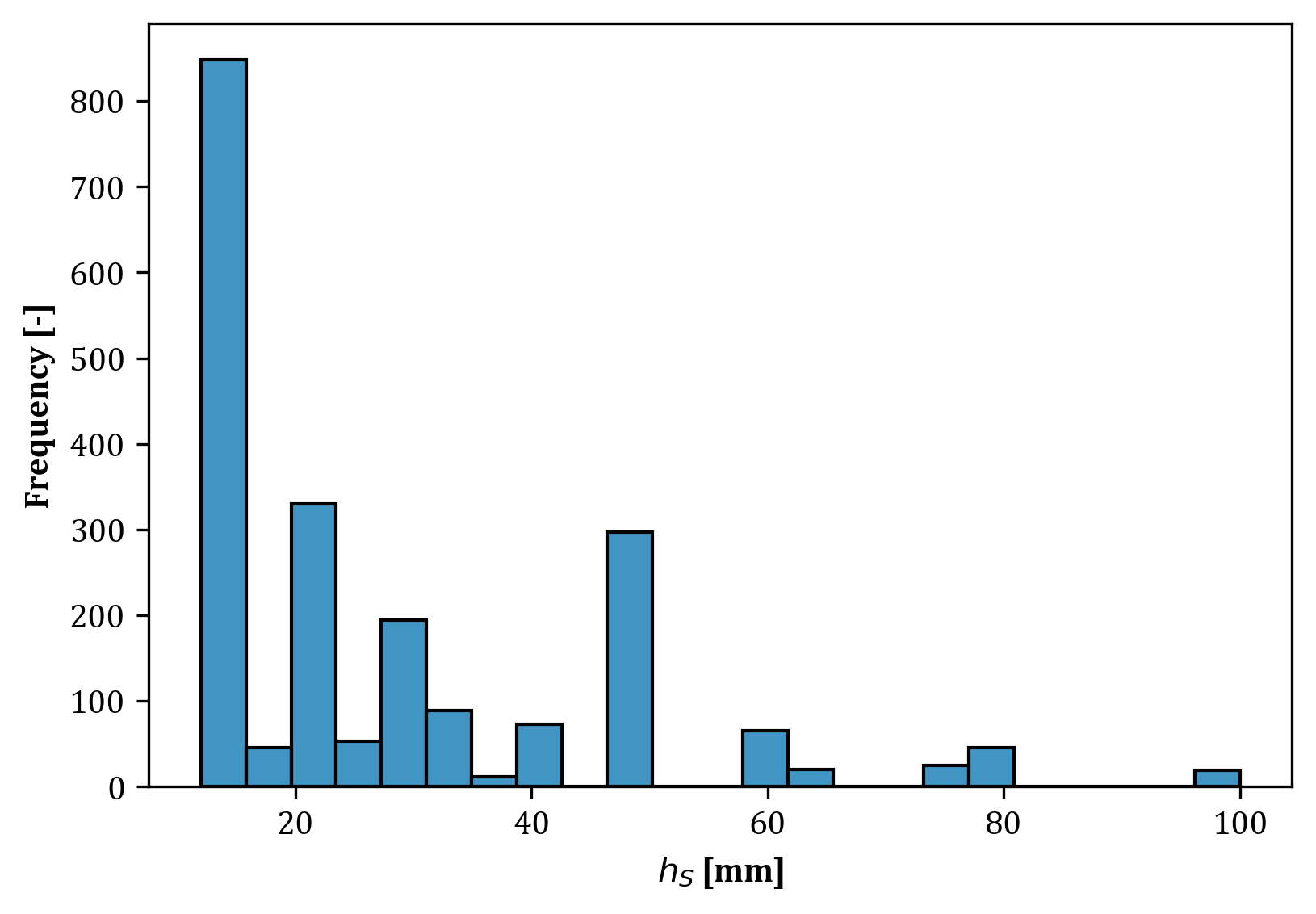}
    \includegraphics[width=0.45\textwidth]{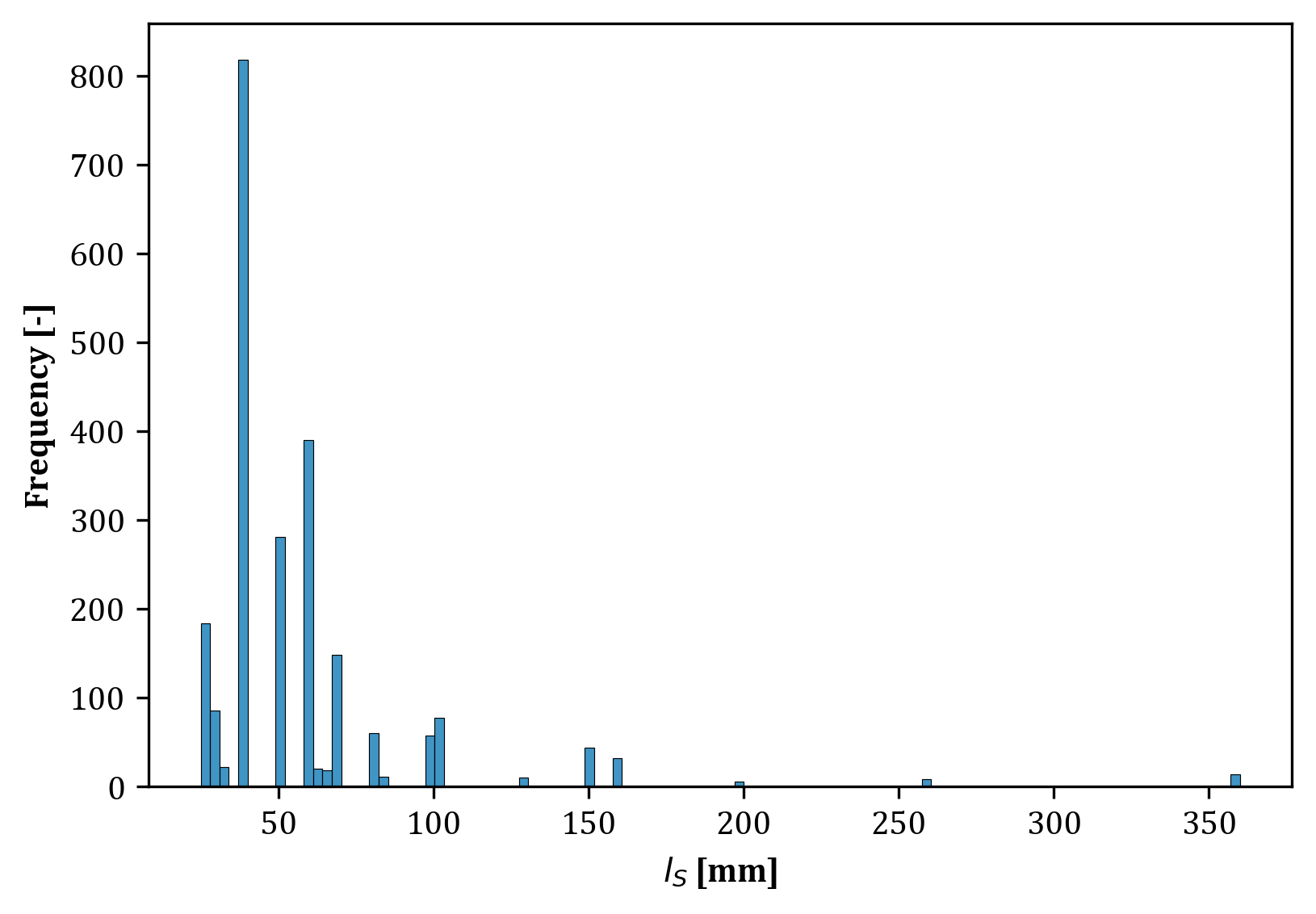}
    \includegraphics[width=0.45\textwidth]{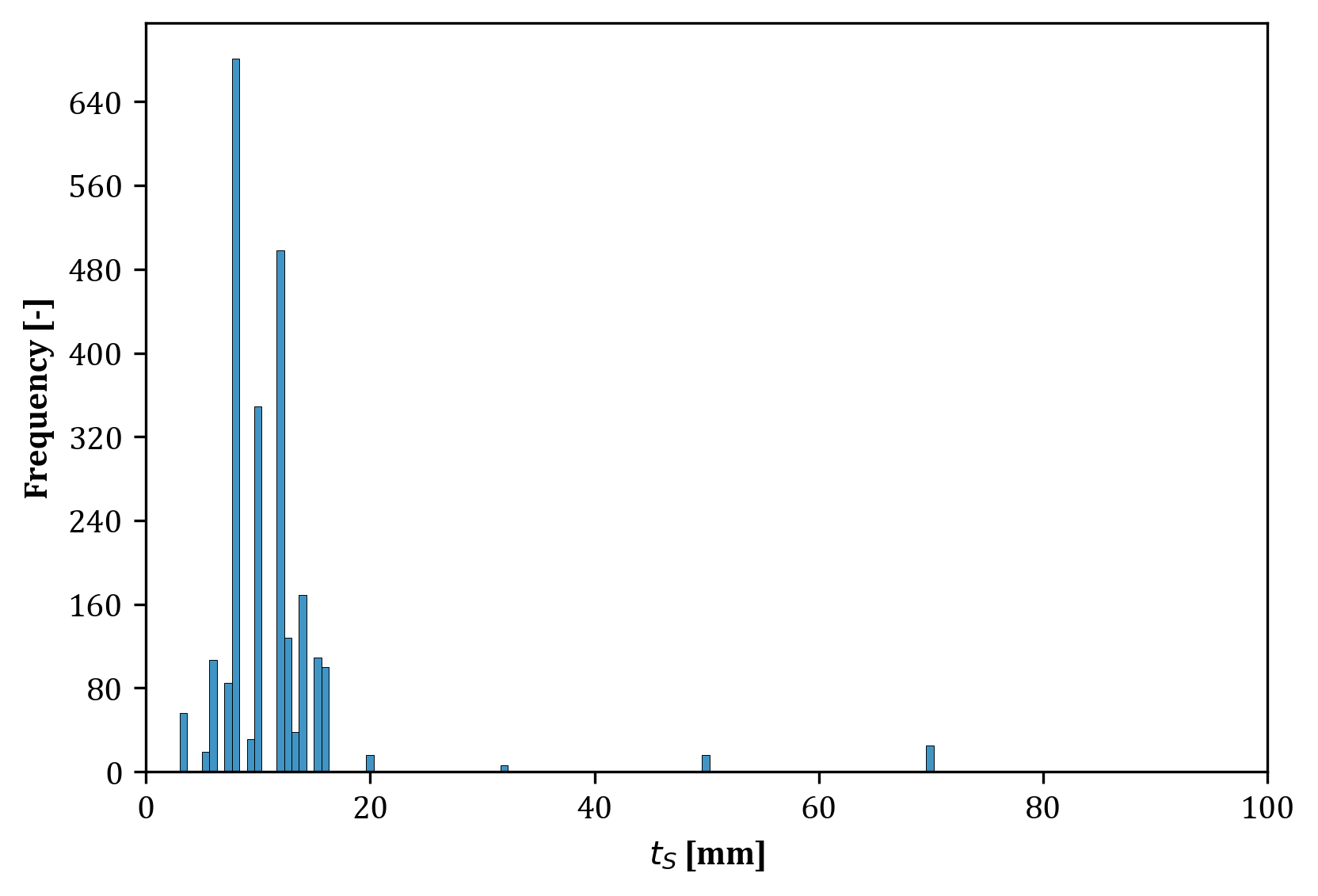}
    \includegraphics[width=0.45\textwidth]{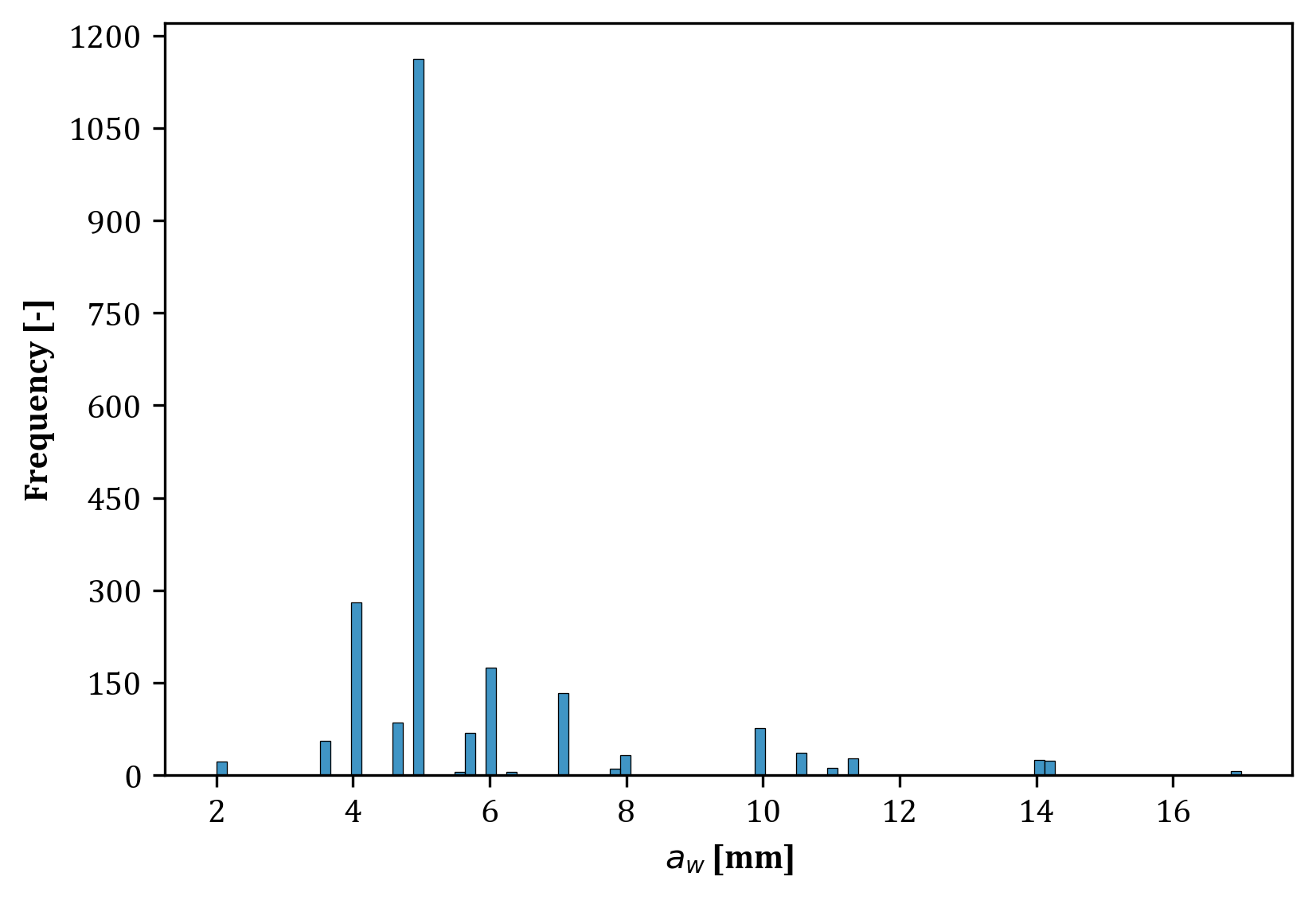}
    \includegraphics[width=0.45\textwidth]{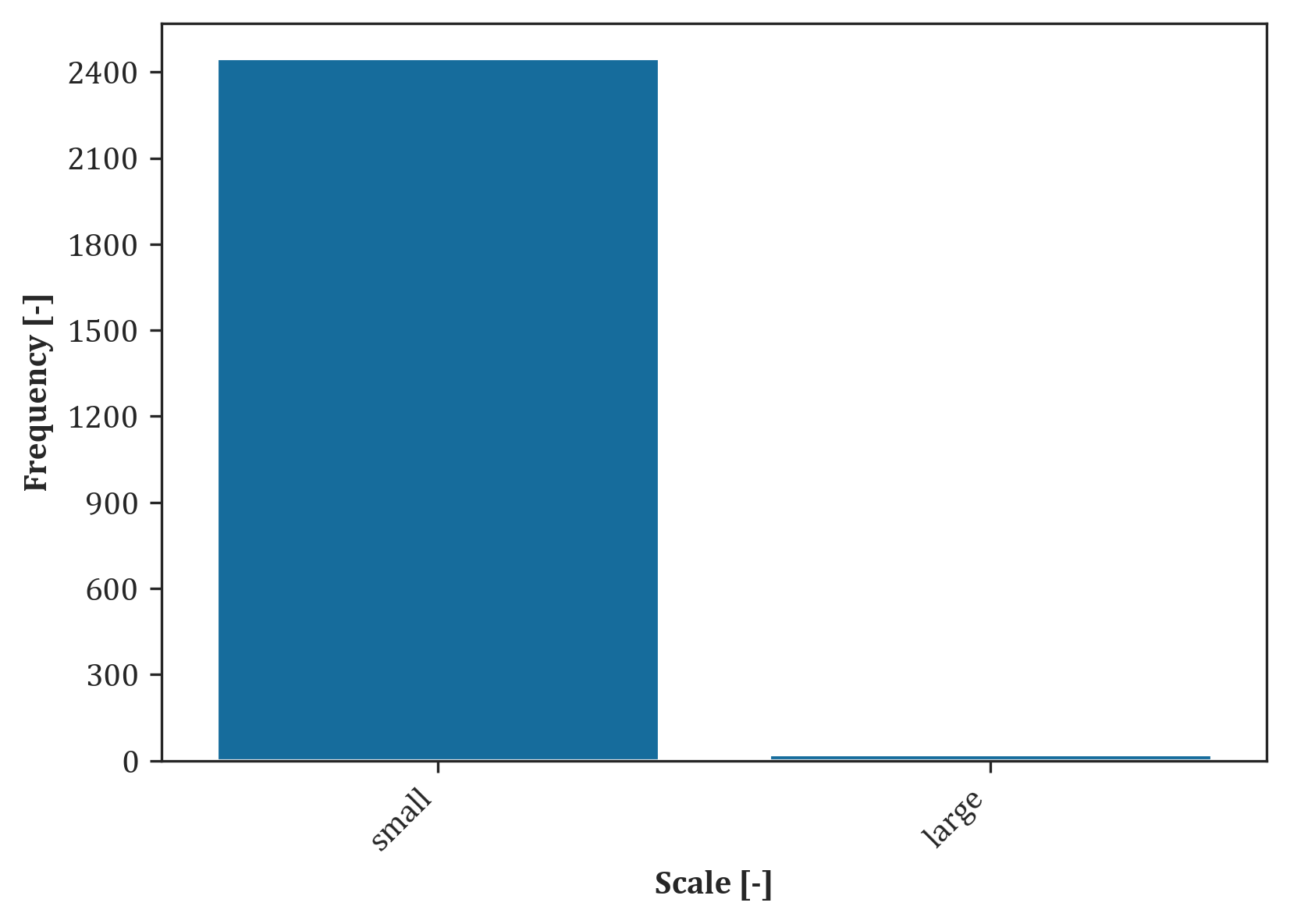}
    \caption{Distributions of global geometric features related to plate and attachment dimensions.}
    \label{fig:eda_cont1}
\end{figure*}

\begin{figure*}[htbp]
    \centering
    \includegraphics[width=0.4\textwidth]{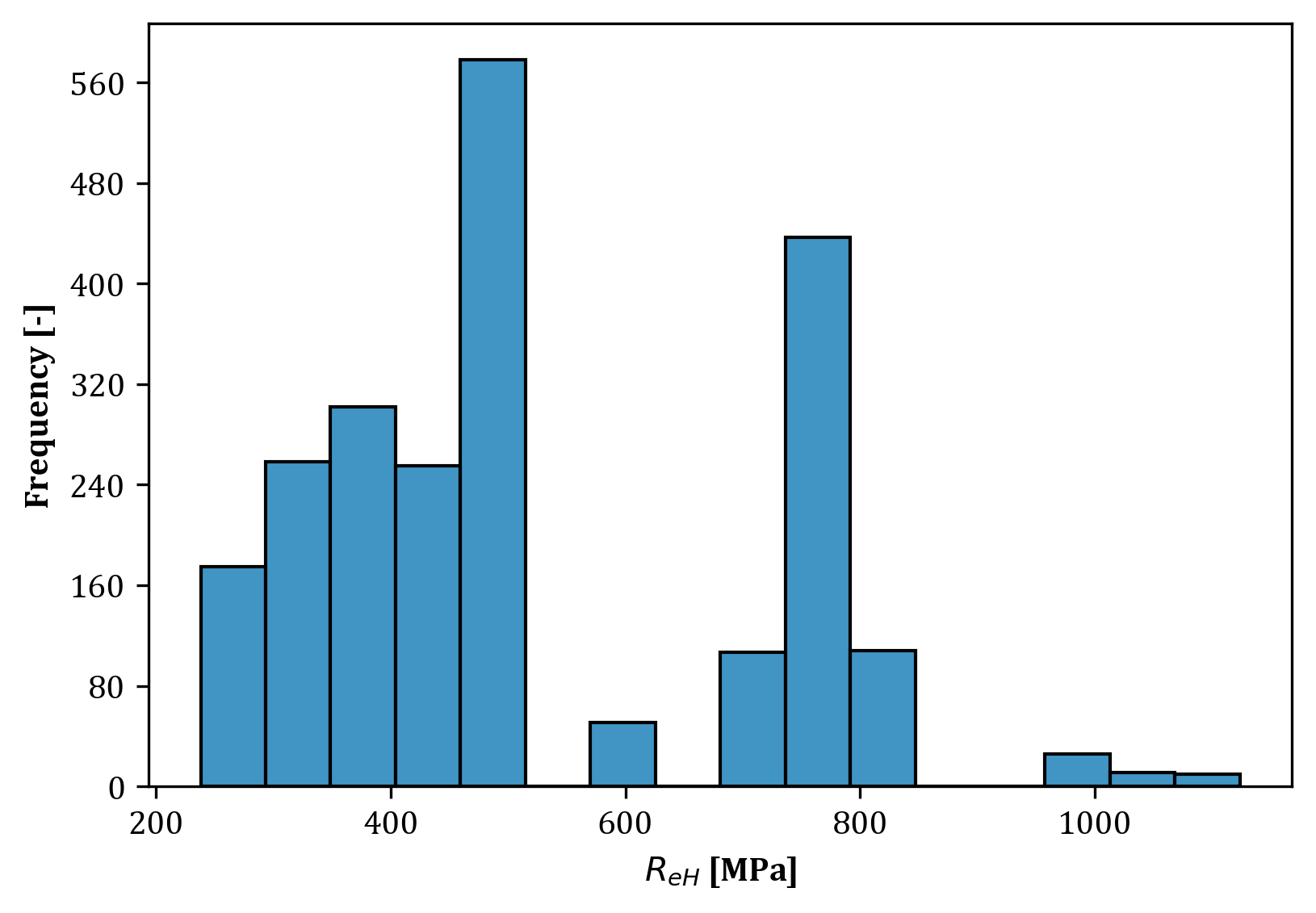}
    \includegraphics[width=0.4\textwidth]{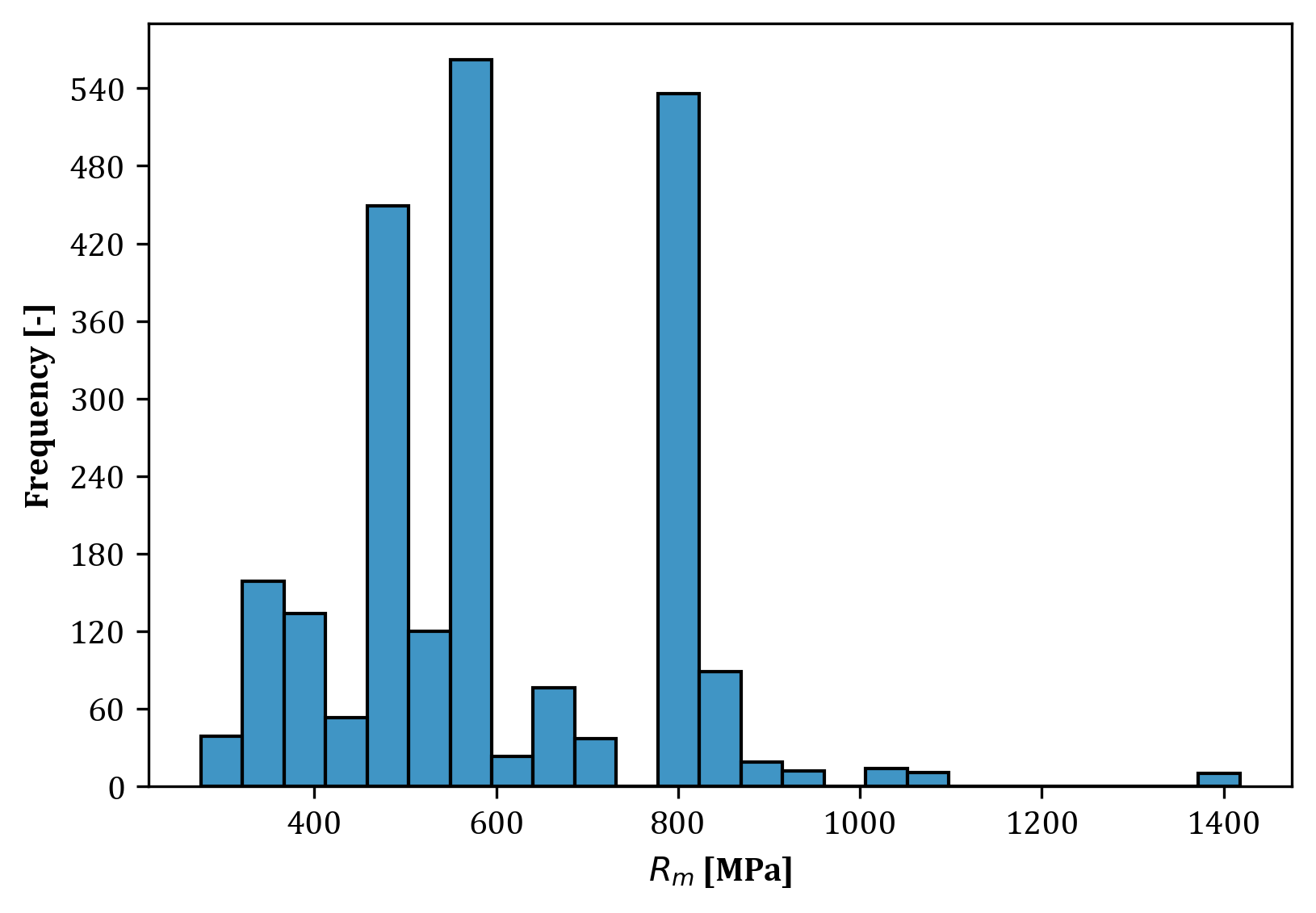}
    \includegraphics[width=0.4\textwidth]{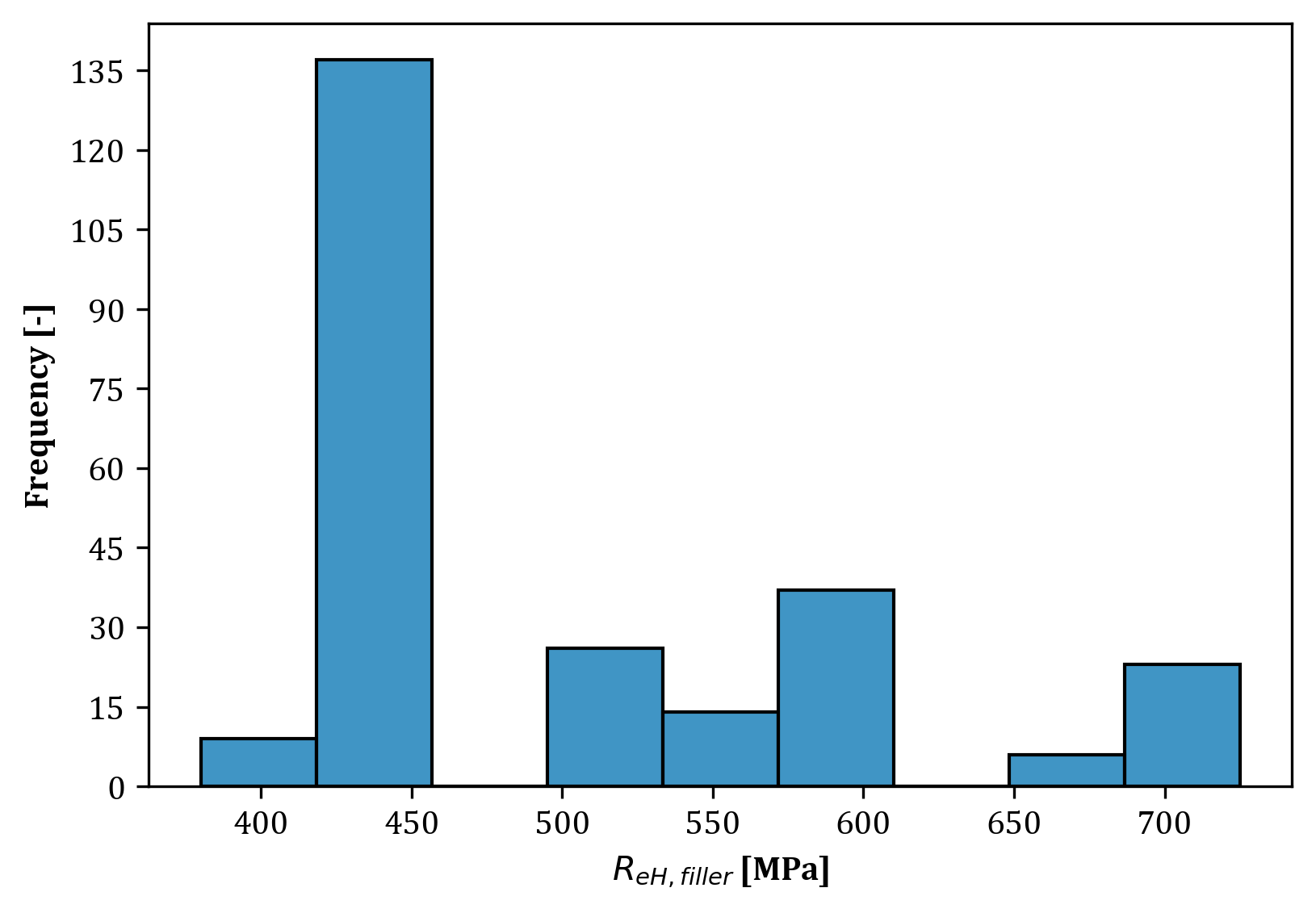}
    \includegraphics[width=0.4\textwidth]{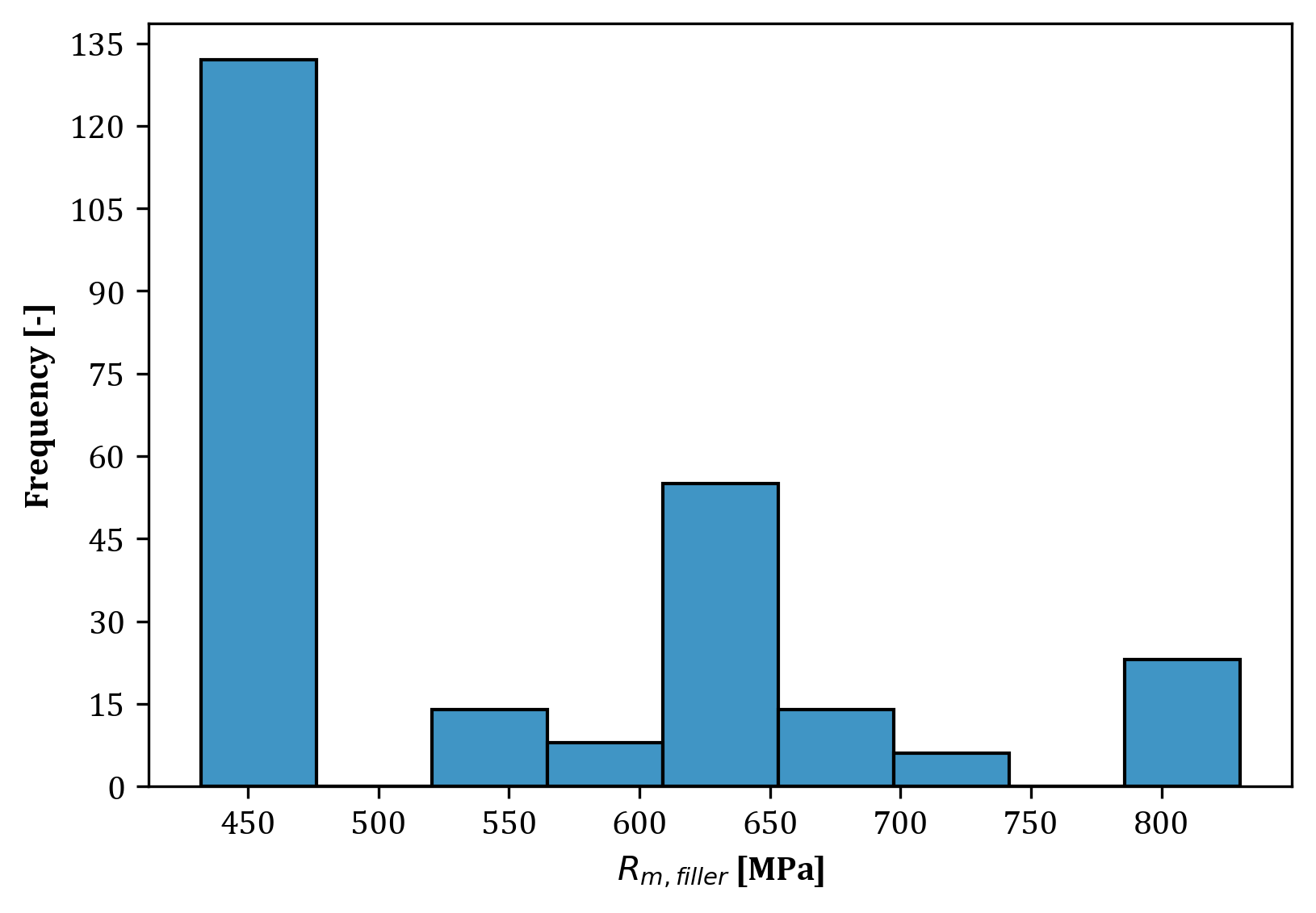}
    \caption{Distributions of mechanical strength characteristics of base and filler material.}
    \label{fig:eda_cont2}
\end{figure*}

The distributional analysis of the continuous geometric, mechanical, and loading-related features provides important insight into the structure and variability of the fatigue dataset (cf. Figure~\ref{fig:eda_cont1} and~\ref{fig:eda_cont2}), revealing substantial heterogeneity across the fatigue test database. The base plate geometry, defined by its length ($l_\mathrm{BP}$), width ($w_\mathrm{BP}$), and thickness ($t_\mathrm{BP}$), exhibits multimodal and positively skewed distributions, with dominant peaks around standard specimen dimensions such as $l_\mathrm{BP} \approx 500$\,mm and $w_\mathrm{BP} \approx 40$\,mm. Similarly, the attachment dimensions ($h_\mathrm{S}$, $l_\mathrm{S}$, $t_\mathrm{S}$) show strong clustering at nominal values, suggesting a prevalence of standardized welded detail configurations. The weld throat thickness $a_\mathrm{w}$ displays a pronounced concentration around 5\,mm, indicating limited variability in weld sizing. 

\begin{figure}[htbp]
    \centering
    \includegraphics[width=0.85\textwidth]{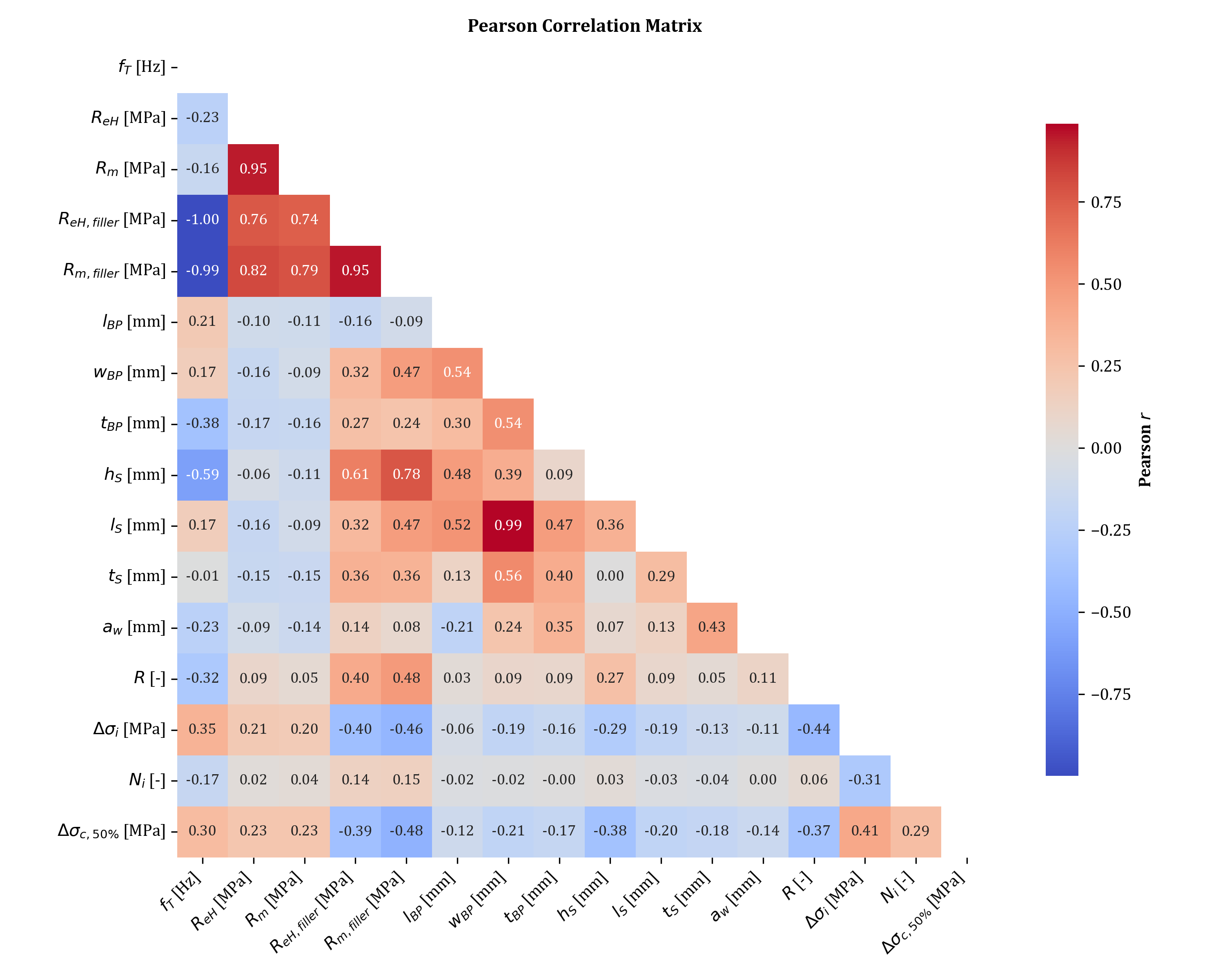}
    \caption{Pearson correlation heatmap of continuous features and fatigue targets.}
    \label{fig:eda_corr}
\end{figure}

Material strength properties, including yield and tensile strengths of both base and filler materials ($R_\mathrm{eH}$, $R_\mathrm{m}$,  $R_{\mathrm{eH},\mathrm{filler}}$, $R_{\mathrm{m},\mathrm{filler}}$), exhibit multiple discrete modes corresponding to typical structural steel classes (e.g., S355, S690, or high-strength steels above 800\,MPa). The testing frequency $f_\mathrm{T}$ and number of weld passes show less pronounced, yet still informative, multimodal patterns. The correlation matrix (Figure~\ref{fig:eda_corr}) confirms intuitive relationships, such as high collinearity between yield and tensile strengths and a notable correlation between geometric features (e.g., $h_\mathrm{S}$, $l_\mathrm{S}$) and the resulting fatigue strength $\Delta \sigma_{C,50\%}$.

\begin{figure*}[htbp]
    \centering
    \includegraphics[width=0.4\textwidth]{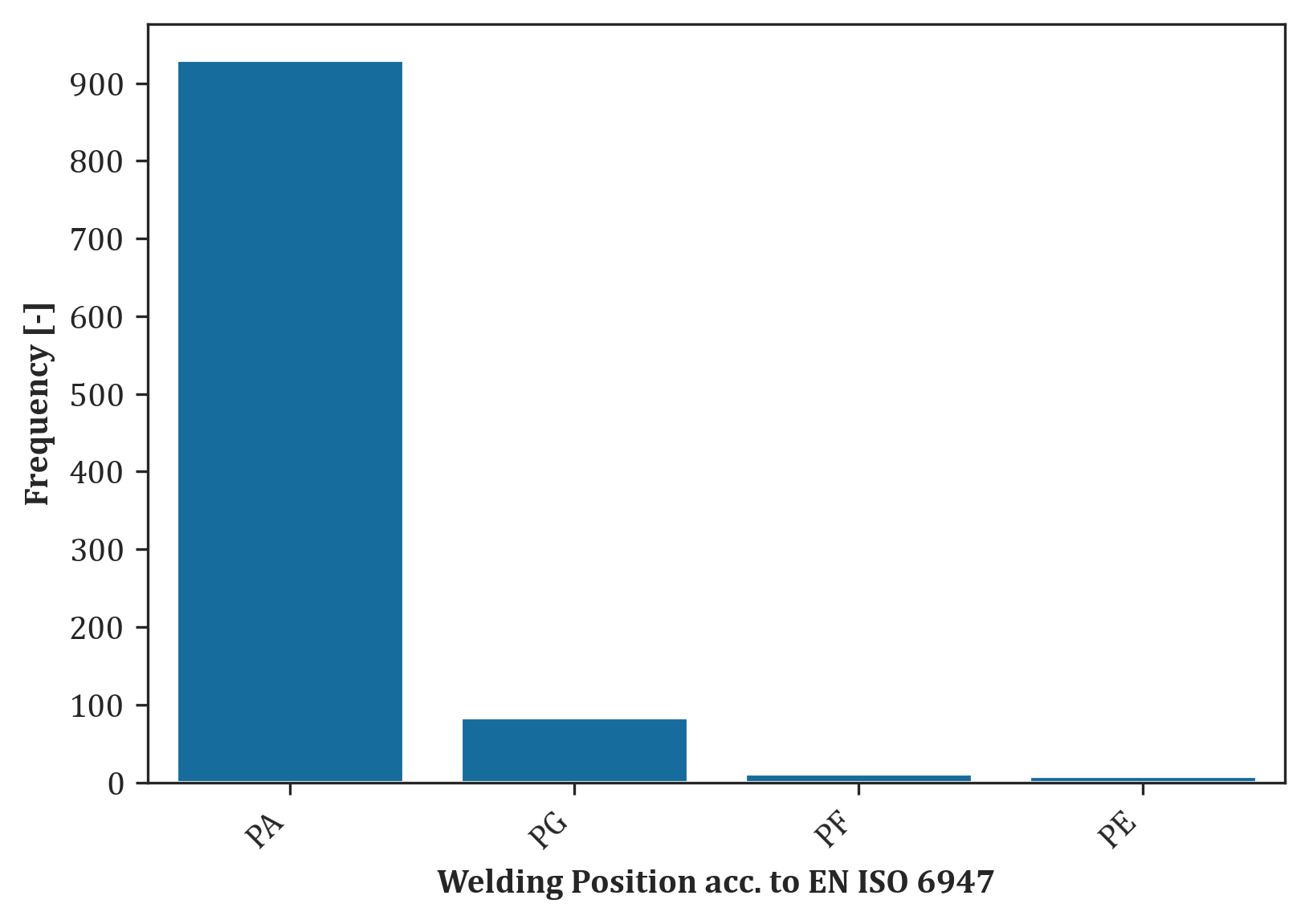}
    \includegraphics[width=0.4\textwidth]{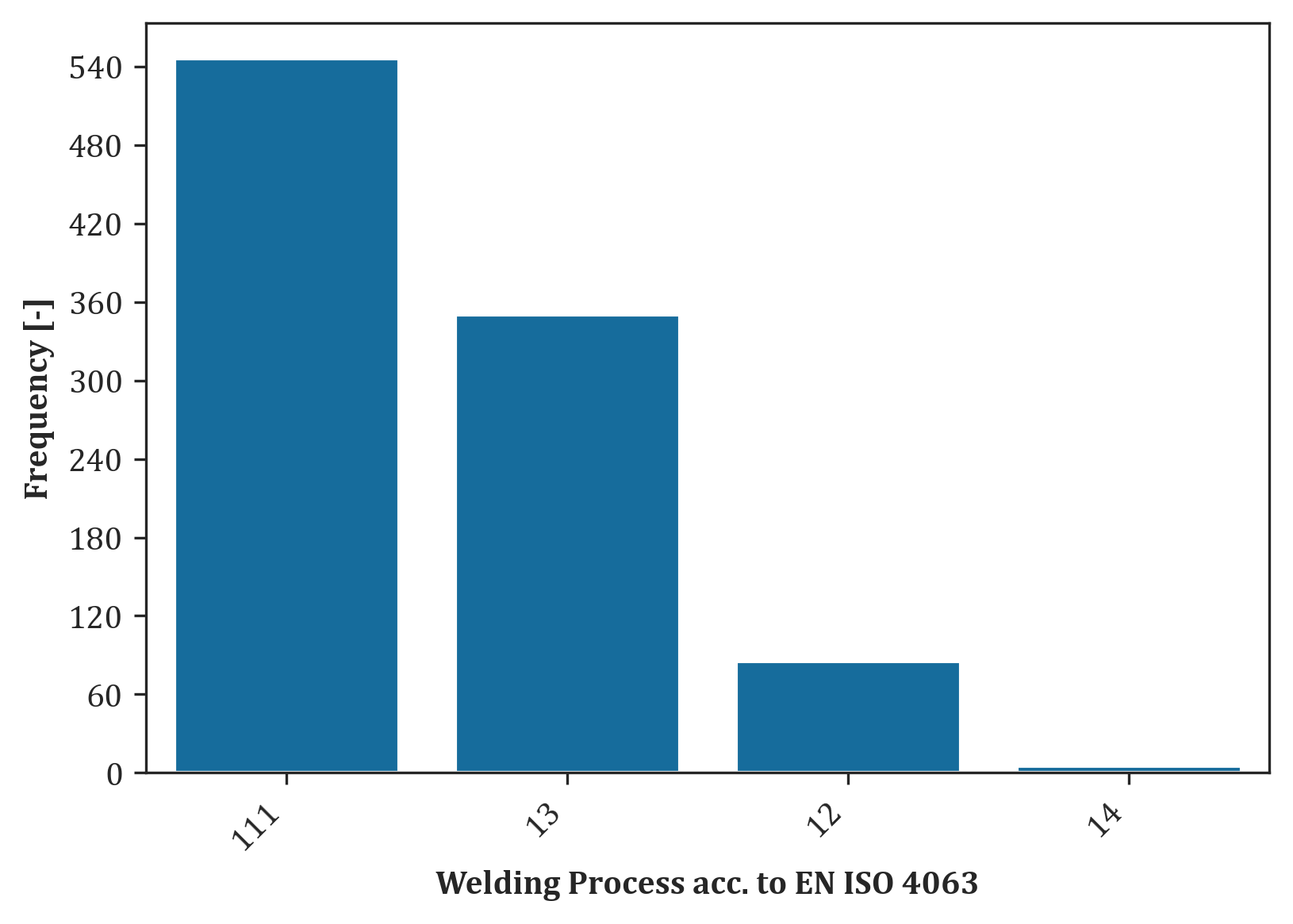}
    \includegraphics[width=0.4\textwidth]{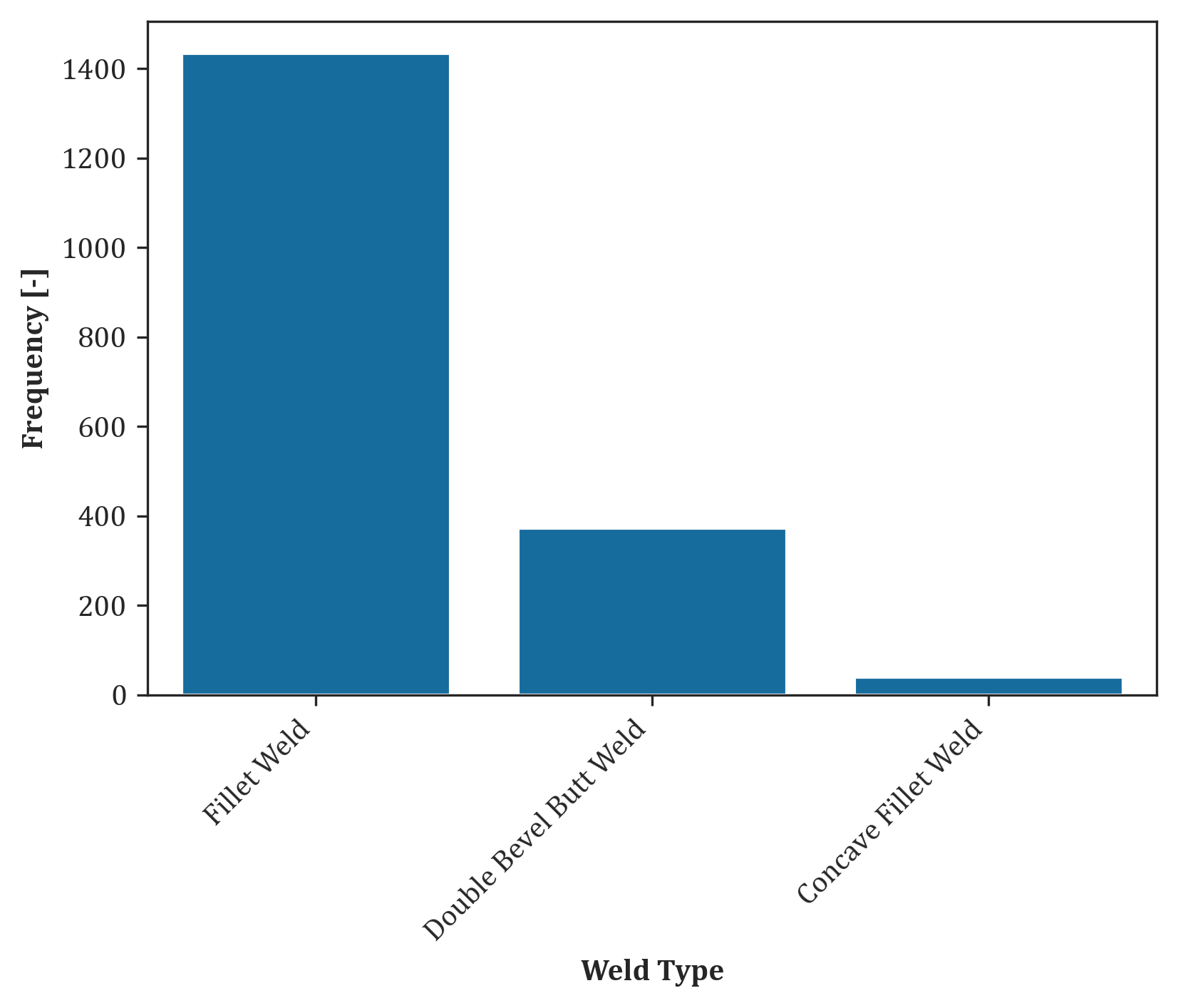}
    \includegraphics[width=0.4\textwidth]{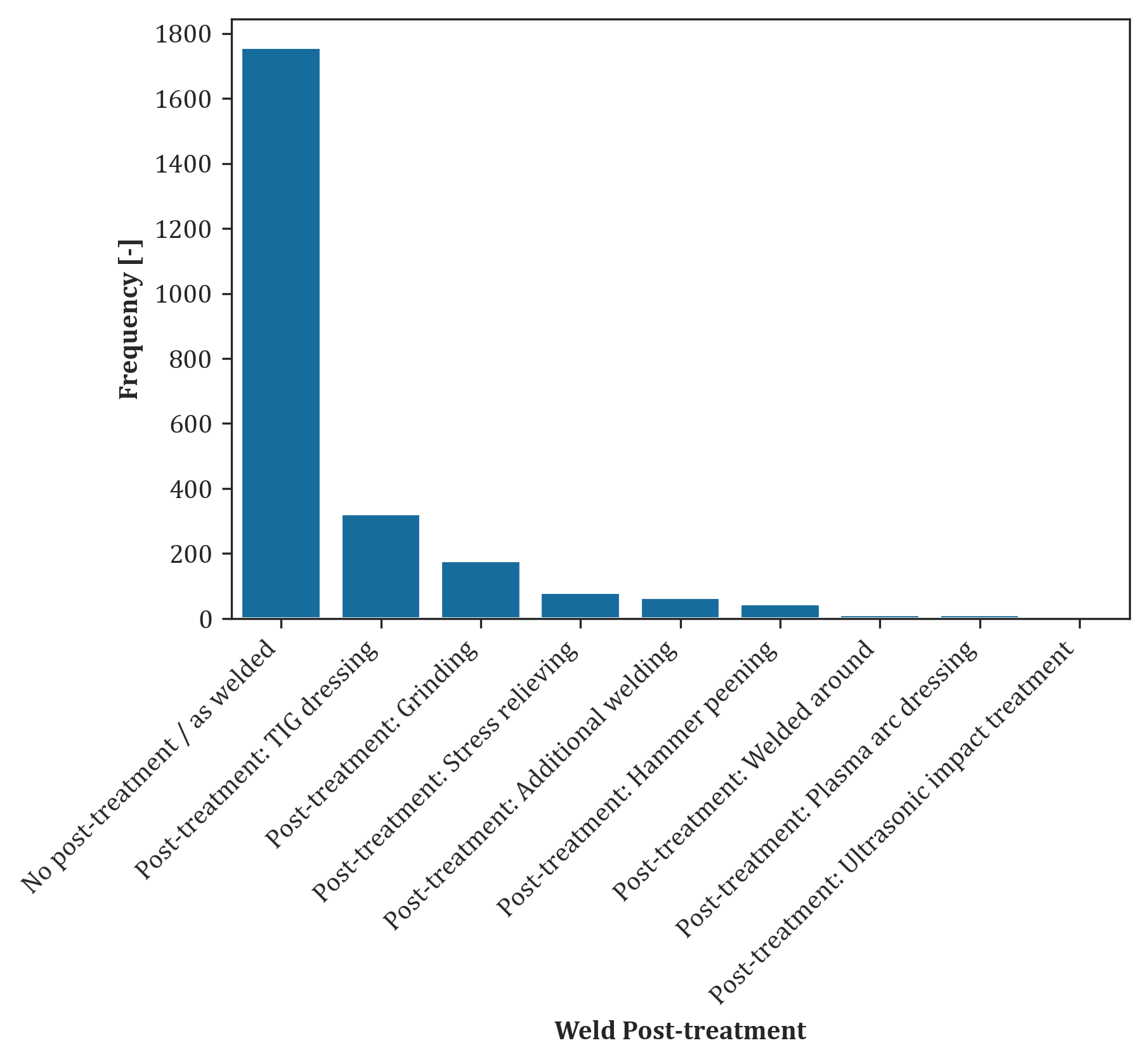}
    \caption{Distributions of welding parameters.}
    \label{fig:eda_cat1}   
\end{figure*}

Load- and stress-related parameters such as the stress ratio $R$ and applied stress range $\Delta \sigma_\mathrm{i}$ show broader distributions. Particularly, the $R$-ratio is dominated by fully reversed and pulsating tension loading ($R = -1$ and $R = 0$). The stress range $\Delta \sigma_\mathrm{i}$ spans a wide interval, with a right-skewed distribution reflecting both high- and low-cycle fatigue scenarios in the finite life fatigue regime.

\begin{figure*}[htbp]
    \centering
     \includegraphics[width=0.48\textwidth]{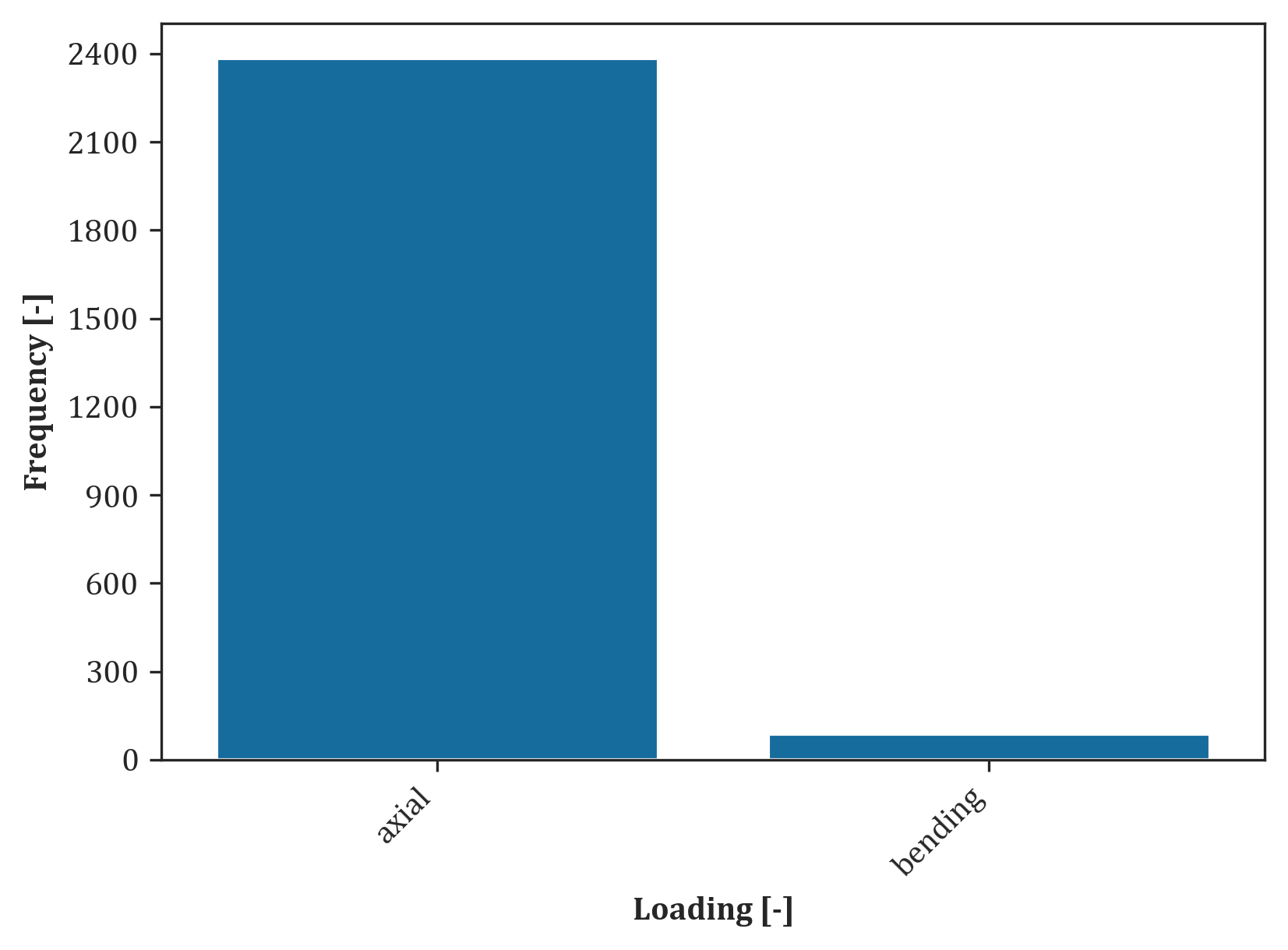}
    \includegraphics[width=0.48\textwidth]{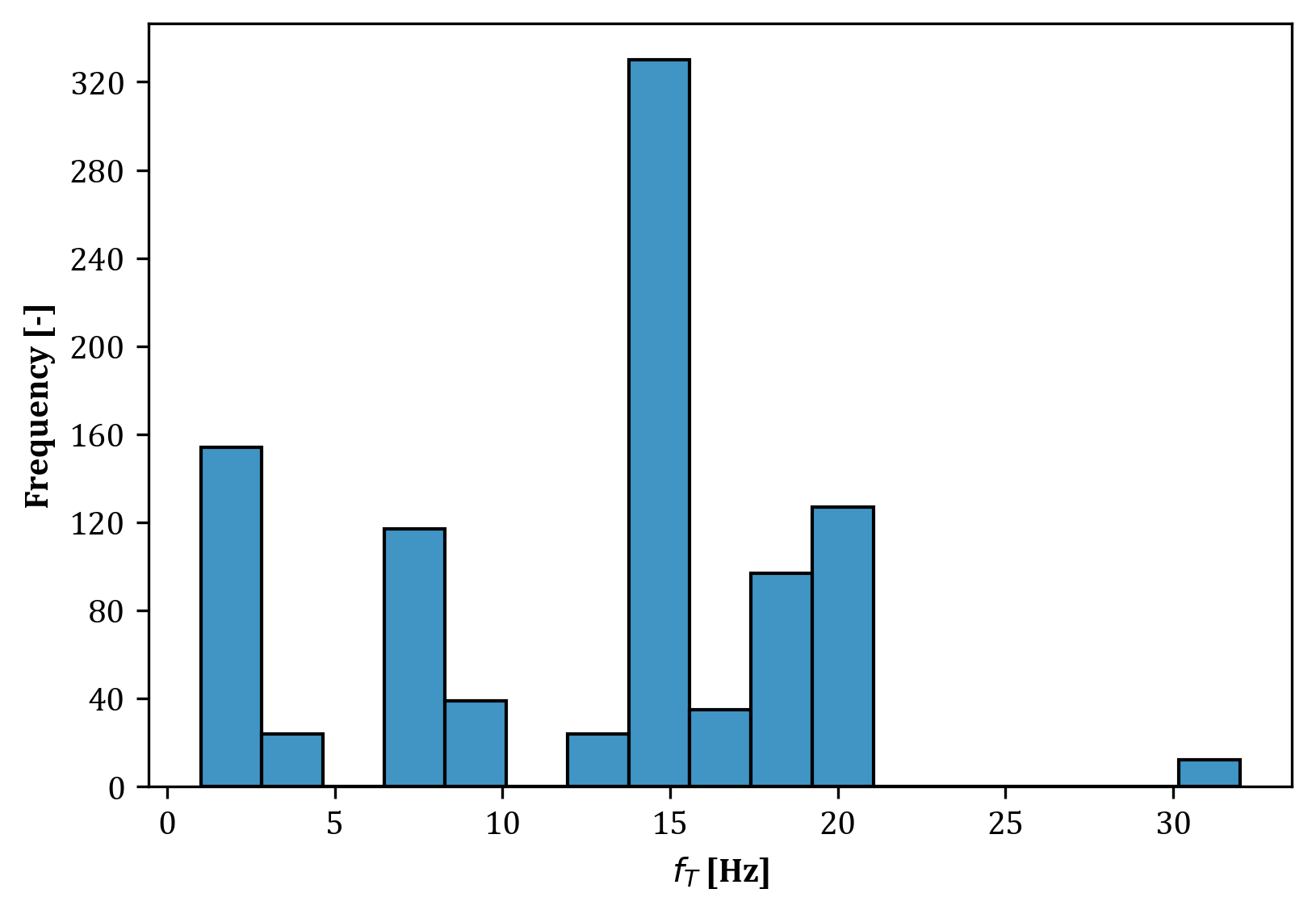}
    \includegraphics[width=0.48\textwidth]{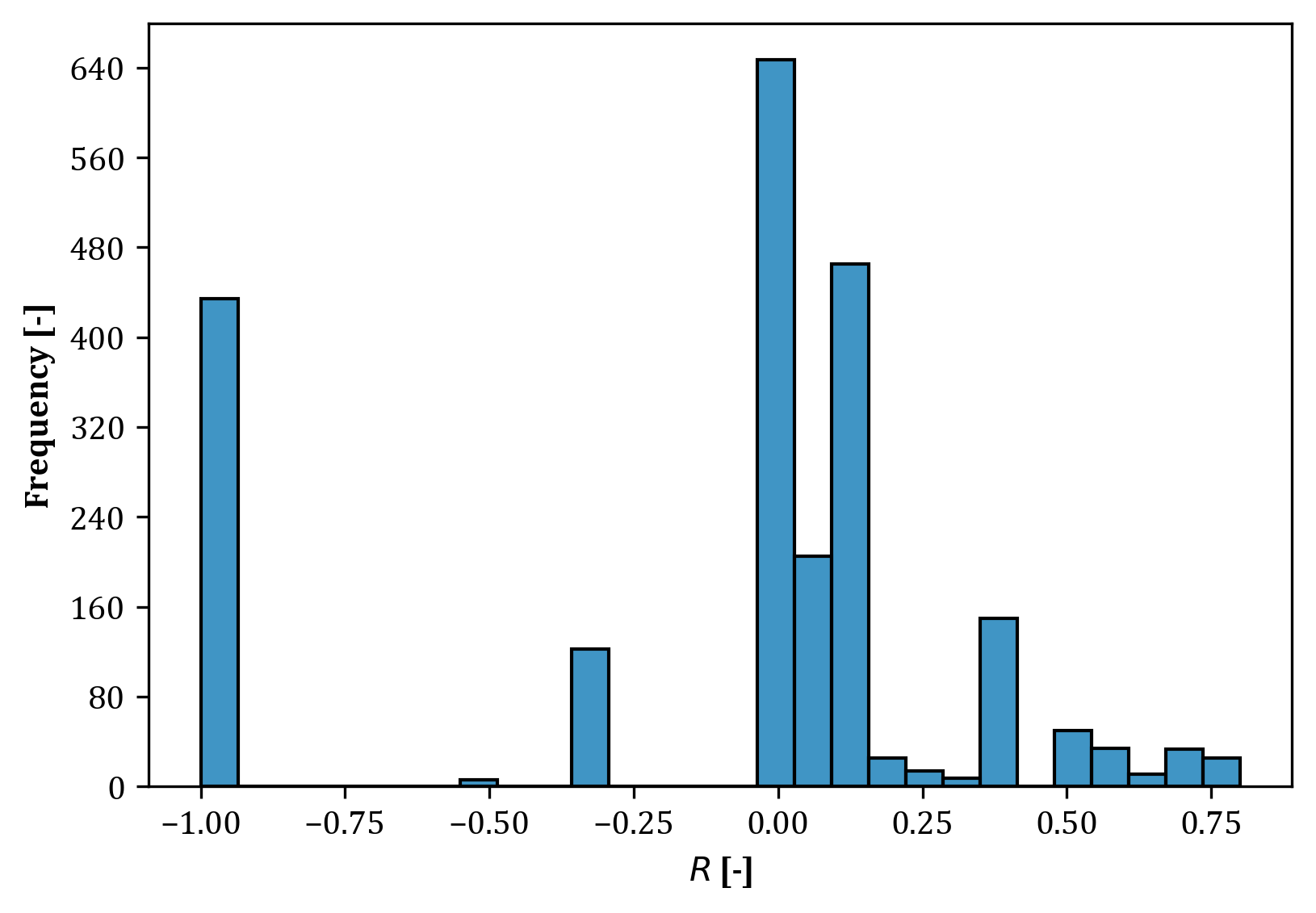}
    \includegraphics[width=0.48\textwidth]{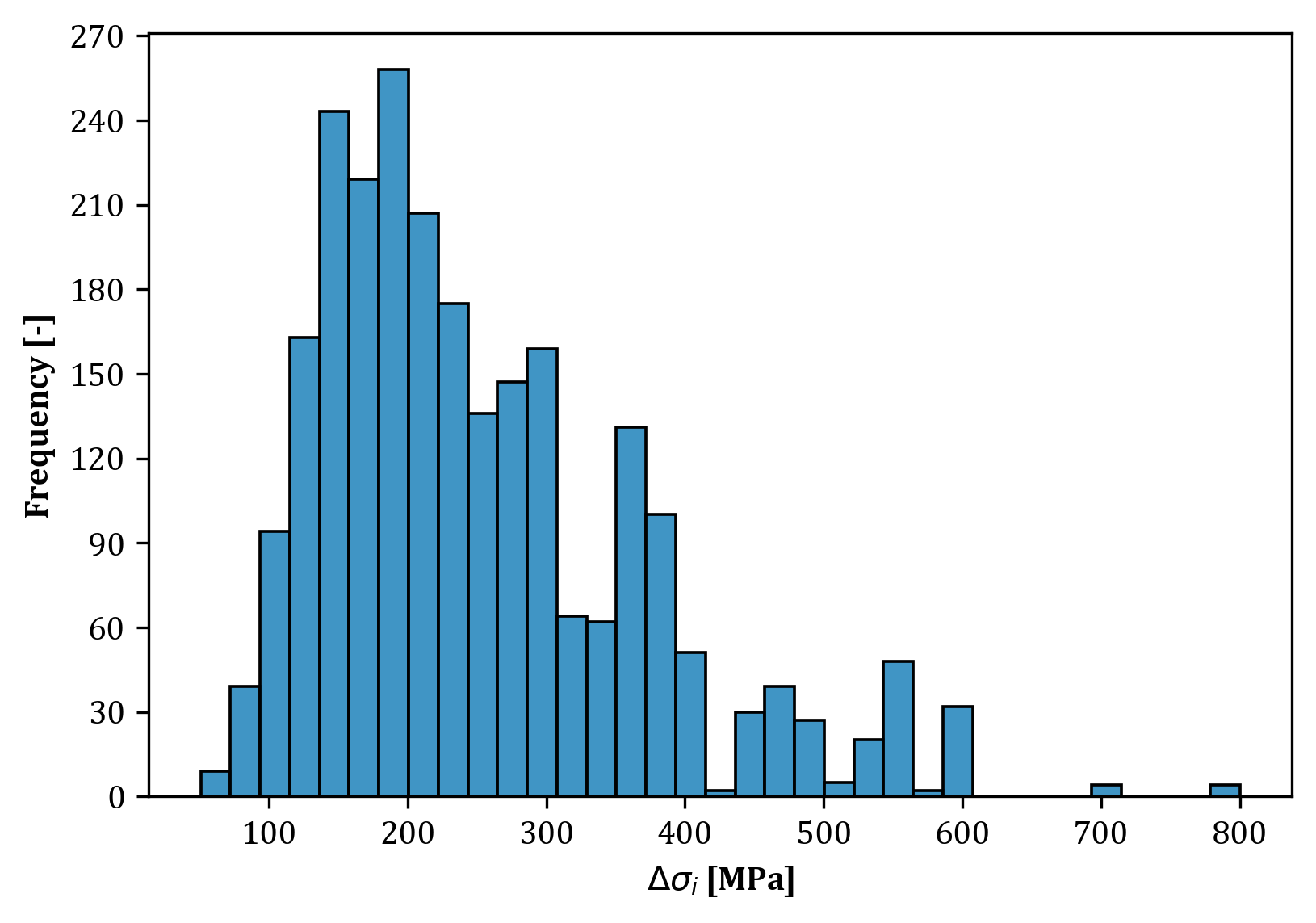}
    \caption{Distributions of load parameters.}
    \label{fig:eda_cat2}
\end{figure*}

Figure~\ref{fig:eda_corr} shows the Pearson correlation matrix between continuous variables. Strong correlations are observed among material strength properties, e.g., between $R_m$ and $R_{m,\mathrm{filler}}$ ($r = 0.95$), and among geometric quantities of the base and attached parts (e.g., $l_S$ vs. $h_S$, $r = 0.78$). The correlation between load amplitude $\Delta \sigma_i$ and fatigue life $N_i$ is negative ($r = -0.44$), reflecting the well-known inverse relationship in S-N fatigue behavior. Interestingly, only moderate correlation is observed between $\Delta \sigma_{c,50\%}$ and other parameters, suggesting complex multivariate dependencies that motivate data-driven modeling. The correlation matrix results give rise to further feature engineering and especially employing the VIF method, cf. Sec.~\ref{sec:VIF_results}.

The dataset exhibits substantial class imbalance across several categorical variables relevant for fatigue life prediction in welded joints.

\begin{itemize}
  \item \textbf{Scale Effects.} The majority of fatigue tests were conducted at small-scale specimens, with large-scale data being underrepresented.
  \item \textbf{Loading Conditions.} Axial loading dominates the dataset, while bending conditions are rare (Fig.~\ref{fig:eda_cat2}, top left).
  \item \textbf{Post-Treatment Methods.} Most samples are in the untreated (as-welded) state. TIG dressing and grinding are the most frequently applied post-treatments, while methods such as ultrasonic impact treatment or full grinding are rarely found (Fig.~\ref{fig:eda_cat1}, bottom right).
  \item \textbf{Welding Parameters.} A similar imbalance is observed for welding positions (EN ISO 6947 PA dominant), welding processes (EN ISO 4063 no. 111 and 13 dominate), and weld types (fillet welds are most frequent) as shown in Fig.~\ref{fig:eda_cat1}.
  \item \textbf{Corrosive Environment and Pretreatment.} The overwhelming majority of tests were performed in a non-corrosive environment, and without pretreatment.
\end{itemize}

These distributions underline the need for robust model training procedures capable of handling imbalanced classes, especially for less frequent treatments and loading scenarios.

\subsubsection{Distribution of Target Variables}
\label{sec:Results_DataCollectionPreprocessing_Targets}

To inform the preprocessing pipeline and ensure appropriate transformation of the regression targets, we examined the empirical distributions of the two key fatigue descriptors: the number of cycles to failure $N_i$ and the fatigue strength at 50\% survival probability $\Delta\sigma_{c,50\%}$.

\begin{figure*}[htbp]
    \centering
    \includegraphics[width=\textwidth]{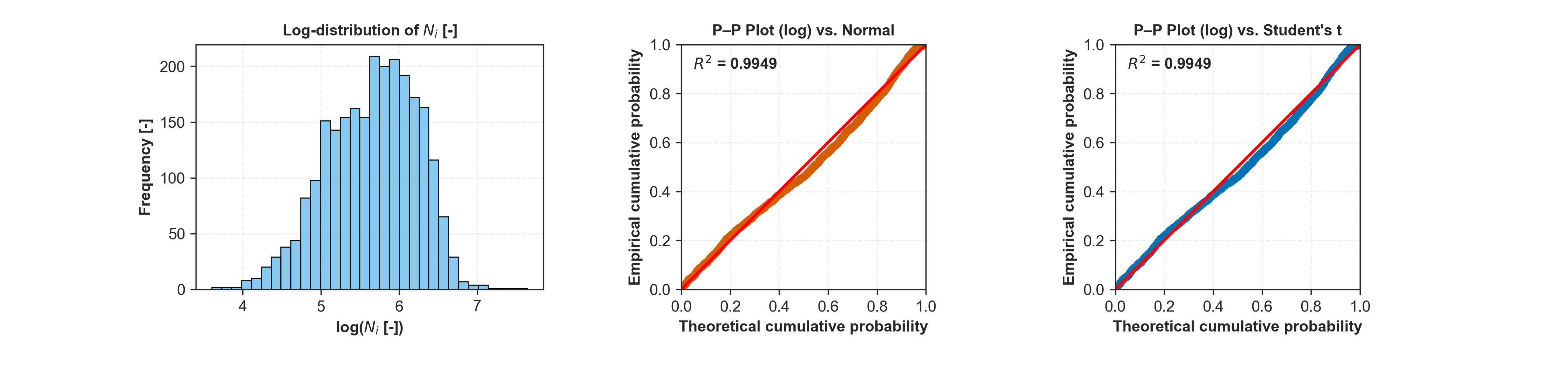}
    \includegraphics[width=\textwidth]{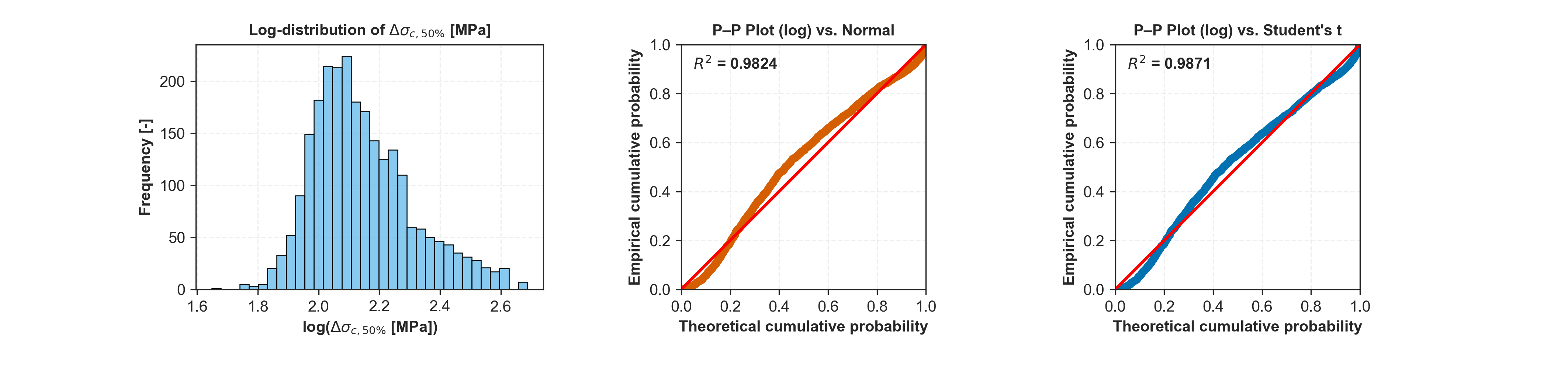}
    \caption{Distributions of target variables $N_i$ (top) and $\Delta\sigma_{c,50\%}$ (bottom) in log-space. Left: histogram with KDE overlay; middle: P–P plot vs. normal distribution; right: P–P plot vs. Student's~t.}
    \label{fig:target_distributions}
\end{figure*}

Figure~\ref{fig:target_distributions} displays histograms and probability–probability (P–P) plots for both targets, each evaluated in linear and log-transformed space. In linear scale, both $N_i$ and $\Delta\sigma_{c,50\%}$ exhibit right-skewed, heavy-tailed distributions, poorly approximated by either Gaussian or Student's~t reference distributions (see e.g. $R^2 = 0.8284$ and $R^2 = 0.9677$ for $N_i$ in linear space). However, log-transformation leads to approximately symmetric, bell-shaped distributions. For $\log_{10}(N_i)$, both the histogram and P–P plots indicate a near-normal distribution with $R^2 = 0.9949$ for the Gaussian and $R^2 = 0.9949$ for the t-distribution fit. A similar improvement is observed for $\log_{10}(\Delta\sigma_{c,50\%})$, where the log-transformed values yield $R^2 = 0.9824$ and $R^2 = 0.9871$, respectively.

These results validate the use of $\log_{10}$-scaling as part of the target transformation strategy. The increased normality in log-space is particularly beneficial for regression algorithms that implicitly assume homoscedastic Gaussian noise or linear residual structures.

\subsection{Feature Engineering, Data Imputation and Golden Feature Creation}
\label{sec:Results_featureEng}

\subsubsection{Data Imputation}
\label{sec:Resuls_DataImputation}

Data imputation was conducted as described in Sec.~\ref{sec:Methods_DataImputation} using domain expertise and statistical methods within an integrated data preprocessing pipeline. After imputation, all features contained no more missing value entries.

\subsubsection{Multicollinearity Assessment and Feature Reduction}
\label{sec:VIF_results}

To mitigate multicollinearity among the continuous input variables, we computed the Variance Inflation Factor (VIF) for all numerical features after imputation but prior to scaling. The VIF quantifies the degree to which each predictor variable is linearly correlated with the remaining features, with a common threshold of $VIF > 5$ indicating problematic redundancy. 

An initial VIF analysis revealed several strongly collinear features, including the tensile and yield strengths from both base and filler materials, as well as geometric variables such as plate width and attachment length. Tab.~\ref{tab:vif_results} shows, that for the initial set of features, four features exhibited VIFs~>~10. As a consequence, features exhibiting high collinearity were systematically excluded or recombined. Specifically, the width and attachment length were condensed into a derived variable denoted as \textit{overhang}, defined as half the difference between plate width and attachment length. Similarly, redundant material parameters were reduced by retaining only the yield strength of the base material as a representative mechanical indicator.
\begin{table}[htbp]
\centering
\caption{Variance Inflation Factors (VIF) Across Feature Selection Rounds}
\label{tab:vif_results}
\begin{tabular}{llr}
\toprule
\textbf{Round} & \textbf{Feature} & \textbf{VIF} \\
\midrule
\multirow{13}{*}{Initial} 
    & $R_{m}$ [MPa]           & 11.85 \\
    & $R_{e_{H}}$ [MPa] & 11.01 \\
    & $w_{B_P}$ [mm]            & 4.92 \\
    & $l_S$ [mm]                & 4.60 \\
    & $l_{B_P}$ [mm]            & 1.48 \\
    & $a_w$ [mm]                & 1.46 \\
    & ...    & ... \\
\midrule
\multirow{12}{*}{After Removal}
    & $w_{B_P}$ [mm]            & 4.90 \\
    & $l_S$ [mm]                & 4.58 \\
    & $l_{B_P}$ [mm]            & 1.47 \\
    & $a_w$ [mm]                & 1.43 \\
    & $t_S$ [mm]                & 1.37 \\
    & $\Delta\sigma_i$ [MPa]    & 1.34 \\
    & $t_{B_P}$ [mm]            & 1.34 \\
    & ...    & ... \\
\bottomrule
\end{tabular}
\end{table}

The revised feature set exhibited VIF values consistently below 5 (cf.~Tab.~\ref{tab:vif_results}), confirming acceptable collinearity levels and a reduced risk of variance inflation in downstream modeling. This refinement step not only ensured statistical robustness but also aligned with engineering relevance by preserving structurally meaningful and independent predictors. The resulting set offers an interpretable and stable foundation for fatigue strength modeling.

\subsubsection{Golden Feature Creation via AutoML} 
\label{sec:Results_featureEng_GoldenFeatures}

During model training, cf.~Sec.~\ref{sec:Results_HypothesesModels}, we employed MLJAR’s automated \emph{golden features} mechanism \cite{mljar}, which generates interaction- and ratio-based derived features. These engineered variables (including sums, differences, fractions, etc.) demonstrated being amongst the best performing models, the created features however are meaningless from an engineering perspective. Tab.~\ref{tab:example_goldenFeatures} presents a few illustrative examples of golden features identified - such as sums and ratios combining post-treatment indicators with geometric parameters. 

\begin{table}[htbp]
\centering
\caption{Selected examples of MLJAR-generated “golden features”}
\label{tab:example_goldenFeatures}
\begin{tabular}{ll}
\toprule
\textbf{Golden Feature} \\
\midrule
Sum of 'as-welded' post-treatment indicator and plate thickness \\
Difference between stress ratio and TIG-dressing indicator \\
Difference of attachment thickness and TIG-dressing indicator \\
Difference of plate length and TIG-dressing indicator \\
\bottomrule
\end{tabular}
\end{table}

While these automatically derived features occasionally yielded modest improvements in cross-validation metrics, closer inspection revealed they often lacked physical meaning, mixed categorical and continuous variables, or violated unit consistency principles (e.g., adding “TIG dressing” flags to plate thickness). Due to this uninterpretability and inconsistency from an engineering standpoint, we ultimately excluded all golden features from the final models, prioritizing transparency, unit coherence, and domain relevance. For the sake of interpretability, the authors decided to omit all models with non-interpretable new features.

\subsection{Machine and Deep Learning Model Performance Across Fatigue Model Hypotheses}
\label{sec:Results_HypothesesModels}

\subsubsection{Model \texorpdfstring{$\mathcal{M}_1$}{M1}: Base Configuration}
\label{sec:Results_HypothesesModels_M1}

\begin{figure*}[htbp]
    \centering
    \includegraphics[width=0.9\textwidth]{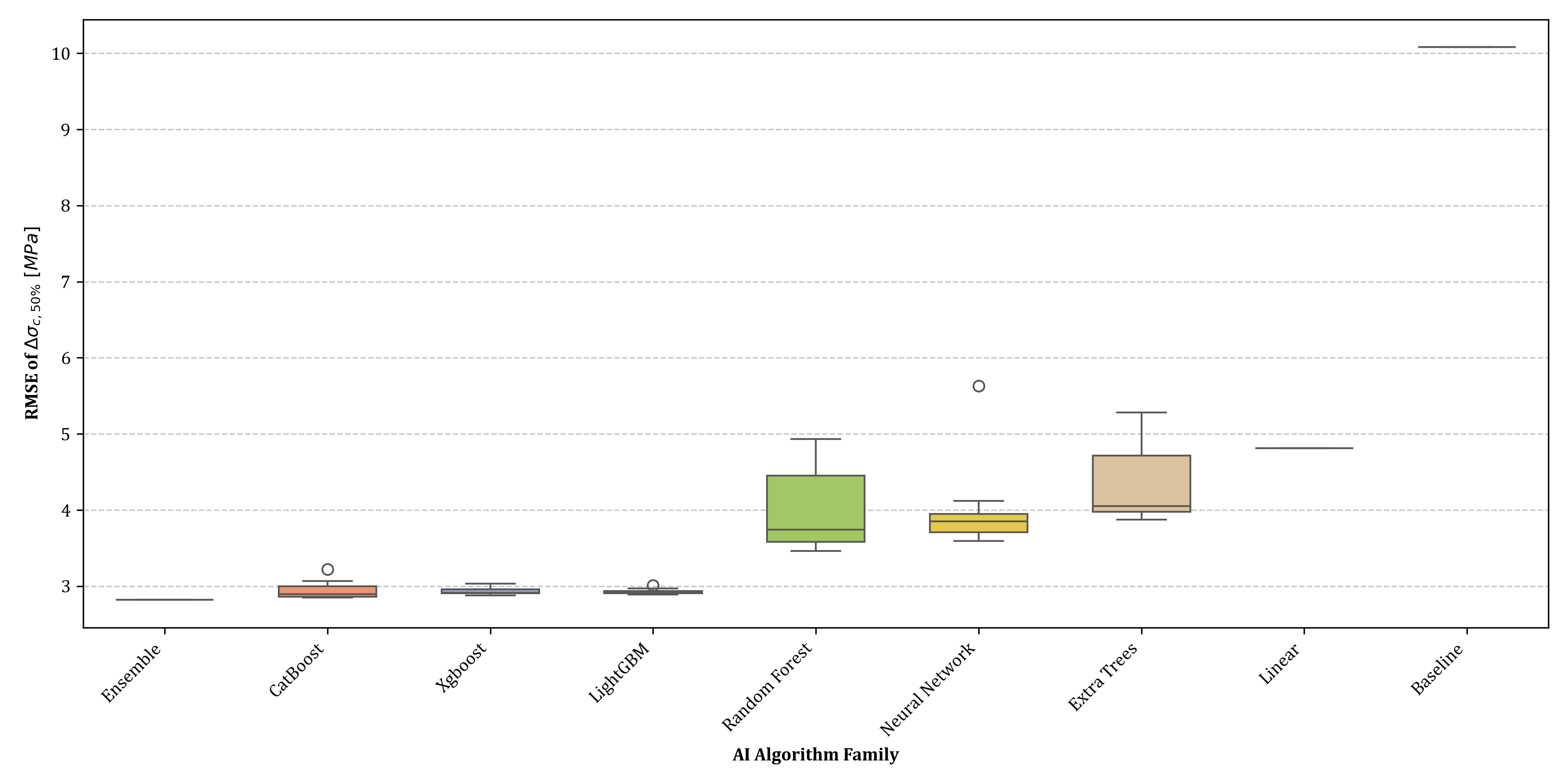}
    \caption{RMSE of different AI algorithm families for predicting $\Delta\sigma_{c,50\%}$ with Model $\mathcal{M}_1$.}
    \label{fig:m1_rmse_boxplot}
\end{figure*}

For predicting $\Delta\sigma_{c,50\%}$, the model $\mathcal{M}1$ yields a training performance of $R^2 = 0.8982$ and RMSE = 21.41 MPa and a test set performance of $R^2 = 0.7792$ and RMSE = 30.2 MPa, with low variance across ensemble-based methods as shown in Figure~\ref{fig:m1_rmse_boxplot}.
 
\begin{figure}[htbp]
\centering
\includegraphics[width=0.9\textwidth]{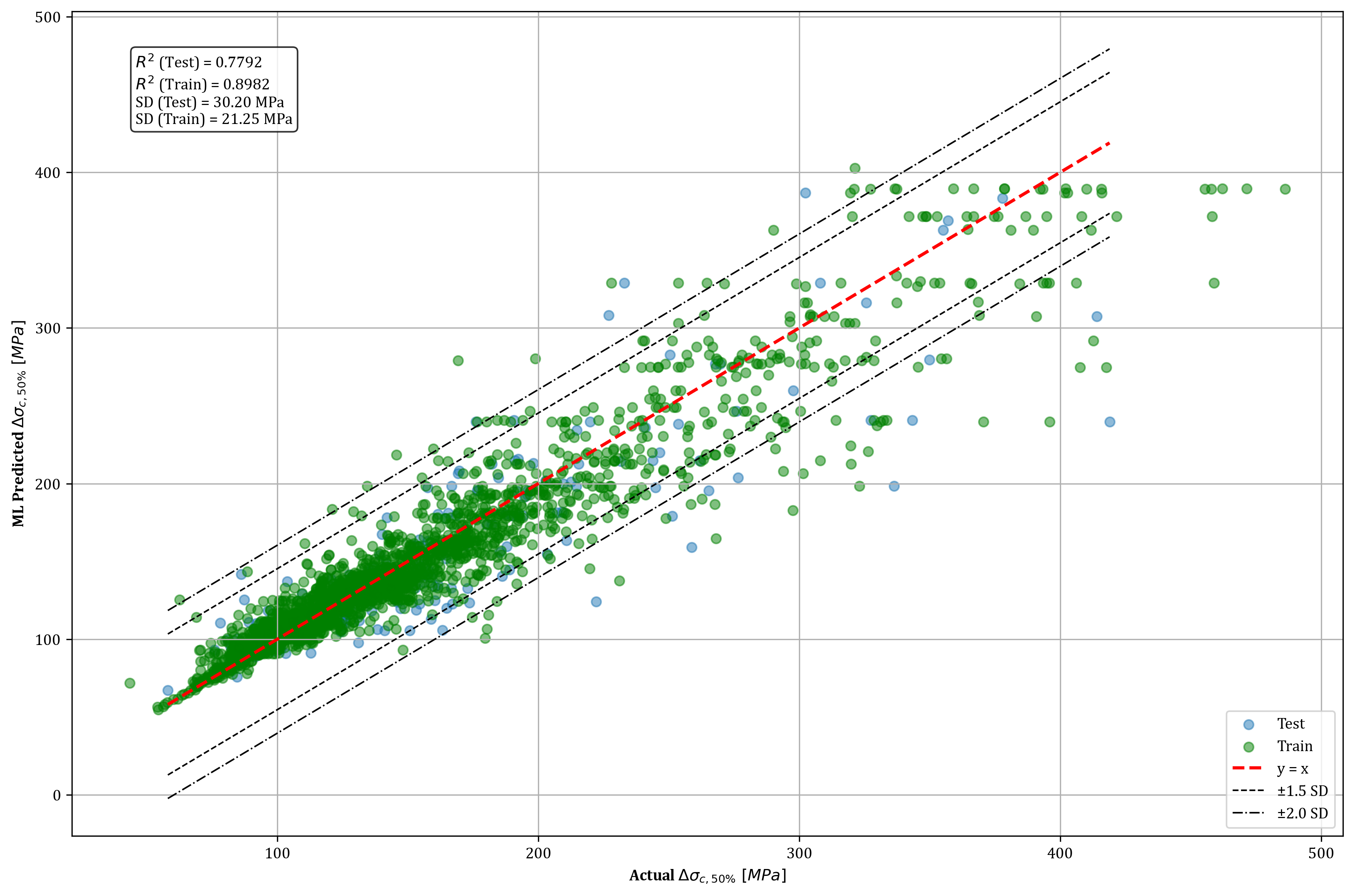}
\caption{Comparison between predicted and actual values of $\Delta\sigma_{c,50\%}$ for Model $\mathcal{M}_1$. }
\label{fig:m1_pred_actual}
\end{figure}

The parity plot (Figure~\ref{fig:m1_pred_actual}) confirms strong agreement between predicted and observed values, with the majority of predictions within $\pm1.5$ standard deviations (dashed line).

\begin{figure*}[htbp]
    \centering
    \begin{subfigure}[b]{0.45\textwidth}
        \includegraphics[width=\textwidth]{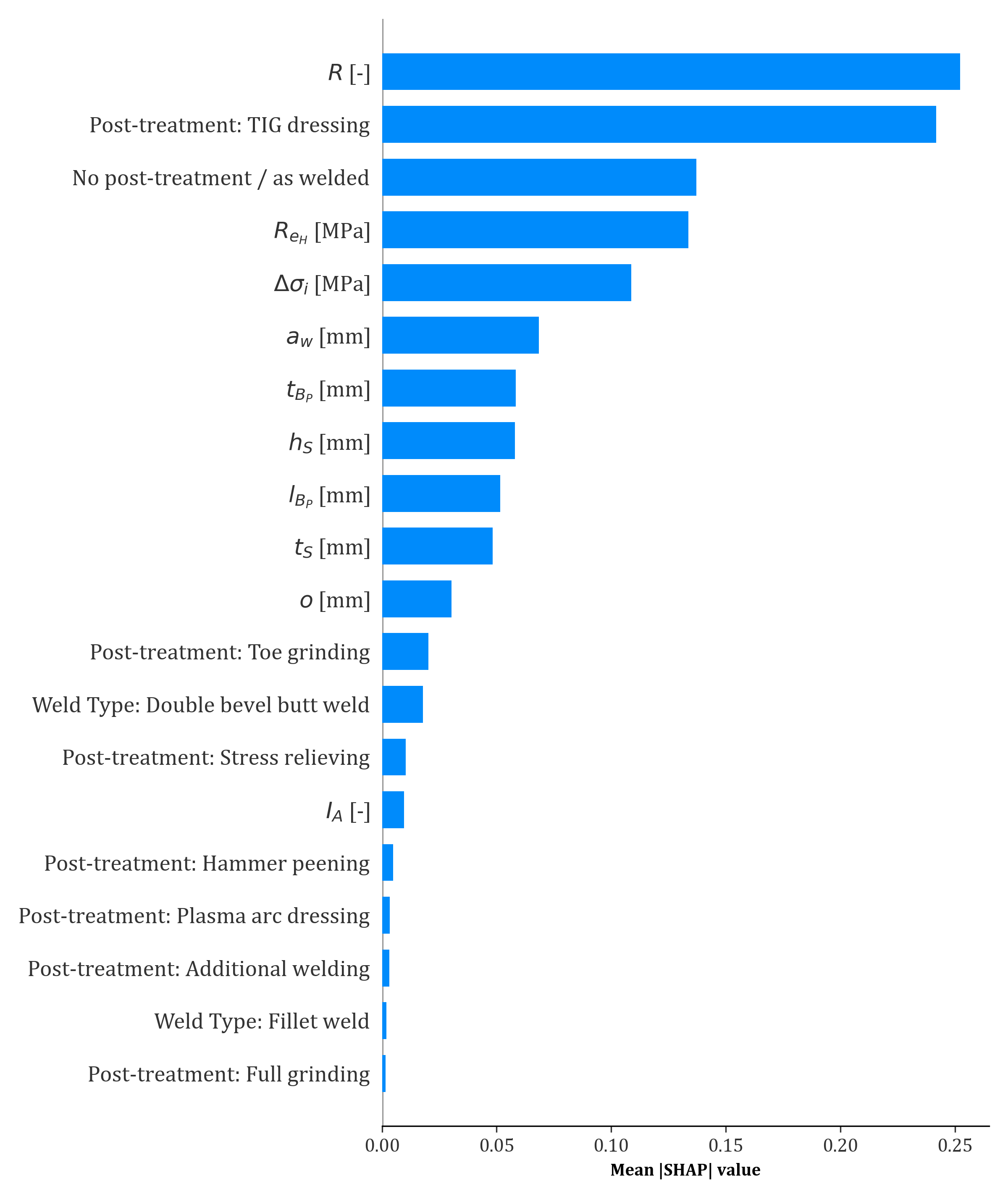}
        \caption{Mean absolute SHAP values}
        \label{fig:m1_shap_bar}
    \end{subfigure}
    \hfill
    \begin{subfigure}[b]{0.45\textwidth}
        \includegraphics[width=\textwidth]{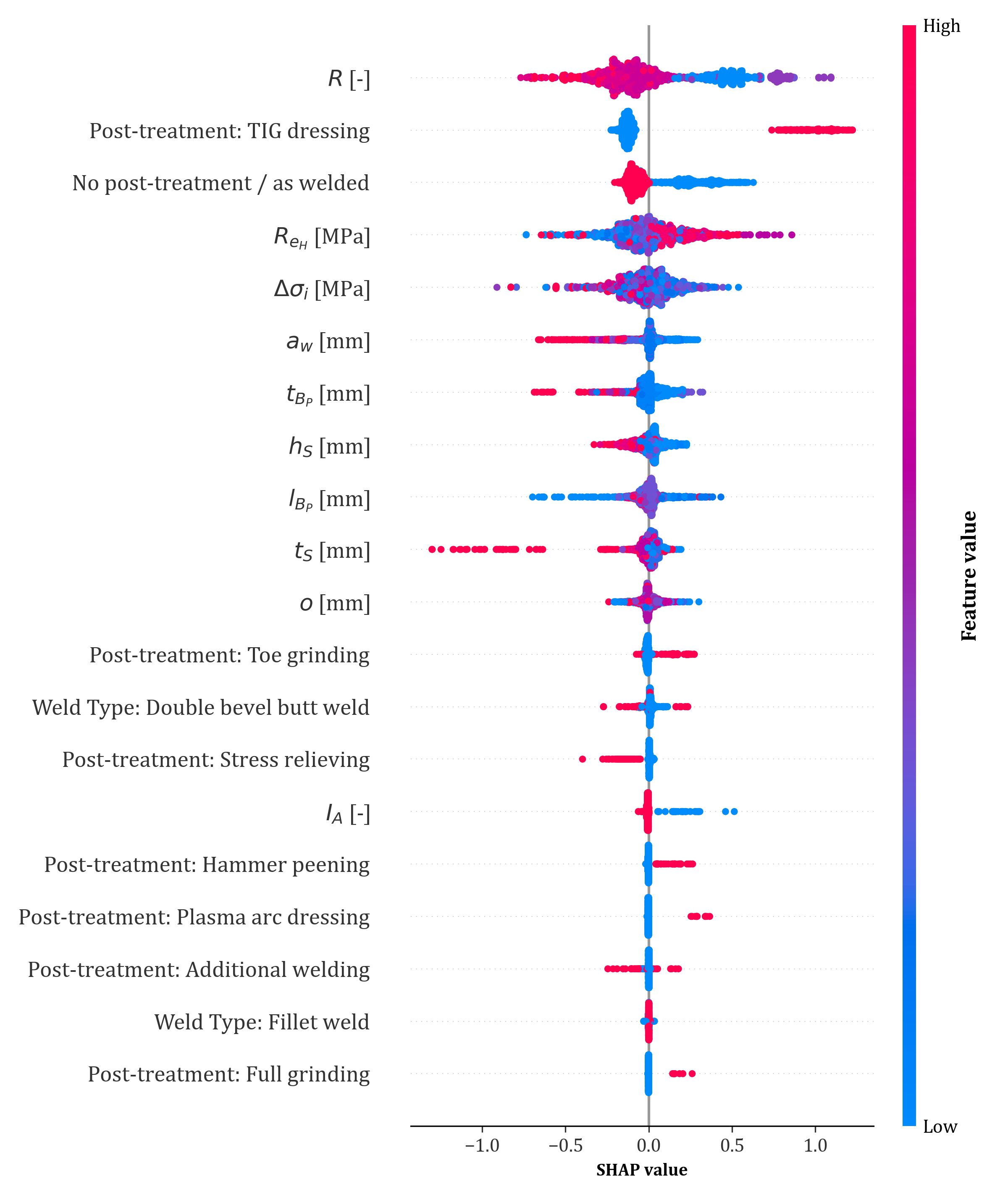}
        \caption{SHAP beeswarm plot}
        \label{fig:m1_shap_beeswarm}
    \end{subfigure}
    \caption{SHAP analysis for Model $\mathcal{M}_1$: (a) Global feature importance, (b) Feature value impact.}
    \label{fig:m1_SHAP}
\end{figure*}

SHAP analysis (Figures ~\ref{fig:m1_shap_bar} and ~\ref{fig:m1_shap_beeswarm}) reveals the $R$-ratio, post-treatment types (notably TIG dressing and as-welded condition), and mechanical parameters such as $R_{eH}$ and $\Delta\sigma_i$ as the most impactful features. The feature $R$-ratio and post-treatment condition are most influential. These results are well aligned with the current state of research on fatigue. It is well known that stress ratio influences fatigue behavior. Post-treatment methods naturally also affect fatigue performance, which is their intended purpose. It is also plausible that yield strengths can impact fatigue resistance. Also, the stress range itself indicating the cycle area in the fatigue life plays a role: The model shows less scatter in the lower area of stress ranges, going along with high cycle fatigue area, which is the important part for real structures. 
From an expert perspective, these findings are therefore also valid. Interestingly, TIG dressing as a post-treatment method appears to have a distinct effect compared to other techniques. Geometric parameters, which often receive considerable attention in research, appear only later in the order of influence.

\subsubsection{Model \texorpdfstring{$\mathcal{M}_2$}{M2}: Position-Augmented}
\label{sec:Results_HypothesesModels_M2}

\begin{figure*}[htbp]
    \centering
    \includegraphics[width=0.9\textwidth]{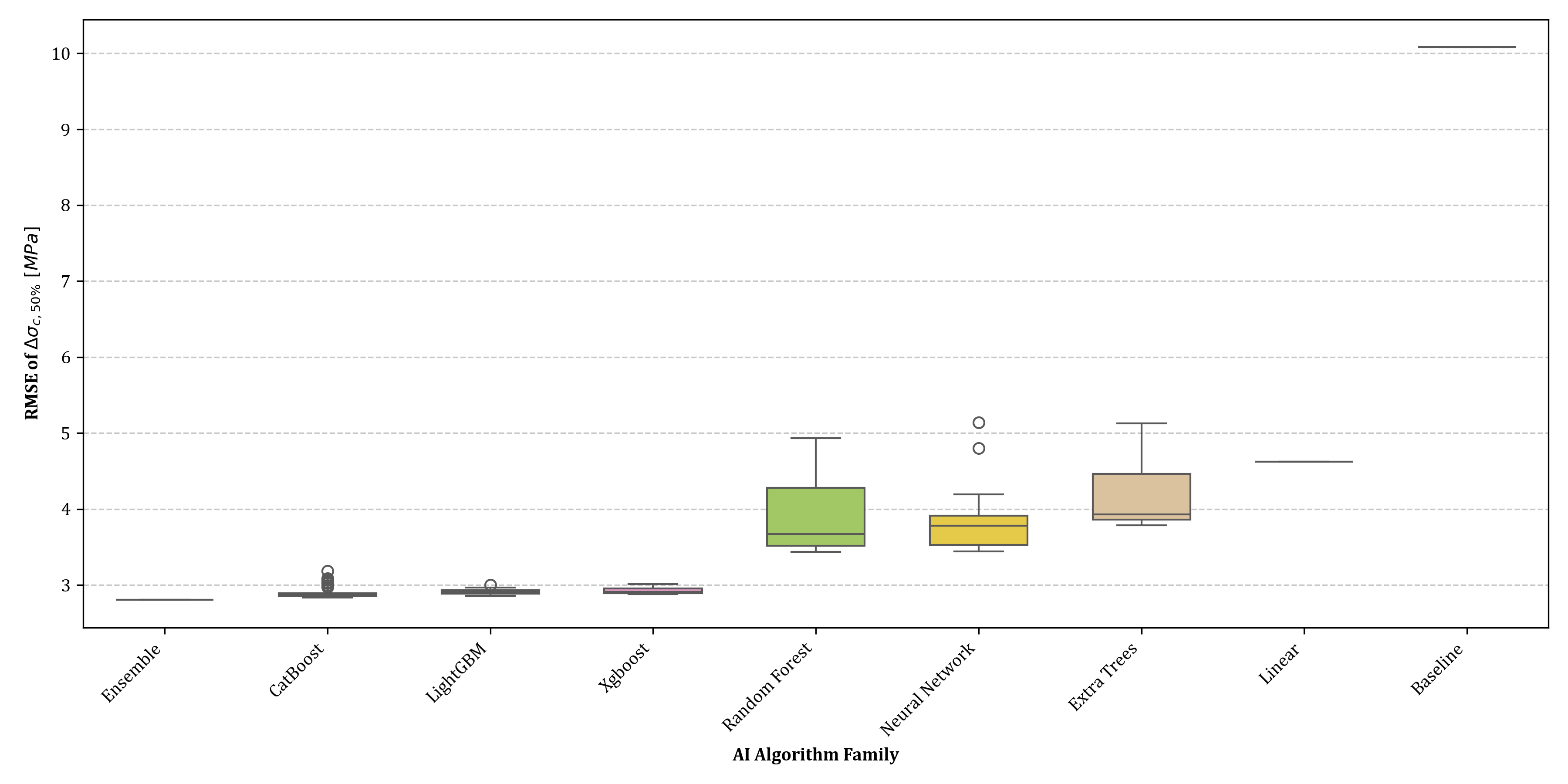}
    \caption{RMSE of different AI algorithm families for predicting $\Delta\sigma_{c,50\%}$ with Model $\mathcal{M}_2$.}
    \label{fig:m2_rmse_boxplot}
\end{figure*}

\begin{figure}[htbp]
\centering
\includegraphics[width=0.9\textwidth]{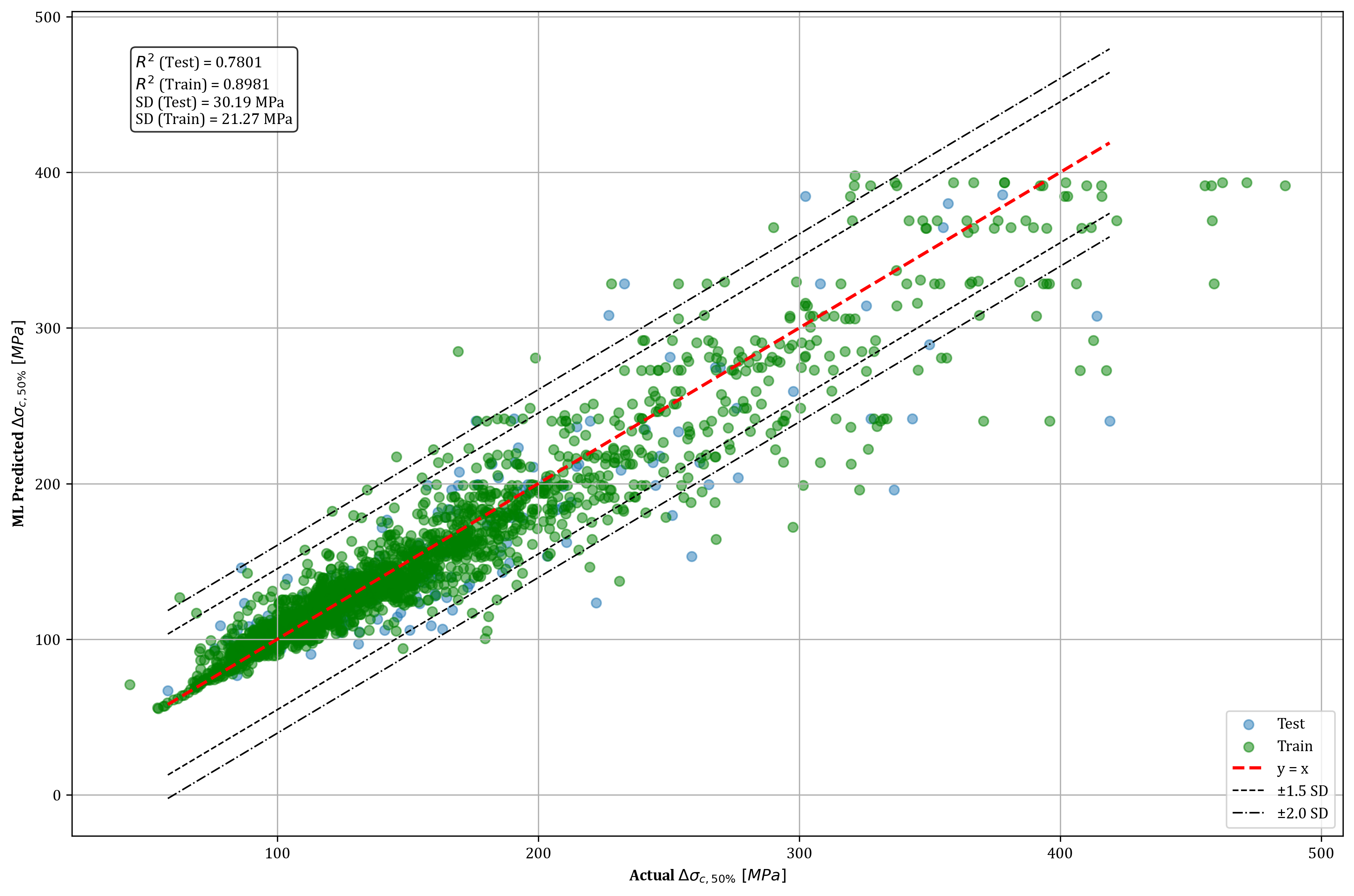}
\caption{Comparison between predicted and actual values of $\Delta\sigma_{c,50\%}$ for Model $\mathcal{M}_2$. }
\label{fig:m2_pred_actual}
\end{figure}

The ensemble model $\mathcal{M}_2$ demonstrated competitive performance across all tested regressors. As illustrated in Figure~\ref{fig:m2_rmse_boxplot}, the "Ensemble" approach achieved the lowest median RMSE among all AI model families within this hypotheses space, followed closely by CatBoost, LightGBM, and XGBoost, which form the top-performing gradient boosting methods. The prediction accuracy is further substantiated in the parity plot (Figure~\ref{fig:m2_pred_actual}), where the model achieved a coefficient of determination of $R^2 = 0.8981$ resp. RMSE = 21.43 MPa on the training set and $R^2 = 0.7801$ resp. RMSE = 30.63 MPa on the test set.

The parity plot (Figure~\ref{fig:m2_pred_actual}) confirms strong agreement between predicted and observed values, with the majority of predictions within $\pm1.5$ standard deviations (dashed line).

\begin{figure*}[htbp]
    \centering
    \begin{subfigure}[b]{0.45\textwidth}
        \includegraphics[width=\textwidth]{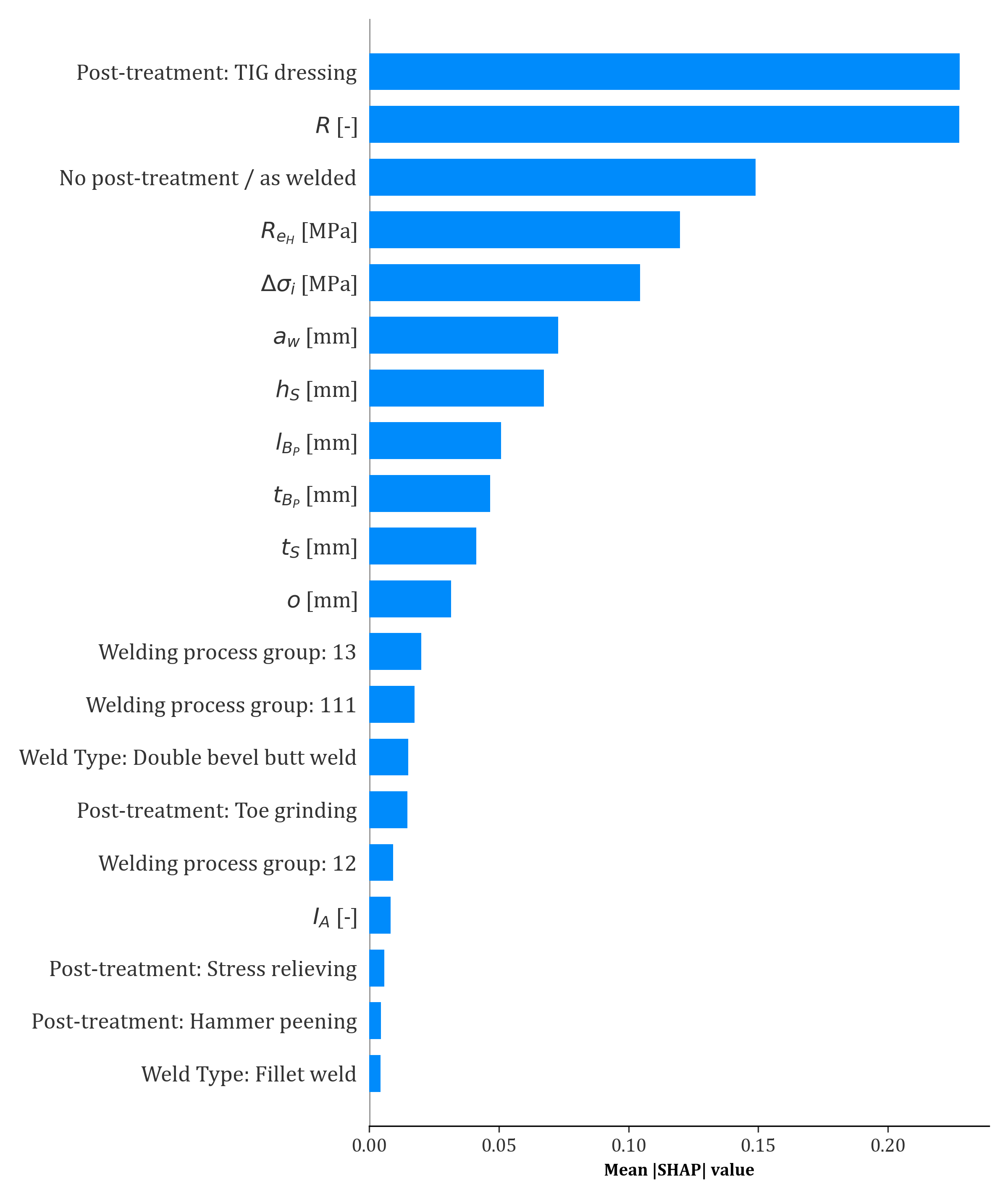}
        \caption{Mean absolute SHAP values}
        \label{fig:m2_shap_bar}
    \end{subfigure}
    \hfill
    \begin{subfigure}[b]{0.45\textwidth}
        \includegraphics[width=\textwidth]{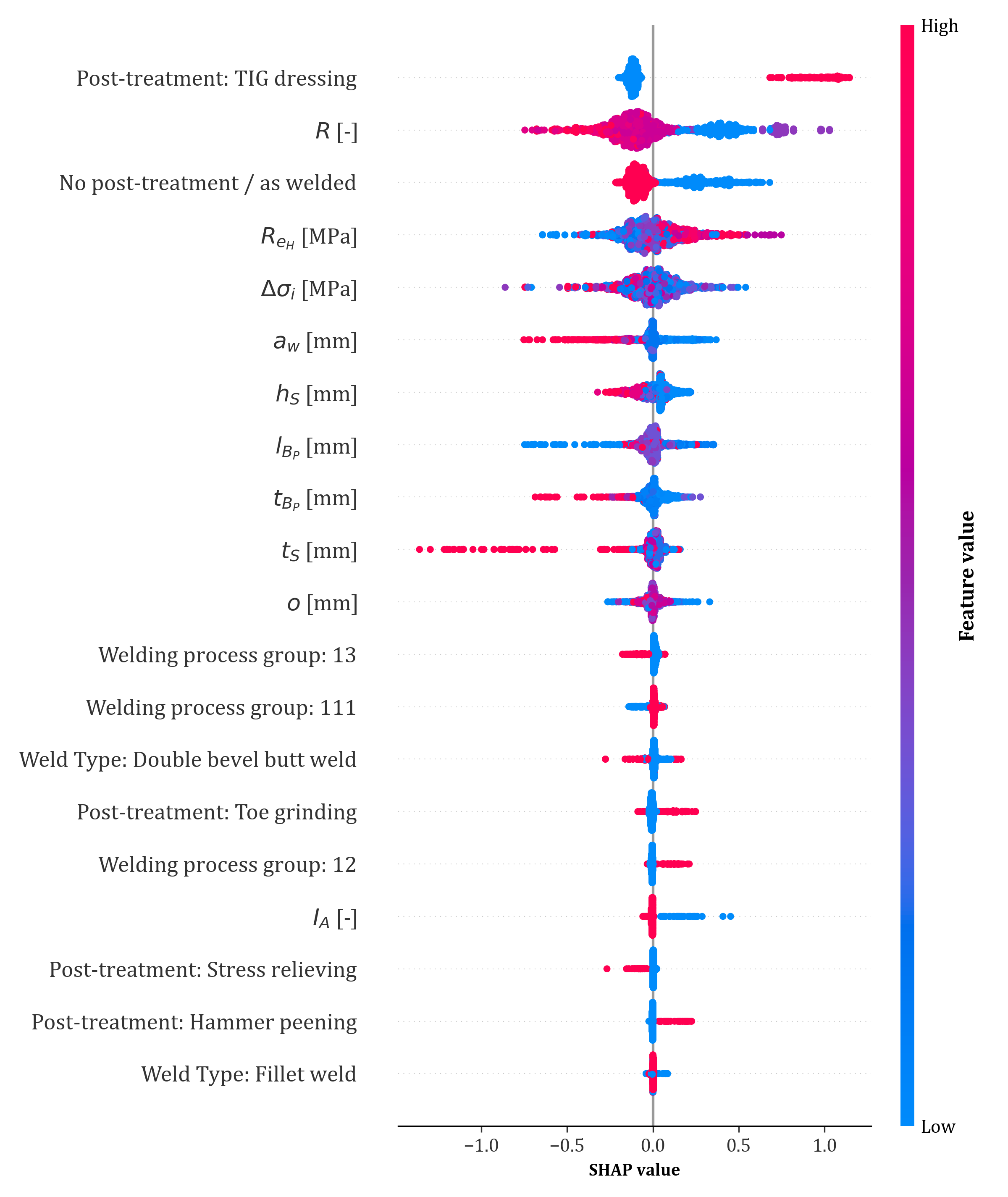}
        \caption{SHAP beeswarm plot}
        \label{fig:m2_shap_beeswarm}
    \end{subfigure}
    \caption{SHAP analysis for Model $\mathcal{M}_2$: (a) Global feature importance, (b) Feature value impact.}
    \label{fig:m2_SHAP}
\end{figure*}

Feature importance analysis via SHAP values is summarized in Figures~\ref{fig:m2_shap_bar} and~\ref{fig:m2_shap_beeswarm}. The most influential predictor was the post-treatment method “TIG dressing,” followed by the stress ratio $R$, and the absence of post-treatment (as-welded). Other relevant features include material yield strength $R_{eH}$, applied stress range $\Delta\sigma_i$, throat thickness $a_w$, and geometric properties such as $h_S$, $l_{BP}$, and $t_S$. The SHAP beeswarm plot reveals clear directional effects, with high $R$-values and TIG dressing contributing positively to higher fatigue classes.
Again, the results are valid from a fatigue-expert view: TIG dressing, a low stress ratio as well as a high yield strength tend to improve the fatigue strength, shown by common research results as well as by the presented model.

\subsubsection{Model \texorpdfstring{$\mathcal{M}_3$}{M3}: Full Process Model}
\label{sec:Results_HypothesesModels_M3}

\begin{figure*}[htbp]
    \centering
    \includegraphics[width=0.9\textwidth]{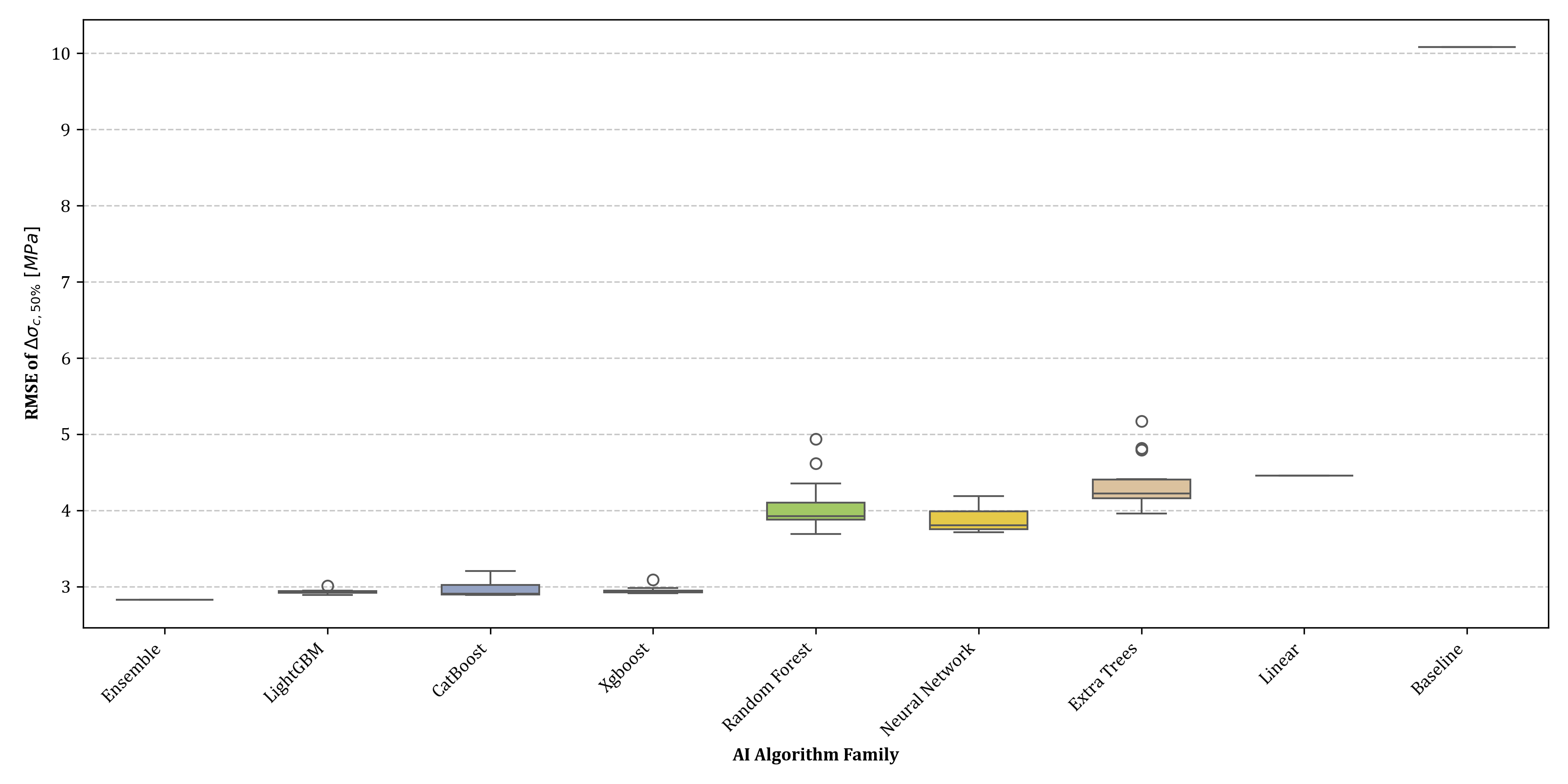}
    \caption{RMSE of different AI algorithm families for predicting $\Delta\sigma_{c,50\%}$ with Model $\mathcal{M}_3$.}
    \label{fig:m3_rmse_boxplot}
\end{figure*}

The performance of Model $\mathcal{M}_3$ - based on the final engineered feature set - demonstrates improved predictive accuracy and robustness across multiple metrics. The RMSE distribution (Figure \ref{fig:m3_rmse_boxplot}) shows that the ensemble model outperforms all individual algorithms, including gradient boosting variants (LightGBM, XGBoost, CatBoost), random forests, neural networks, and linear models. The ensemble achieves a median RMSE well below that of the baseline and other algorithmic families, indicating enhanced generalization.

\begin{figure}[htbp]
\centering
\includegraphics[width=0.9\textwidth]{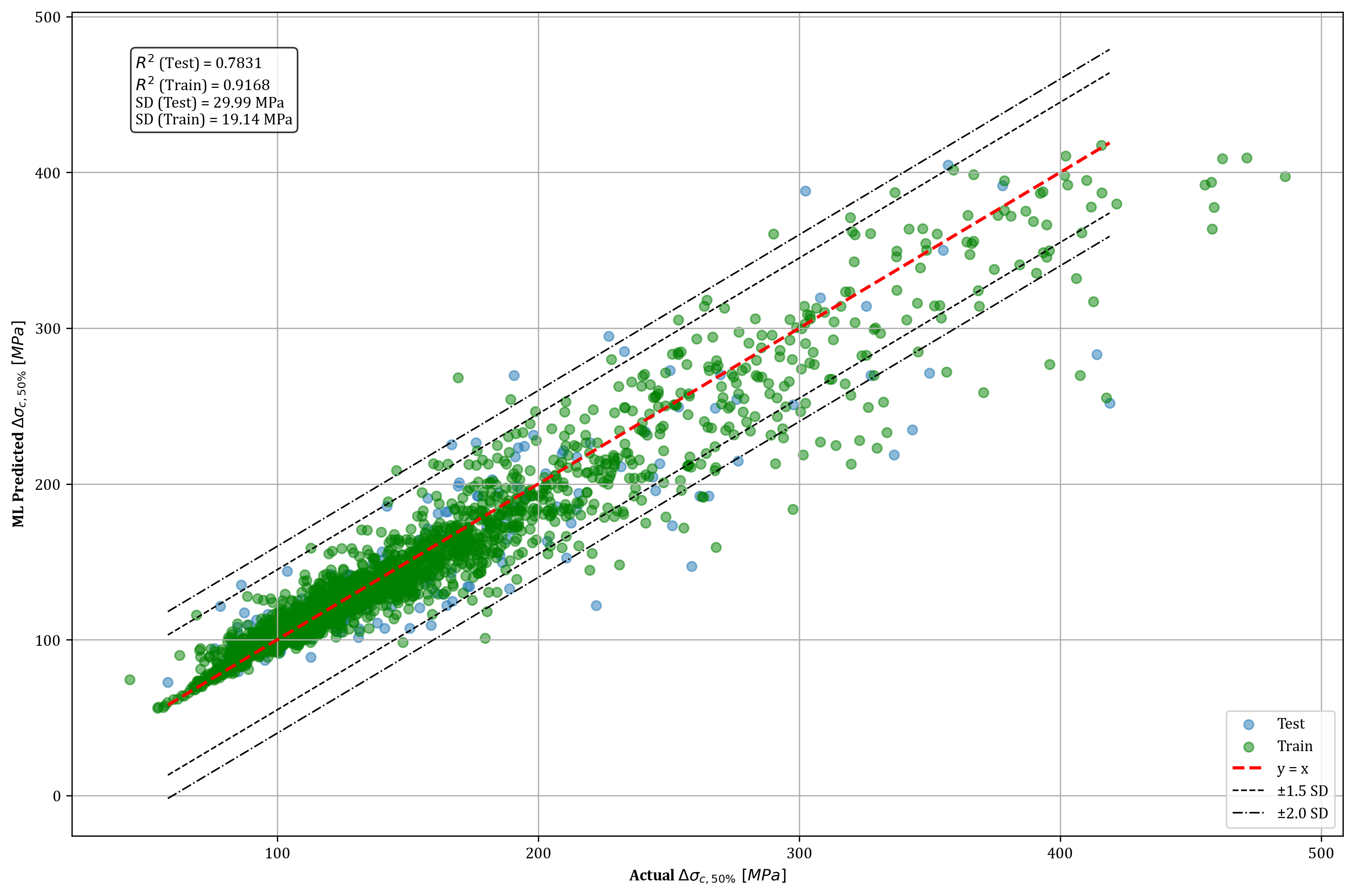}
\caption{Comparison between predicted and actual values of $\Delta\sigma_{c,50\%}$ for Model $\mathcal{M}_3$. }
\label{fig:m3_pred_actual}
\end{figure}

The parity plot of predicted versus actual values for $\Delta\sigma_{c,50\%}$ (Figure \ref{fig:m3_pred_actual}) confirms the high accuracy, with a coefficient of determination of $R^2 = 0.9168$ resp. RMSE = 19.36 MPa on the training set and $R^2 = 0.7831$ resp. RMSE = 30.42 MPa on the test set. The residuals remain mostly within the $\pm1.5$ standard deviation confidence bounds, reflecting stable predictive variance. The observed slight overprediction trend at higher stress levels is consistent with previously discussed model biases.

\begin{figure*}[htbp]
    \centering
    \begin{subfigure}[b]{0.45\textwidth}
        \includegraphics[width=\textwidth]{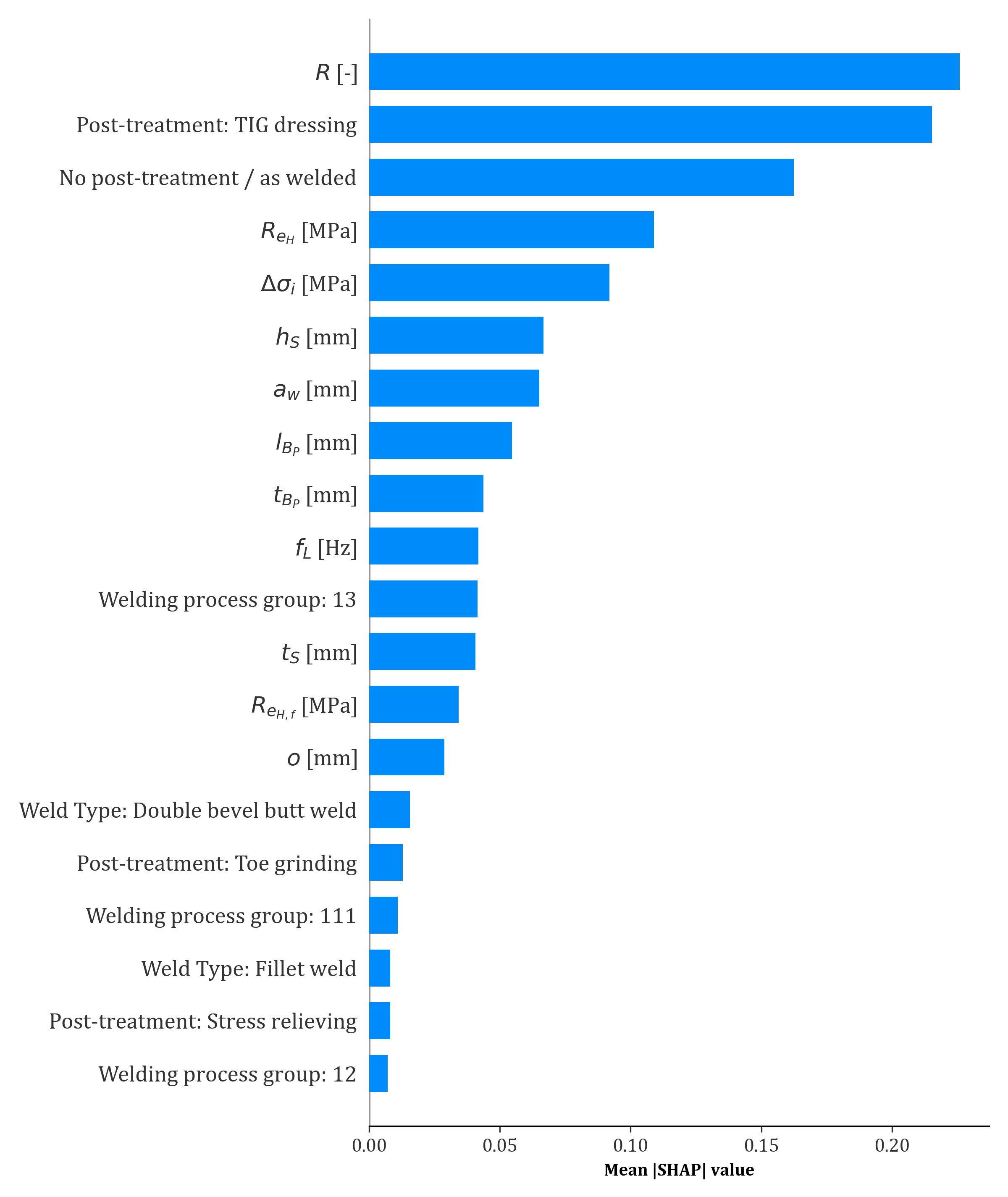}
        \caption{Mean absolute SHAP values}
        \label{fig:m3_shap_bar}
    \end{subfigure}
    \hfill
    \begin{subfigure}[b]{0.45\textwidth}
        \includegraphics[width=\textwidth]{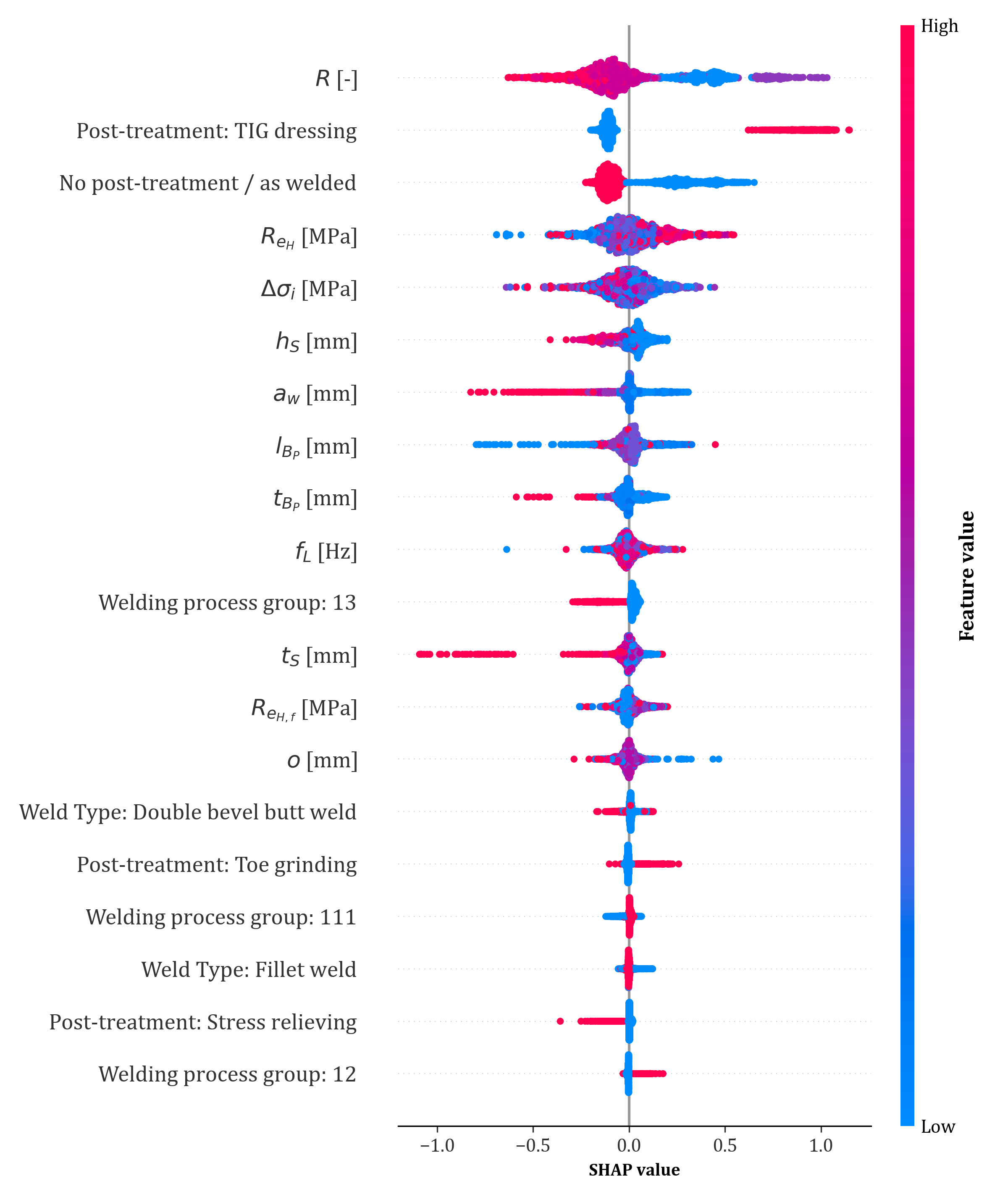}
        \caption{SHAP beeswarm plot}
        \label{fig:m3_shap_beeswarm}
    \end{subfigure}
    \caption{SHAP analysis for Model $\mathcal{M}_3$: (a) Global feature importance, (b) Feature value impact.}
    \label{fig:m3_SHAP}
\end{figure*}

SHAP-based feature importance analysis (Figures \ref{fig:m3_shap_bar} and \ref{fig:m3_shap_beeswarm}) identifies the most influential factors in the model's prediction. The stress ratio $R$ and post-treatment method “TIG dressing” exhibit the highest mean SHAP values, followed by base material yield strength ($R_{eH}$) and initial stress range $\Delta\sigma_i$. Notably, geometric descriptors such as attachment height $h_S$ and weld throat thickness $a_w$ also emerge as significant contributors, underscoring the relevance of local detailing to fatigue strength. Again, also these results comply with common known influencing factors on fatigue strength of welded details.  

\subsubsection{Model Performance Comparison}
\label{sec:Results_HypothesesModels_Mcomparison}
Table~\ref{tab:model_comparison} summarizes the predictive performance of the three models \(\mathcal{M}_1\), \(\mathcal{M}_2\), and \(\mathcal{M}_3\). On the full range of $\Delta\sigma_{c,50\%}$, all models perform similarly with test-set $R^2\approx0.78$ and RMSE$_\text{Test}\approx30\,$MPa. Model~$\mathcal{M}_3$ achieves the lowest training error (RMSE$_\text{Train}=19.36\,$MPa, $R^2_\text{Train}=0.9168$) but only marginally outperforms the others on the test set. Restricting the analysis to the relevant band regarding high cycle fatigue e.g. in bridges ($\Delta\sigma_{c,50\%}\le150\,$MPa) reduces RMSE for all models by over 50\%, with $\mathcal{M}_3$ again performing best in-sample (RMSE$_\text{Train}=9.43\,$MPa) and being closely matched by $\mathcal{M}_1$ on the test set (13.03 vs.\ 13.17~MPa). The nearly identical metrics of $\mathcal{M}_1$ and $\mathcal{M}_2$ across both ranges underscores the robustness of the core features used.

\begin{table}[htbp] 
\caption{Performance comparison of Models $\mathcal{M}_1$‒$\mathcal{M}_3$}
\centering
\label{tab:model_comparison}
\setlength{\tabcolsep}{6pt}  
\begin{tabular}{lcccccc}
\toprule
Range of $\Delta\sigma_{c,50\%}$ & \multicolumn{3}{c}{\textbf{Full}} & \multicolumn{3}{c}{\textbf{0 – 150 MPa}} \\
\cmidrule(lr){2-4} \cmidrule(lr){5-7}
\textbf{Metric}
& $\boldsymbol{\mathcal{M}_1}$ & $\boldsymbol{\mathcal{M}_2}$ & $\boldsymbol{\mathcal{M}_3}$
& $\boldsymbol{\mathcal{M}_1}$ & $\boldsymbol{\mathcal{M}_2}$ & $\boldsymbol{\mathcal{M}_3}$ \\
\midrule
$R^2_{\mathrm{Train}}$
& 0.8982 & 0.8981 & 0.9168
& 0.7180 & 0.7168 & 0.7791 \\
$R^2_{\mathrm{Test}}$
& 0.7792 & 0.7801 & 0.7831
& 0.5539 & 0.5264 & 0.5440 \\
RMSE Train [MPa]
& 21.41 & 21.43 & 19.36
& 10.65 & 10.67 & 9.427 \\
RMSE Test [MPa]
& 30.69 & 30.63 & 30.42
& 13.03 & 13.43 & 13.17 \\
MAE Train [MPa]
& 12.97 & 12.96 & 11.91
&  7.425 &  7.413 &  6.781 \\
MAE Test [MPa]
& 18.98 & 19.05 & 18.97
& 9.363 & 9.684 &  9.678 \\
\bottomrule
\end{tabular}
\end{table}

\begin{figure}
    \centering
    \includegraphics[width=0.95\linewidth]{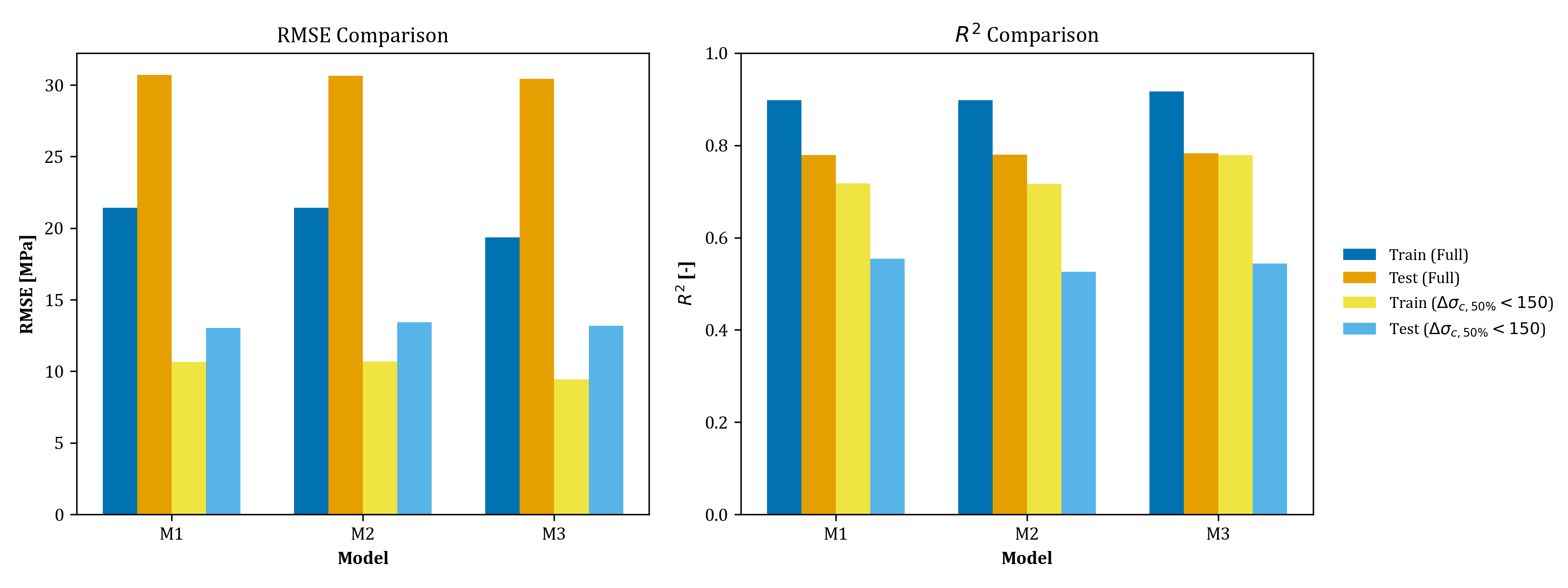}
    \caption{Comparative performance of the three AutoML-derived models depending on the numerical range of $\Delta\sigma_{c,50\%}$}
    \label{fig:modelcomparison}
\end{figure}

A grafical representation of the model results is provided in Fig.~\ref{fig:modelcomparison}.

\section{Discussion}
\label{sec:Results_Discussion}

Transforming the raw 78-column fatigue database into an AutoML-ready asset demanded substantial effort. Extensive exploratory analysis uncovered numerous inconsistencies -units, typos, duplicated records—and more than 30\% overall missingness, requiring tailored imputation schemes that respected physical bounds and default design values. Categorical variables were harmonised through manual ontology mapping (e.g., unifying more than 40 post-treatment spellings), while continuous features underwent log- and Yeo–Johnson transforms to stabilise variance and satisfy model assumptions. Redundant or highly collinear predictors were pruned via VIF screening and expert checks, reducing dimensionality by roughly one‐third without sacrificing domain coverage. In total, this data-wrangling pipeline accounted for the majority of project time—but it was indispensable for achieving stable, interpretable AutoML outcomes.

Table~\ref{tab:model_comparison} provides a comprehensive comparison of the three AutoML‐derived models across both the full range and the relevant range of the target regarding high cycle fatigue ($\Delta\sigma_{c,50\%}\le150\,$MPa). Across the full range, all models exhibit strong training performance (\(R^2_{\mathrm{Train}} \approx 0.90\)) and slightly degraded but consistent test metrics (\(R^2_{\mathrm{Test}} \approx 0.78\), RMSE $\approx30–31$~MPa, MAE $\approx 19$~MPa). Model \(\mathcal{M}_3\), with its expanded feature set, achieves the highest training \(R^2\) (0.9168) and lowest training error, but only marginally improves test performance over \(\mathcal{M}_1\) and \(\mathcal{M}_2\), suggesting limited gains from the additional variables.

When focusing solely on the 0–150~MPa domain - most representative of design‐relevant fatigue strengths - the coefficient of determination $R^2$ decreases significantly across all models while the RMSE and MAE metrics massively improve. Here, training \(R^2\) drops to around 0.72 for \(\mathcal{M}_1\) and \(\mathcal{M}_2\), increasing to 0.78 for \(\mathcal{M}_3\). Test \(R^2\) remains in the 0.53–0.55 range, with RMSE values stabilizing between 13.0 and 13.4~MPa, and MAE between 9.3 and 9.7~MPa. These results indicate robust and relatively similar performance across all configurations, with \(\mathcal{M}_3\) showing slightly better training fit in the reduced domain but no meaningful improvement in test accuracy. Notably, the comparable RMSE and MAE values across the models suggest that, despite its complexity, \(\mathcal{M}_3\) does not provide significantly superior generalization over simpler alternatives.

The SHAP‐based feature importance analysis across all models consistently emphasizes the stress ratio \(R\), yield strength \(R_{eH}\), stress range \(\Delta\sigma_i\), and categorical indicators of post‐weld treatment—especially “TIG dressed” and “as welded”—as the most influential predictors. These results align with fatigue mechanics: stress ratio and stress range are fundamental loading parameters affecting fatigue life, yield strength defines material resistance, and TIG dressing improves weld toe geometry and residual stress distribution. Geometric factors such as plate width, stiffener height, and throat thickness appear as secondary influencers, corroborating their role in local stiffness and stress concentration. The consistency of this feature hierarchy across multi‐model SHAP analyses reinforces the physical plausibility and domain relevance of the predictive models.

From an engineering standpoint, \(\mathcal{M}_2\) provides the optimal trade‐off: it achieves nearly identical generalization performance compared to \(\mathcal{M}_3\), while avoiding unnecessary complexity and overfitting. Its slightly higher RMSE in the reduced domain is offset by similar predictive reliability and interpretability. Thus, \(\mathcal{M}_2\) is recommended as the preferred model: it balances accuracy, stability, and transparency—key criteria for deployment in fatigue assessment pipelines. For practitioners, this means robust fatigue‐strength predictions with clear, actionable insights on critical variables and minimal risk of over‐fit artifacts.

\section{Conclusions and Outlook}
\label{sec:Conclusions}
The paper investigated the discovery of fatigue strength models via feature engineering and automated explainable machine learning showing the example of the welded transverse stiffener.

This study demonstrates the following statements:

\begin{itemize}
    \item This paper successfully develops and applies a \textbf{unified AutoML + XAI pipeline for fatigue strength prediction} for fatigue strength prediction of welded transverse stiffener details, combining domain-informed feature engineering with SHAP-based explainability resp. interpretability.

    \item \textbf{Ensemble gradient-boosting models} (\textsc{XGBoost}, \textsc{LightGBM}) deliver the best performance in predicting the medium fatigue strength with 50\% failure probability ($\Delta\sigma_{c,50\%}$), outperforming neural-network and linear baselines. This is also in line with the common sense in the AI community as tree models being the gold standard for structured data, however leaving a ca. 10\% coefficient of variation for model prediction errors.
    
    \item \textbf{Optimal model hypothesis parametrization}: All three models (\(\mathcal{M}_1\)–\(\mathcal{M}_3\)) achieved test RMSE $\approx~30–31$~MPa and \(R^2_{\mathrm{Test}}\) $\approx~0.78$ over the full \(\Delta\sigma_{c,50\%}\) range. Restricting to the engineering-relevant 0 - 150~MPa domain reduced RMSE to $13–13.4$~MPa and \(R^2_{\mathrm{Test}}\) to 0.53–0.55, confirming applicability for real-world structural steel scenarios.

    \item \textbf{Simplicity outweighs complexity:} Although \(\mathcal{M}_3\) included additional engineered feature interactions, it delivered only minimal training performance gains without improving generalization. In contrast, \(\mathcal{M}_2\) matched accuracy while preserving model parsimony, thus avoiding overfitting. The model selection results hold for both investigated ranges of the target variable. Most interestingly, a critical domain-centered assessment of the automatedly generated "golden features" delivered their physical inconsistencies, and justify their compelte rejection - an aspect rarely documented in structural AI work.
    
    \item \textbf{Physically meaningful key influencing variables} identified by SHAP analysis include the stress ratio $R$, yield strength $R_{eH}$, stress range $\Delta\sigma_{i}$, and especially post-weld treatment via \emph{TIG dressing}, aligning well with established fatigue theory. Geometric factors such as plate width, throat thickness, and stiffener height contribute meaningfully though secondary.
    
    \item \textbf{Practical engineering relevance}: Highly accurate predictions of fatigue classes mostly lie under the area of 150~MPa, matching the operational range of typical civil engineering welded details and enabling immediate practical use.
    
    \item \textbf{Explainability as safeguard}: SHAP values revealed that auto-generated "golden features" were of limited benefit and sometimes inconsistent in units; their exclusion preserved physical coherence and generalisability.
    
  \item \textbf{Recommended model for engineering practice:} Given its balance of predictive accuracy, interpretability, and operational simplicity, \(\mathcal{M}_2\) is the preferred choice for integration into structural fatigue assessment workflows.

\end{itemize}

\textbf{Limitations of this work} stem from: (i) a dataset with a significant amount of missing values in its raw form, which necessitated extensive imputation and may limit representativeness of certain scenarios; (ii) feature imputation strategies were chosen based on domain expertise and were not systematically evaluated for their impact on model stability or performance—alternative imputation schemes (e.g., multiple imputation, KNN, or iterative methods) may produce different results; and (iii) while explainable AutoML methods (e.g., SHAP, feature importance rankings) provided consistent insights, our analysis did not include uncertainty quantification for feature attributions or model predictions, limiting the assessment of reliability under data variability.

\textbf{Future work} will (i) incorporate uncertainty quantification via Bayesian or quantile regressors, (ii) look into mixture-of-expert-architectures in combination with carefully chosen loss function (analogously to \cite{kraus2024strength}), (iii) embed physics-informed regularisation rooted in fracture-mechanics constraints, and (iv) pursue transfer learning to additional fatigue details such as the butt-welded joint or longitudinal welds.  Integration into digital twins and structural-health-monitoring pipelines is envisaged to enable continuous, interpretable fatigue assessment in practice.


\section*{Declaration of Competing Interest}
The authors declare that they have no known competing financial interests or personal relationships that could have appeared to influence the work reported in this paper.

\section*{Declaration of Generative AI and AI-assisted technologies in the writing process} \label{sec:statements}
During the preparation of this paper, the authors did use generative AI and AI-assisted technologies in the form of openAI's chatGPT and its gpt-4-0125-preview model via API as well as "perplexity.ai", each with training status of  knowledge 2024.

\section*{Declaration of Contribution}
\textbf{Michael A. Kraus}: Conceptualisation, Methodology, Formal analysis, Investigation, Data curation, Software, Validation, Interpretation of Results, Visualization,  Writing - original draft, Writing - review \& editing

\textbf{Helen Bartsch}: Conceptualisation, Methodology, Formal analysis, Investigation, Data curation, Validation, Interpretation of Results, Visualization,  Writing - original draft, Writing - review \& editing \\

\section*{Code and Data Availability}
The raw data for this project may be obtained with a Data Use Agreements resp. Researchers interested in access to the data may contact Dr. Helen Bartsch. The code will be made publicly available via Github after publication. 

\section*{Acknowledgements}
The authors kindly acknowledge the financial support provided by TU Darmstadt and RWTH Aachen.

\bibliographystyle{elsarticle-num} 

\bibliography{Manuscript_hb} 

\begin{thebibliography}{10}
\expandafter\ifx\csname url\endcsname\relax
  \def\url#1{\texttt{#1}}\fi
\expandafter\ifx\csname urlprefix\endcsname\relax\def\urlprefix{URL }\fi
\expandafter\ifx\csname href\endcsname\relax
  \def\href#1#2{#2} \def\path#1{#1}\fi

\bibitem{prEN2023}
{FprEN 1993–1-9:2025}, {Eurocode 3: Design of steel structures – Part
  1–9: Fatigue} (2025).

\bibitem{gurney1962}
T.~Gurney, C.~Woodley, Investigation into the fatigue strength of welded beams,
  part iii: high tensile steel beams with stiffeners welded to the web, Br.
  Weld. J. 9~(9) (1962) 533--539.

\bibitem{fisher1974}
J.~Fisher, P.~Albrecht, B.~Yen, D.~Klingerman, M.~McNamee, Fatigue strength of
  steel beams with welded stiffeners and attachments - report 147, Tech. rep.,
  National Cooperative Highway Research Program, Washington D.C., USA (1974).

\bibitem{duerr2017}
A.~Dürr, Ömer Bucak, J.~Roth, Size effect of as-welded and post-weld treated
  construction details under fatigue loading, in: EUROSTEEL 2017, September
  13--15, 2017, Copenhagen, Denmark, University of Applied Sciences Munich,
  Dept. of Civil Engineering, Copenhagen, Denmark, 2017, andre.duerr@hm.edu.

\bibitem{alden2020}
R.~Aldén, Z.~Barsoum, T.~Vouristo, M.~Al-Emrani, Robustness of the hfmi
  techniques and the effect of weld quality on the fatigue life improvement of
  welded joints, Welding in the World 64~(9) (2020) 1947--1956.

\bibitem{gkatzogiannis2021}
S.~Gkatzogiannis, J.~Schubnell, P.~Knoedel, M.~Farajian, T.~Ummenhofer,
  M.~Luke, Investigating the fatigue behaviour of small scale and real size
  hfmi-treated components of high strength steels, Engineering Failure Analysis
  123 (2021) 105300, available online 17 February 2021.
\newblock \href {https://doi.org/10.1016/j.engfailanal.2021.105300}
  {\path{doi:10.1016/j.engfailanal.2021.105300}}.

\bibitem{braun2020}
M.~Braun, R.~Scheffer, W.~Fricke, S.~Ehlers, Fatigue strength of fillet-welded
  joints at subzero temperatures, Fatigue Fract Eng Mater Struct 43 (2020)
  403--416.
\newblock \href {https://doi.org/10.1111/ffe.13163}
  {\path{doi:10.1111/ffe.13163}}.

\bibitem{ferraz2024}
G.~Ferraz, B.~Karabulut, B.~Rossi, Fatigue resistance and reliability
  assessment of hot dip galvanised plates with welded transversal stiffeners,
  Engineering Failure Analysis 163 (2024) 108577, available online 16 June
  2024.
\newblock \href {https://doi.org/10.1016/j.engfailanal.2024.108577}
  {\path{doi:10.1016/j.engfailanal.2024.108577}}.

\bibitem{Bartsch2023b}
H.~Bartsch, M.~Feldmann, An experimental investigation into the influence of
  incorrect root gaps in welded-in stiffeners on fatigue performance,
  International Journal of Fatigue 175~(10) (2023) 107773.
\newblock \href {https://doi.org/10.1016/j.ijfatigue.2023.107773}
  {\path{doi:10.1016/j.ijfatigue.2023.107773}}.

\bibitem{nascimento2023}
S.~Nascimento, J.~O. Pedro, U.~Kuhlmann, Numerical study on steel plate girders
  — intermediate transverse stiffeners axial force, Ce/Papers (2023).
\newblock \href {https://doi.org/10.1002/cepa.2522}
  {\path{doi:10.1002/cepa.2522}}.

\bibitem{loschner2024}
D.~Löschner, I.~Engelhardt, T.~Nitschke-Pagel, T.~Ummenhofer, Study on the
  applicability of a modified strain approach to predict the fatigue life of
  hfmi-treated transverse stiffeners under variable amplitude loading, Welding
  in the World 68 (2024) 1259--1270.
\newblock \href {https://doi.org/10.1007/s40194-024-01746-0}
  {\path{doi:10.1007/s40194-024-01746-0}}.

\bibitem{Han1995}
Y.-L. Han, Artificial neural network technology as a method to evaluate the
  fatigue life of weldments with welding defects, International Journal of
  Pressure Vessels and Piping~(63) (1995).

\bibitem{Pleune2000}
T.~T. Pleune, O.~K. Chopra, Using artificial neural networks to predict the
  fatigue life of carbon and low-alloy steels, Nuclear Engineering and
  Design~(197) (2000).

\bibitem{Genel2004}
K.~Genel, Application of artificial neural network for predicting strain-life
  fatigue properties of steels on the basis of tensile tests, International
  Journal of Fatigue~(26) (2004) 1027--1035.

\bibitem{ghahremani2013}
A.~Ghahremani, et~al., Fatigue testing and structural health monitoring of
  retrofitted web stiffeners on steel highway bridges, Transportation Research
  Record Journal of the Transportation Research Board (2013).
\newblock \href {https://doi.org/10.3141/2360-04} {\path{doi:10.3141/2360-04}}.

\bibitem{zhu2018}
S.-P. Zhu, H.-Z. Huang, Y.~Li, Y.~Liu, Y.~Yang, Fatigue crack propagation
  prediction using an artificial neural network model, Engineering Failure
  Analysis 92 (2018) 92--105.

\bibitem{wang2019}
D.~Wang, D.~Zhang, S.~Ge, Fatigue damage prediction for offshore steel
  structures using support vector machine, Ocean Engineering 173 (2019)
  321--328.

\bibitem{zhao2020}
W.~Zhao, Y.~Li, X.~Gao, Fatigue crack growth prediction in steel structures
  using extreme gradient boosting, Engineering Fracture Mechanics 238 (2020)
  107258.

\bibitem{yang2020}
Y.~Yang, S.-P. Zhu, H.-Z. Huang, Z.~Luo, A hybrid machine learning approach to
  fatigue life prediction, Reliability Engineering \& System Safety 197 (2020)
  106802.

\bibitem{lee2020}
H.~Lee, S.~Ahn, B.-H. Choi, Machine learning for fatigue prediction of steel
  bridges, Applied Sciences 10~(5) (2020) 1703.

\bibitem{zhang2021}
W.~Zhang, X.~Guo, W.~Zhong, A deep learning approach for fatigue life
  prediction of welded joints, Engineering Structures 230 (2021) 111726.

\bibitem{li2021}
X.~Li, W.~Zhang, X.~Guo, Transfer learning for fatigue strength prediction of
  steel welded joints, Engineering Applications of Artificial Intelligence 100
  (2021) 104179.

\bibitem{braun2022comparison}
M.~Braun, L.~Kellner, Comparison of machine learning and stress concentration
  factors-based fatigue failure prediction in small-scale butt-welded joints,
  Fatigue \& fracture of engineering materials \& structures 45~(11) (2022)
  3403--3417.

\bibitem{wu2022}
J.~Wu, X.~Guo, W.~Zhang, Graph neural networks for fatigue life prediction of
  steel joints, Computer-Aided Civil and Infrastructure Engineering 37~(2)
  (2022) 239--257.

\bibitem{liu2023a}
Y.~Liu, X.~Wang, Z.~Chen, Transformer networks for time-series fatigue data
  analysis in steel structures, Structural Health Monitoring 22~(1) (2023)
  252--270.

\bibitem{Pan2024}
R.~Pan, J.~Gao, L.~Meng, F.~Heng, H.~Yang, A new approach to multiaxial fatigue
  life prediction: A multi-dimensional multi-scale composite neural network
  with multi-depth, Engineering Fracture Mechanics~(310) (2024).

\bibitem{Zhang2024}
P.~Zhang, K.~Tang, A.~Wang, H.~Wu, Z.~Zhong, Neural network integrated with
  symbolic regression for multiaxial fatigue life prediction, International
  Journal of Fatigue~(188) (2024).

\bibitem{chen2019}
Z.~Chen, H.~Li, H.~Wang, Probabilistic fatigue life prediction of bridge
  components using bayesian neural networks, Journal of Bridge Engineering
  24~(9) (2019) 04019089.

\bibitem{tan2020}
Z.~Tan, J.~Tao, X.~Han, Z.~Zuo, Deep belief network for fatigue reliability
  assessment of steel bridges, Advances in Structural Engineering 23~(4) (2020)
  526--539.

\bibitem{wang2022}
H.~Wang, H.~Li, Z.~Chen, Federated learning for collaborative fatigue analysis
  of steel structures, Automation in Construction 134 (2022) 104055.

\bibitem{zhang2023}
W.~Zhang, X.~Guo, J.~Wu, Attention-based neural network for multiaxial fatigue
  life prediction in steel components, International Journal of Fatigue 168
  (2023) 107484.

\bibitem{balmer2024design}
V.~Balmer, S.~V. Kuhn, R.~Bischof, L.~Salamanca, W.~Kaufmann, F.~Perez-Cruz,
  M.~A. Kraus, Design space exploration and explanation via conditional
  variational autoencoders in meta-model-based conceptual design of pedestrian
  bridges, Automation in Construction 163 (2024) 105411.

\bibitem{bucher2023performance}
M.~J.~J. Bucher, M.~A. Kraus, R.~Rust, S.~Tang, Performance-based generative
  design for parametric modeling of engineering structures using deep
  conditional generative models, Automation in Construction 156 (2023) 105128.

\bibitem{silva2021}
A.~Silva-Campillo, et~al., Design criteria for scantling of longitudinal and
  transverse connections in the torsion box under fatigue loading, Polish
  Maritime Research (2021).
\newblock \href {https://doi.org/10.2478/pomr-2021-0028}
  {\path{doi:10.2478/pomr-2021-0028}}.

\bibitem{liu2023}
B.~Liu, H.~Xin, J.~Liu, M.~Veljkovic, Fatigue performance evaluation of
  stiffener-to-deck plate weld in orthotropic steel decks based on generative
  adversarial network, Ce/Papers (2023).
\newblock \href {https://doi.org/10.1002/cepa.2730}
  {\path{doi:10.1002/cepa.2730}}.

\bibitem{park2021}
J.~Park, S.-H. Sim, B.~F. Spencer~Jr, Reinforcement learning for optimizing
  inspection schedules of fatigue-prone steel components, Computer-Aided Civil
  and Infrastructure Engineering 36~(7) (2021) 927--942.

\bibitem{kraus2020physik}
M.~A. Kraus, A.~Taras, Physik-informierte k{\"u}nstliche intelligenz zur
  berechnung und bemessung im stahlbau, Stahlbau 89~(10) (2020) 824--832.

\bibitem{kim2021}
B.~Kim, S.~Cho, Physics-informed neural networks for fatigue damage estimation
  in steel beams, Structural Health Monitoring 20~(5) (2021) 2308--2323.

\bibitem{Zhou2023}
T.~Zhou, S.~Jiang, T.~Han, S.-P. Zhu, Y.~Cai, A physically consistent framework
  for fatigue life prediction using probabilistic physics-informed neural
  network, International Journal of Fatigue~(166) (2023).

\bibitem{Jing2024}
G.~Jing, et~al., Physically hierarchical neural network for low cycle fatigue
  life prediction of compacted graphite cast iron based on small data,
  International Journal of Fatigue~(188) (2024).

\bibitem{Dong2025}
Y.~Dong, X.~Yang, D.~Chang, Q.~Li, Predicting fatigue life of multi-defect
  materials using the fracture mechanics-based physics-informed neural network
  framework, International Journal of Fatigue~(190) (2025).

\bibitem{Feng2024}
F.~Feng, T.~Zhu, B.~Yang, S.~Zhou, S.~Xiao, A physics-informed neural network
  approach for predicting fatigue life of slm 316l stainless steel based on
  defect features, International Journal of Fatigue~(188) (2024).

\bibitem{Li2024}
X.~Li, Z.~Fu, J.~Shu, B.~Ji, B.~Ji, A modified physics-informed neural network
  to fatigue life prediction of deck-rib double-side welded joints,
  International Journal of Fatigue~(189) (2024).

\bibitem{Bartosak2025}
M.~Bartosák, et~al., Using physics-informed neural networks to predict the
  lifetime of laser powder bed fusion processed 316l stainless steel under
  multiaxial low-cycle fatigue loading, International Journal of Fatigue~(190)
  (2025).

\bibitem{Halamka2023}
J.~Halamka, M.~Bartosak, M.~Spaniel, Using hybrid physics-informed neural
  networks to predict lifetime under multiaxial fatigue loading, Engineering
  Fracture Mechanics~(289) (2023).

\bibitem{He2023}
G.~He, Y.~Zhao, C.~Yan, Multiaxial fatigue life prediction using
  physics-informed neural networks with sensitive features, Engineering
  Fracture Mechanics~(289) (2023).

\bibitem{ma2023}
Y.~Ma, B.~Wang, A.~Chen, Inspection data-based prediction on fatigue crack of
  orthotropic steel deck using interpretable machine learning method, Fatigue
  \& Fracture of Engineering Materials \& Structures 47~(10) (2024) 3874--3893.

\bibitem{sun2023}
Z.~Sun, J.~Santos, E.~Caetano, Data‐driven prediction and interpretation of
  fatigue damage in a road‐rail suspension bridge considering multiple loads
  (2022).
\newblock \href {https://doi.org/10.1002/stc.2997}
  {\path{doi:10.1002/stc.2997}}.

\bibitem{li2023}
Q.~Li, Z.~Song, Ensemble-learning-based prediction of steel bridge deck defect
  condition (2022).
\newblock \href {https://doi.org/10.3390/app12115442}
  {\path{doi:10.3390/app12115442}}.

\bibitem{serradilla2023}
O.~Serradilla, E.~Zugasti, C.~Cernuda, A.~Aranburu, J.~R. de~Okariz,
  U.~Zurutuza, Interpreting remaining useful life estimations combining
  explainable artificial intelligence and domain knowledge in industrial
  machinery (2020).
\newblock \href {https://doi.org/10.1109/FUZZ48607.2020.9177537}
  {\path{doi:10.1109/FUZZ48607.2020.9177537}}.

\bibitem{movsessian2023}
A.~Movsessian, D.~Cava, D.~Tcherniak, Interpretable machine learning in damage
  detection using shapley additive explanations (2021).
\newblock \href {https://doi.org/10.31224/osf.io/96yf5}
  {\path{doi:10.31224/osf.io/96yf5}}.

\bibitem{peralta2021}
J.~J. Peralta, H.~Fritz, G.~I. Dadoulis, K.~Dragos, An explainable artificial
  intelligence approach for damage detection in structural health monitoring
  (2021).

\bibitem{peralta2022}
J.~J. Peralta, H.~Fritz, G.~I. Dadoulis, K.~Dragos, K.~Smarsly, Automated
  decision making in structural health monitoring using explainable artificial
  intelligence (2021).

\bibitem{Bartsch2018}
H.~Bartsch, M.~Feldmann, Assessment of fatigue tests to review detail
  categories of ec3, in: Proceedings of the IABMAS Conference, CRC Press,
  Melbourne, Australia, 2018, pp. 2220--2227.

\bibitem{Feldmann2019}
M.~Feldmann, H.~Bartsch, U.~Kuhlmann, K.~Drebenstedt, T.~Ummenhofer,
  B.~Seyfried, Auswertung von ermüdungsversuchsdaten zur Überprüfung von
  kerbfallklassen nach ec3-1-9, Stahlbau 88~(10) (2019) 1004--1017.
\newblock \href {https://doi.org/10.1002/stab.201900066}
  {\path{doi:10.1002/stab.201900066}}.

\bibitem{Bartsch2019}
H.~Bartsch, B.~Hoffmeister, M.~Feldmann, Investigations on the fatigue behavior
  of end plate connections with prestressed bolts, Procedia Structural
  Integrity 19 (2019) 395--404.
\newblock \href {https://doi.org/10.1016/j.prostr.2019.12.043}
  {\path{doi:10.1016/j.prostr.2019.12.043}}.

\bibitem{Bartsch2020a}
H.~Bartsch, B.~Hoffmeister, M.~Feldmann, Fatigue analysis of welds and bolts in
  end plate connections of i-girders, International Journal of Fatigue 138
  (2020) 105674.
\newblock \href {https://doi.org/10.1016/j.ijfatigue.2020.105674}
  {\path{doi:10.1016/j.ijfatigue.2020.105674}}.

\bibitem{Bartsch2020b}
H.~Bartsch, K.~Drebenstedt, B.~Seyfried, M.~Feldmann, U.~Kuhlmann,
  T.~Ummenhofer, Analysis of fatigue test data to reassess en 1993-1-9 detail
  categories, Steel Construction 13~(4) (2020) 280--293.
\newblock \href {https://doi.org/10.1002/stco.202000019}
  {\path{doi:10.1002/stco.202000019}}.

\bibitem{Bartsch2021a}
H.~Bartsch, M.~Feldmann, Reassessment of fatigue detail categories of bolts and
  rods according to ec 3-1-9, Journal of Constructional Steel Research 180
  (2021) 106588.
\newblock \href {https://doi.org/10.1016/j.jcsr.2021.106588}
  {\path{doi:10.1016/j.jcsr.2021.106588}}.

\bibitem{Bartsch2021b}
H.~Bartsch, M.~Feldmann, Revision of fatigue detail categories of plain members
  and mechanically fastened joints according to ec 3-1-9, Journal of
  Constructional Steel Research 179~(10) (2021) 106549.
\newblock \href {https://doi.org/10.1016/j.jcsr.2021.106549}
  {\path{doi:10.1016/j.jcsr.2021.106549}}.

\bibitem{Bartsch2021c}
H.~Bartsch, M.~Feldmann, Numerical and databased investigations on the fatigue
  resistance of load-carrying cruciform joints with gaps, Journal of
  Constructional Steel Research 185 (2021) 106843.
\newblock \href {https://doi.org/10.1016/j.jcsr.2021.106843}
  {\path{doi:10.1016/j.jcsr.2021.106843}}.

\bibitem{Bartsch2022a}
H.~Bartsch, S.~Citarelli, M.~Feldmann, Investigations on the fatigue behaviour
  of welded-in stiffeners with gaps, Journal of Constructional Steel Research
  189 (2022) 107075.
\newblock \href {https://doi.org/10.1016/j.jcsr.2021.107075}
  {\path{doi:10.1016/j.jcsr.2021.107075}}.

\bibitem{BartschNN2024}
H.~Bartsch, J.~Voelkel, M.~Feldmann,
  \href{https://onlinelibrary.wiley.com/doi/abs/10.1002/stco.202400029}{Developing
  artificial neural networks to estimate the fatigue strength of structural
  steel details using the new european database}, Steel Construction n/a~(n/a)
  (2025).
\newblock \href
  {http://arxiv.org/abs/https://onlinelibrary.wiley.com/doi/pdf/10.1002/stco.202400029}
  {\path{arXiv:https://onlinelibrary.wiley.com/doi/pdf/10.1002/stco.202400029}},
  \href {https://doi.org/https://doi.org/10.1002/stco.202400029}
  {\path{doi:https://doi.org/10.1002/stco.202400029}}.
\newline\urlprefix\url{https://onlinelibrary.wiley.com/doi/abs/10.1002/stco.202400029}

\bibitem{tukey1977exploratory}
J.~W. Tukey, Exploratory data analysis, Reading, Mass., Reading, MA, 1977.

\bibitem{little2019statistical}
R.~J.~A. Little, D.~B. Rubin, Statistical analysis with missing data, 3rd
  Edition, Vol. 793, John Wiley \& Sons, 2019.

\bibitem{guyon2003introduction}
I.~Guyon, A.~Elisseeff, An introduction to variable and feature selection,
  Journal of Machine Learning Research 3~(Mar) (2003) 1157--1182.

\bibitem{vanderplas2016python}
J.~VanderPlas, Python data science handbook: Essential tools for working with
  data, O'Reilly Media, Inc., 2016.

\bibitem{james2023introduction}
G.~James, D.~Witten, T.~Hastie, R.~Tibshirani, J.~Taylor, An introduction to
  statistical learning: With applications in python, Springer Nature, 2023.

\bibitem{hyndman2018forecasting}
R.~J. Hyndman, G.~Athanasopoulos, Forecasting: principles and practice, OTexts,
  2018.

\bibitem{smyth1997stacked}
P.~Smyth, D.~Wolpert, Stacked density estimation, Advances in neural
  information processing systems 10 (1997).

\bibitem{montgomery2021introduction}
D.~C. Montgomery, E.~A. Peck, G.~G. Vining, Introduction to linear regression
  analysis, John Wiley \& Sons, 2021.

\bibitem{breiman2001random}
L.~Breiman, Random forests, Machine learning 45 (2001) 5--32.

\bibitem{hastie2009elements}
T.~Hastie, The elements of statistical learning: data mining, inference, and
  prediction (2009).

\bibitem{geurts2006extremely}
P.~Geurts, D.~Ernst, L.~Wehenkel, Extremely randomized trees, Machine learning
  63 (2006) 3--42.

\bibitem{ke2017lightgbm}
G.~Ke, Q.~Meng, T.~Finley, T.~Wang, W.~Chen, W.~Ma, Q.~Ye, T.-Y. Liu, Lightgbm:
  A highly efficient gradient boosting decision tree, Advances in neural
  information processing systems 30 (2017).

\bibitem{chen2016xgboost}
T.~Chen, C.~Guestrin, Xgboost: A scalable tree boosting system, in: Proceedings
  of the 22nd acm sigkdd international conference on knowledge discovery and
  data mining, 2016, pp. 785--794.

\bibitem{prokhorenkova2018catboost}
L.~Prokhorenkova, G.~Gusev, A.~Vorobev, A.~V. Dorogush, A.~Gulin, Catboost:
  unbiased boosting with categorical features, Advances in neural information
  processing systems 31 (2018).

\bibitem{goodfellow2016deep}
I.~Goodfellow, Y.~Bengio, A.~Courville, Y.~Bengio, Deep learning, Vol.~1, MIT
  press Cambridge, 2016.

\bibitem{bishop2006pattern}
C.~M. Bishop, N.~M. Nasrabadi, Pattern recognition and machine learning,
  Vol.~4, Springer, 2006.

\bibitem{mljar}
A.~P\l{}o\'{n}ska, P.~P\l{}o\'{n}ski,
  \href{https://github.com/mljar/mljar-supervised}{Mljar: State-of-the-art
  automated machine learning framework for tabular data. version 0.10.3}
  (2021).
\newline\urlprefix\url{https://github.com/mljar/mljar-supervised}

\bibitem{kraus2024strength}
M.~A. Kraus, R.~Bischof, H.~Riedel, L.~Schmeiser, A.~Pauli, I.~Stelzer,
  M.~Drass, Strength lab ai: a mixture-of-experts deep learning approach for
  limit state analysis and design of monolithic and laminate structures made of
  glass, Glass Structures \& Engineering 9~(3) (2024) 607--655.

\end{thebibliography}

\end{document}